# The Growth of Galaxy Stellar Haloes Over $0.2 \leq z \leq 1.1$

DEVIN J. WILLIAMS,[1] IVANA DAMJANOV,[1] MARCIN SAWICKI,[1] HARRISON SOUCHEREAU,[2,1] LINGJIAN CHEN,[1] GUILLAUME DESPREZ,[1] ANGELO GEORGE,[1] MARIANNA ANNUNZIATELLA,[3] STÉPHANE ARNOUTS,[4] STEPHEN GWYN,[5] DANILO MARCHESINI,[6] AND ANNA SAJINA[6]

[1]*Institute for Computational Astrophysics and Department of Astronomy & Physics, Saint Mary's University, 923 Robie Street, Halifax, NS B3H 3C3, Canada*
[2]*Department of Astronomy, Yale University, New Haven, CT 06511, USA*
[3]*Centro de Astrobiología (CSIC-INTA), Ctra de Torrejón a Ajalvir, km 4, E-28850 Torrejón de Ardoz, Madrid, Spain*
[4]*Aix Marseille Université, CNRS, CNES, LAM, 38 rue Frédéric Joliot-Curie, Marseille 13388 cedex 13, France*
[5]*NRC Herzberg Astronomy and Astrophysics, 5071 West Saanich Road, Victoria, BC V9E 2E7, Canada*
[6]*Department of Physics and Astronomy, Tufts University, Medford, MA 02155, USA*

## ABSTRACT

Galaxies are predicted to assemble their stellar haloes through the accretion of stellar material from interactions with their cosmic environment. Observations that trace stellar halo buildup probe the processes that drive galaxy size and stellar mass growth. We investigate stellar halo assembly over $0.2 \leq z \leq 1.1$ in a mass-complete ($M_\star \geq 10^{9.5} M_\odot$) sample of 242,456 star-forming and 88,421 quiescent galaxies (SFGs and QGs) from the CLAUDS and HSC-SSP surveys. We extract galaxy rest-frame $g$-band surface brightness ($\mu_g$) profiles to study faint, extended emission in galaxy outskirts. We examine trends in galaxy assembly by analyzing the median $\mu_g$ profiles in different SFG and QG $M_\star$ ranges with decreasing redshift and connecting evolution in galaxy $\mu_g$ profiles with the underlying stellar mass growth in galaxies. Since $z = 1.1$, the majority of evolution in the median $\mu_g$ profiles of galaxies ($\sim$64% in SFGs and $\sim$71% in QGs) occurs throughout their stellar halo regions (2-10$R_e$). More massive galaxies assemble stellar halo material more rapidly at $0.2 \leq z \leq 1.1$. Over this period, QGs grow a larger fraction of their stellar haloes than SFGs at fixed $M_\star$ (factor of $\sim$1.2). Although star formation can account for the stellar halo growth observed in low-mass SFGs ($10^{9.5} M_\odot \leq M_\star < 10^{10.5} M_\odot$), high-mass SFGs ($M_\star \geq 10^{10.5} M_\odot$) and both low- and high-mass QGs require an additional assembly mechanism. Our results suggest accretion via minor mergers drives additional stellar halo growth in these galaxies. The contribution from accretion is larger in more massive galaxies (over $M_\star \geq 10^{9.5} M_\odot$), and QGs exhibit larger fractional increases to their ex-situ fractions over $0.2 \leq z \leq 1.1$ than SFGs at fixed $M_\star$.

## 1. INTRODUCTION

The evolution of galaxies over cosmic timescales is a result of both internally- and externally-driven processes (or *nature* vs. *nurture*; Hickson 1997; De Lucia et al. 2019; Shi et al. 2024). Galaxies form new stars from their supply of dynamically cold gas, and this internally regulated star formation fuels the stellar mass and structural growth of galaxies throughout their lifetimes (e.g., Bergin & Tafalla 2007; McKee & Ostriker 2007). Once the star formation is quenched (see e.g., Man & Belli 2018 for an extensive list of quenching mechanisms), a galaxy will continue to undergo passive evolution. Passive evolution of QGs includes changes to galaxy structure (e.g., size and morphology) and stellar properties (e.g., metallicity or colour) due to secular processes (e.g., bar dynamics, disk heating, growth of pseudobulges; Kormendy 1993; Kormendy & Kennicutt 2004; Merrifield et al. 2001; Khoperskov & Bertin 2017; Géron et al. 2024) and the aging of its underlying stellar population (Bouwens & Silk 1996; Franx et al. 2000; Strateva et al. 2001).

Galaxies can also experience a more active form of evolution where interactions with neighbouring galaxies in their cosmic environment drive changes to their structure and stellar content. These interactions include close encounters (e.g., flybys or harassment) and mergers where both galaxies combine into a single system (e.g., Moore et al. 1996; Lambas

Email: devin.williams@smu.ca

et al. 2012; Huang et al. 2013; Ownsworth et al. 2014). The outcomes of these interactions can vary significantly depending on the properties of the galaxies involved. Different types of mergers are defined based on the mass ratio between the host ($M_h$) and merging ($M_m$) galaxy (i.e. $\alpha_\star = M_m/M_h$) and whether the merging galaxies are gas-rich (a *wet* merger) or gas-poor (a *dry* merger; e.g., Bell et al. 2006; Conselice et al. 2022; Huško et al. 2022)

During a major merger ($\alpha_\star \geq 0.25$), galaxies accrete significant amounts of stars and/or gas which can get deposited throughout their outer regions or funnelled into their inner regions (e.g., Lambas et al. 2012; Zhu et al. 2022; Montenegro-Taborda et al. 2023). Major mergers can induce significant morphological transformations often resulting in dispersion-dominated elliptical morphologies (Martig et al. 2009; Jackson et al. 2022; Graham 2023). In comparison, during minor mergers ($\alpha_\star < 0.25$), galaxies accrete stellar material from smaller companions and distribute it primarily throughout their outer regions resulting in much more subtle morphological transformations than major mergers (e.g., Lambas et al. 2012; Hilz et al. 2013; Jackson et al. 2019, 2022). This extended accretion from minor mergers drives a greater increase in galaxy sizes than in total stellar mass (e.g., $R_e \propto M^{2-2.5}$), in contrast to the roughly one-to-one increase (i.e. $R_e \propto M$) expected from major mergers (e.g., Naab et al. 2009; Bezanson et al. 2009; Hopkins et al. 2010; Trujillo et al. 2011; Kaviraj et al. 2014).

In wet mergers, the abundant gas can trigger an enhancement of star formation (e.g., a starburst) and AGN activity in the host galaxy due to gas inflows caused by tidal forces generated during the interaction (e.g., Ellison et al. 2013, 2018, 2020; Wilkinson et al. 2022; Li et al. 2023). Based on predictions from the Horizon-AGN cosmological simulation wet minor mergers play a significant role in assembling high-mass disk galaxies due to their ability to enhance disk components rather than destroy them (Jackson et al. 2020). Studies of massive ($M_\star \gtrsim 10^{11} M_\odot$) disk galaxies in the local universe (i.e. $z < 0.1$) confirm this prediction (Kaviraj 2014; Jackson et al. 2022). These studies find tidal features and star formation rate (SFR) enhancements in massive disk galaxies that are best explained by previous interactions with small gas-rich companions.

In dry mergers, star formation enhancement is minimal due to the scarcity of gas involved. Based on observations of massive ($M_\star \gtrsim 10^{11} M_\odot$) QGs and early-type galaxies (ETGs) at high and low redshifts (e.g., $0 < z < 2$), dry major mergers are driving the stellar mass assemblies of these massive quiescent systems (e.g., Bell et al. 2006; van Dokkum et al. 2010; Bernardi et al. 2011). This is due to the large amounts of accretion required to explain the high stellar masses of these galaxies and the fact that wet gas-rich mergers would lead to a significant enhancement in SFRs. Similarly, the accelerated size growth observed in the QG population from high to low redshift (e.g., a factor of 2.5-4 increase in size since $z$ = 1-2) has been attributed to dry minor mergers based on the larger size increase expected from minor mergers and the inability of dry mergers to increase SFR significantly (Daddi et al. 2005; Trujillo et al. 2007; Trujillo & Bakos 2013a; van der Wel et al. 2014; Damjanov et al. 2019; George et al. 2024).

Despite our progress in understanding how different interactions impact the galaxies involved, more work is needed to better understand which types of mergers drive the majority of galaxy size and stellar mass growth at different redshifts and test predictions of the contribution from accretion in different types of galaxies (e.g., Naab et al. 2009; Trujillo et al. 2011; Ownsworth et al. 2014; Peschken et al. 2020; Cannarozzo et al. 2022; Giri et al. 2023). At $z = 0$, accreted stellar mass fractions (i.e. *ex-situ* fractions) in low-mass galaxies ($M_\star \lesssim 10^{9.5-10} M_\odot$) are predicted to be fairly small (~1-20%), but can be significantly larger (~50-90%) in massive ($M_\star \sim 10^{12} M_\odot$) galaxies (e.g., Cook et al. 2016; Tacchella et al. 2019; Cannarozzo et al. 2022; Huško et al. 2022). Additionally, different simulations disagree on whether merger-driven growth is increased in QGs relative to SFGs at fixed mass (e.g., see Fig. 5 in Rodriguez-Gomez et al. 2016 and Fig. 2 in Davison et al. 2020).

Due to long dynamical timescales in galaxy outskirts, signatures of a galaxy's past interactions with its environment can be seen throughout its stellar halo regions (e.g., Martínez-Delgado et al. 2008; Iodice et al. 2017; Dey et al. 2023; Jensen et al. 2024; Perrotta et al. 2024). Studying the assembly of this stellar halo material in galaxies thus enables us to probe the impact of mergers and quantify the contribution from accretion to galaxy stellar mass growth. One method of analyzing this faint extended stellar halo emission in galaxies is to measure their radial surface brightness ($\mu$) profiles (referred to as galaxy *light* or $\mu$ profiles interchangeably throughout this text; Buitrago et al. 2017; Huang et al. 2018; Wang et al. 2019; Spavone et al. 2020, 2021; Gilhuly et al. 2022).

Cosmological galaxy simulations identify galaxy light profiles as efficient tracers of assembly activity (e.g., Hopkins et al. 2010; Hilz et al. 2013; Hirschmann et al. 2015; Cook et al. 2016). Flatter gradients in light profiles of simulated galaxies have been attributed to merger-driven stellar mass growth, a result of accreted stellar material representing a larger fraction of the total stellar mass contained at larger galactocentric distances (Hilz et al. 2013; Pillepich et al. 2014; Cook et al. 2016; Rodriguez-Gomez et al. 2016). Steeper $\mu$ gradients are instead predicted to arise from assembly via star formation as lower SFRs in galaxy outskirts lead to fewer stars formed at larger radii (e.g., Cook et al. 2016; Chamba et al. 2022; Trujillo et al. 2020). Observa-tionally, galaxies with flatter $\mu$ gradients have been shown to exhibit higher ex-situ fractions based on small samples of massive ($M_\star \sim 10^{12} M_\odot$) ETGs in the local universe ($z \sim 0.03$; Spavone et al. 2021). Ex-situ fractions in this study are estimated by fitting a galaxy's $\mu$ distribution with three Sérsic components and taking the fraction of total light within the two outer components to be the galaxy's accreted fraction. Additionally, in galaxies of $M_\star > 10^{10} M_\odot$ at $z \leq 0.1$, D'Souza et al. (2014) find that low-concentration galaxies (disk-like galaxies, majority of which are SFGs) exhibit steeper gradients than high-concentration galaxies (spheroid-like galaxies, majority of which are QGs).

Integrated galaxy light profiles within specific radial regions (e.g., 0-1 $R_e$, or 10-100 kpc), coupled with the conversion between $L$ and $M_\star$ via an assumption of galaxy $[M_\star/L]$ ratios, provide the estimates of stellar mass contained within different galaxy physical components (e.g., bulges vs. stellar haloes). Based on predictions from cosmological simulations, the stellar mass contained beyond a certain radius (e.g., $2R_e$, 20 kpc) can be used as a proxy for the non-observable ex-situ fraction of a galaxy (e.g., Elias et al. 2018; Merritt et al. 2020). In previous studies of the most massive ($M_\star \gtrsim 10^{11} M_\odot$) QGs over a limited redshift range ($0.3 < z < 0.67$), the stellar mass fraction beyond 10 kpc agrees with predicted ex-situ fractions from the Illustris and IllustrisTNG simulations (Buitrago et al. 2017; Huang et al. 2018). However, these studies do not consider lower mass QGs ($M_\star \lesssim 10^{11} M_\odot$) or the SFG population.

In this work, we aim to address limitations of previous studies on galaxy stellar haloes that have been constrained to massive galaxy samples ($M_\star \gtrsim 10^{11} M_\odot$; e.g., van Dokkum et al. 2010; Ownsworth et al. 2014; Buitrago et al. 2017; Huang et al. 2013, 2018), or low redshift intervals ($z \lesssim 0.1$; e.g., D'Souza et al. 2014; Merritt et al. 2016a; Iodice et al. 2016, 2017; Spavone et al. 2017, 2020, 2021; Ann & Park 2018; Gilhuly et al. 2022). We study galaxy stellar halo assembly over a wide redshift range ($0.2 \leq z \leq 1.1$) in a large mass-complete sample ($M_\star \geq 10^{9.5} M_\odot$) of 242,456 SFGs and 88,421 QGs from the CLAUDS (Sawicki et al. 2019) and HSC-SSP (Aihara et al. 2022) surveys.

We extract galaxy light profiles from deep multi-wavelength (~4000-10000 Å rest-frame) photometric observations to analyze the extended stellar halo emission in galaxy outskirts and connect the evolution in galaxy light profiles with the underlying stellar mass growth in galaxies. We compute median light profiles of different SFG and QG $M_\star$ ranges and compare the evolution observed in their gradients and integrated quantities with decreasing redshift. We investigate the contributions to galaxy stellar halo growth from different assembly mechanisms (e.g., star formation vs. accretion), and test predictions from cosmological simulations regarding merger-driven growth in galaxies.

This paper is organized as follows. Sec. 2 describes the photometric datasets and catalogues we use. In Sec. 3 we outline the computational procedures we use to extract individual galaxy light profiles. Sec. 4 details our methods of computing median light profiles of different galaxy subpopulations and quantifying their evolution over redshift. We present our main results in Sec. 5, with further discussion and comparisons with predictions from cosmological simulations in Sec. 6. We summarize our main conclusions in Sec. 7. Throughout this work, magnitudes are quoted in the AB system and a ΛCDM cosmological model with $\Omega_M$ = 0.3, $\Omega_\Lambda$ = 0.7, and $H_0$ = 70 km s$^{-1}$Mpc$^{-1}$ is assumed.

## 2. DATA AND SAMPLE SELECTION

To investigate the faint stellar haloes of galaxies we must be able to detect very low surface brightness features ($\mu$ levels of ~30-31 mag/arcsec$^2$; e.g., Trujillo & Fliri 2016; Merritt et al. 2020; Li et al. 2021; Genina et al. 2023). Furthermore, we require a sufficiently large sample that covers a wide range of redshift and stellar mass to study trends in galaxy assembly over large timescales. To fulfill these requirements, we use the combined datasets and photometric catalogues of two large-area imaging surveys - the Hyper Suprime-Cam Subaru Strategic Program (HSC-SSP, Aihara et al. 2018) and the CFHT Large Area U-band Deep Survey (CLAUDS, Sawicki et al. 2019).

In Sec. 2.1 and Sec. 2.2 we describe the two surveys and the galaxy properties obtained from the photometric catalogues. In Sec. 2.3 we outline our final galaxy sample selection, and in Sec. 4.2 we discuss surface brightness limits reached in median $\mu$ profiles.

### 2.1. *HSC-SSP and CLAUDS Surveys*

We use observations from the Deep and UltraDeep layers of the third Public Data Release (PDR3[1]; Aihara et al. 2022) of HSC-SSP. Images were taken with the Hyper Suprime-Cam on the 8.2m Subaru Telescope (NAOJ) using a set of broadband photometric filters ($g$, $r$, $i$, $z$, and $y$) that span a rest-frame wavelength range of ~4000-10000 Å. The Deep layer (26 deg$^2$) includes observations from four large and widely separated fields (XMM-LSS, E-COSMOS, ELAIS-N1, and DEEP2-3). The UltraDeep layer (3.5 deg$^2$) consists of two smaller fields embedded within a particular Deep field (SXDS inside XMM-LSS and COSMOS inside E-COSMOS). The large areas covered by the multiple fields ensure a wide variety of galaxy environments are sampled which helps alleviate the effect of cosmic variance (Driver & Robotham 2010).

[1] PDR3 Data from HSC-SSP can be obtained at: https://hsc-release.mtk.nao.ac.jp/doc/index.php/data-access__pdr3/.

The CLAUDS[2] survey covers the same Deep (18.6 deg$^2$) and UltraDeep (1.36 deg$^2$) fields as HSC-SSP and extends the wavelength coverage into the ultraviolet (UV) regime. CLAUDS observations were taken by the MegaCam imager on the 3.6m Canada-France-Hawaii Telescope (CFHT) using two UV-band filters $u$ and $u^*$ (rest-frame ∼3000-4100 Å). The addition of CLAUDS $U$-band data helps improve the accuracy of galaxy photometric redshift measurements (Sawicki et al. 2019; Desprez et al. 2023). Throughout this text, we use "$U$-band" to refer to both (or either) of the $u$ and $u^*$ MegaCam filters (as described in Sawicki et al. 2019) and refer to the combined set of CLAUDS+HSC-SSP filters as $U + grizy$.

The individual images we use are 4200x4200 pixels in size (∼0.2 deg$^2$ with CLAUDS+HSC-SSP pixel scale of 0.168″/pixel). Images have been processed for scientific analysis through the image processing pipelines of both surveys (for HSC-SSP pipeline details see Bosch et al. 2018, and for CLAUDS see Gwyn 2008; Sawicki et al. 2019). We select HSC-SSP images produced using the global sky subtraction procedure as it better preserves the faint emission in galaxy outskirts (Aihara et al. 2019, 2022). Both surveys achieve excellent seeing quantified as the full-width half maximum (FWHM) of the point-spread functions (PSFs). The median seeing FWHM of each $U+grizy$ filter is 0.92″, 0.83″, 0.77″, 0.66″, 0.78″, and 0.70″, respectively. In Sec. 3.1.2 we describe our procedure for correcting for the effects of filter PSFs.

### 2.2. *Galaxy Properties and Additional Datasets*

We select galaxies for our sample from the CLAUDS+HSC-SSP photometric catalogues of Desprez et al. 2023 (`hscPipe`/`Phosphoros` versions) which cover a combined ∼18 deg$^2$ over the four Deep fields of both surveys (Sec. 2.1). In addition to CLAUDS+HSC-SSP $U + grizy$ coverage, the catalogues use auxiliary data from the VIDEO (Jarvis et al. 2013) and UltraVISTA (McCracken et al. 2012) surveys which provide near-infrared wavelength coverage from the VIRCAM instrument ($Y$, $J$, $H$, and $Ks$ bands) over a combined 5.5 deg$^2$ across the XMM-LSS and E-COSMOS fields. Additionally, longer wavelength data from the SHIRAZ Survey (Annunziatella et al. 2023, IRAC 3.6$\mu$m and 4.5$\mu$m) covers 17 deg$^2$ across three of the four Deep fields (all but XMM-LSS).

Photometric properties for galaxies in our sample are derived from `hscPipe` cmodel fluxes and are corrected for Galactic extinction. Galaxy photometric redshifts are computed via spectral energy distribution (SED) fitting via `Phosphoros` by Euclid Collaboration et al. (2020); Desprez et al. (2023) using 2″ aperture photometry in the six $U + grizy$ bands. Galaxy templates used during the fitting come from the library of Ilbert et al. (2013), which consists of 33 SEDs representing different types of galaxies including spirals, ellipticals, and starbursts. Errors on photometric redshifts are on the order of $\sigma_{(1+z)} \sim 0.04$, but precision gradually declines with magnitude ($\sigma_{MAD} \sim 0.03$ at $m_i \leq 22.5$, and $\sigma_{MAD} \sim 0.09$ at $m_i \leq 26$) as does the outlier fraction ($\eta \sim 4\%$ to $\eta \sim 29\%$). Outliers are defined as having $|\Delta z| \geq 0.15$ where $\Delta z = (z_{phot} - z_{spec})/(1 + z_{spec})$.

Additional galaxy properties are computed via `LePhare` SED fitting by Chen et al. (in prep.) and compiled in a value-added CLAUDS+HSC-SSP catalogue. Specifically, we obtain galaxy stellar masses ($M_\star$), star-formation rates (SFR, $M_\odot$/yr), and absolute magnitudes from these `LePhare` runs. We use absolute magnitudes to compute global rest-frame $(U-g)$ colours of galaxies in our sample which we use to estimate stellar mass-to-light ratios ($M_\star/L$, Sec. 6.1.1). To improve the accuracy of derived properties, Chen et al. (in prep.) only includes galaxies inside the longer wavelength SHIRAZ coverage which are above a certain detection limit in both IRAC filters (3.6$\mu$m and 4.5$\mu$m). Thus we omit galaxies found in the XMM-LSS field from our analysis (reducing total area to ∼12 deg$^2$).

During the `LePhare` runs, Chen et al. (in prep.) fixes the galaxy redshift to the value obtained from `Phosphoros`. The authors construct a library of galaxy models from SED templates obtained from Bruzual & Charlot (2003). They assume a Chabrier initial mass function (IMF; Chabrier 2003), and include metallicities of $Z \sim 0.004 - 0.2$ and exponentially decaying star-formation histories (SFR $\propto e^{-\tau}$) with timescales of $\tau$ = 0.01-3 Gyr (0.01, 0.03, 0.1, 0.2, 0.3, 0.5, 1.0, 1.5, and 3 Gyr). Three different dust attenuation laws are applied following the methodology of Moutard et al. (2016).

Galaxies in our sample are classified as star-forming (SFG) or quiescent (QG) based on their position on a rest-frame $(NUV - r)$ vs. $(r - K)$ colour-colour diagram ($NUVrK$; Arnouts et al. 2013). $NUVrK$ diagrams, similar to UVJ diagrams (Williams et al. 2009), use different colour excesses as proxies for the level of star formation in a galaxy. $NUVrK$ diagrams are better at distinguishing between the reddening effects of dust and stellar ageing than UVJ diagrams by extending further into infrared wavelengths (Moutard et al. 2018, 2020). The $NUVrK$ classification procedure used on galaxies in our sample, described in Chen et al. (in prep.), employs machine learning to minimize the contamination between the two populations (i.e. SFGs and QGs). The procedure is optimized by training on the galaxy sample of Weaver et al. (2023), where all galaxies with log(sSFR) $< -12$ are classified as quiescent. The boundary separating QGs and SFGs in the $NUVrK$ diagram changes as a smooth function of stellar mass and redshift.

[2]Data from CLAUDS can be obtained at: https://www.clauds.net/available-data.

### 2.3. *Final Galaxy Sample Selection*

We obtain an initial sample of galaxies that lie within the E-COSMOS, ELAIS-N1, and DEEP2-3 fields where both CLAUDS and HSC-SSP observations overlap (Fig. 3 of Sawicki et al. 2019 shows pointings of both surveys). We limit galaxies to apparent $i$-band magnitudes of $m_i \leq 25$ AB as the photometric redshifts become unreliable at fainter magnitudes due to larger photometric uncertainties (Desprez et al. 2023). We limit the stellar mass range for galaxies in our sample to $M_\star \geq 10^{9.5} M_\odot$. We choose this lower $M_\star$ limit based on stellar mass completeness of 90% from the study of satellite galaxy number density distributions and their detection limits within the CLAUDS+HSC-SSP datasets by Chen (2019).

We restrict our sample to the photometric redshift range of $0.2 \leq z \leq 1.1$. We omit redshifts of $z < 0.2$ as the uncertainties on the photometric redshifts ($\sigma_{(1+z)} \sim 0.04$) are close to the redshift values. We determine the upper redshift limit based on the wavelength coverage of the $U + grizy$ filters (rest-frame ∼3000-10000 Å) and our decision to trace rest-frame $g$-band emission in galaxies across our full redshift range (Sec. 4.1). Emission at these wavelengths (e.g., rest-frame ∼5000-6000 Å) traces the populations of low-mass stars in galaxies which form the bulk of their stellar mass (e.g., Pagel & Edmunds 1981; Bruzual & Charlot 2003; Moutard et al. 2018; Huang et al. 2018). At redshifts of $z > 1.1$ this rest-frame $g$-band emission is no longer captured by the reddest filter available (i.e. the HSC-SSP $y$-band filter).

We refine our sample by applying a series of quality cuts to different CLAUDS+HSC-SSP catalogue parameters following the recommendation of Desprez et al. (2023) and Chen et al. (in prep.). We eliminate contamination from point source objects (e.g., stars and quasars) misclassified as galaxies by applying `isStar`=False and `isCompact`=False. We remove galaxies whose images have optical defects that interfere with photometry measurements such as satellite trails or nearby bright star masks by setting `isOutsideMask`=1. We remove galaxies with unreliable photometry measurements in the form of failed cmodel magnitudes due to negative fluxes by setting `CMODEL_FAIL_FLAG`< 2.

Table 1 summarizes the sample limits and quality cuts applied to the initial sample and the number of remaining galaxies after each step. Following these restrictions, 330,877 galaxies remain in our final sample, of which 242,456 (73%) are classified as SFGs and 88,421 (27%) as QGs.

**Table 1.** Summary of the sample limits and quality cuts (discussed in Sec. 2.3) applied to the initial sample of galaxies retrieved from the CLAUDS+HSC-SSP photometric catalogues (Desprez et al. 2023). An additional flag beyond those shown is applied (`isOutsideMask`=1), but the number of galaxies removed is included in the initial sample value.

| **Sample Limit / Quality Cut** | **Value Range** | **# Galaxies Remaining** |
|---|---|---|
| **Initial sample** | - | **4,875,177** |
| Magnitude | $m_i \leq 25$ AB mag | **2,927,983** |
| Stellar Mass | $M_\star \geq 10^{9.5} M_\odot$ | **1,722,263** |
| Redshift | $0.2 \leq z \leq 1.1$ | **397,331** |
| `isStar` | = False | **358,646** |
| `isCompact` | = False | **344,786** |
| `CMODEL_FAIL_FLAG` | < 2 | **330,877** |

## 3. INDIVIDUAL LIGHT PROFILES OF CLAUDS+HSC-SSP GALAXIES

In this section, we discuss our procedure for extracting individual galaxy $\mu$ profiles from their HSC-SSP $grizy$ images. Throughout Sec. 3.1 we explain several corrections we apply to galaxy images to account for different forms of light contamination. In Sec. 3.2 we describe the methodology behind our profile extraction procedure and introduce the computational tool we use for bulk extractions.

### 3.1. *Image Corrections*

#### 3.1.1. *Source Masking and Background Subtraction*

To accurately measure the low $\mu$ levels in galaxy outskirts we must correct for light contamination from foreground and background sources in the image as well as any sky-subtracted background noise that remains (e.g., Szomoru et al. 2012; Trujillo & Fliri 2016; Gilhuly et al. 2022). This correction is important as the excess emission can artificially brighten galaxy $\mu$ profile outskirts (e.g., ∼1-2 mag/arcsec$^2$; Li et al. 2021). In this work, we use the python package `GalPRIME`[3] to apply source masking and background subtraction to galaxy images. Here we briefly describe both procedures, illustrating the steps in Fig. 1.

The source masking procedure starts with a raw galaxy cutout (panel 1A) and creates a segmentation map of all other objects in the image where pixels measured to be $1\sigma$ (`NSIGMA` = 1) above the median background intensity are flagged as belonging to an object. We create object masks (panel 1B) that include all sources that have at least 11 connected pixels (`NPIX` = 11). We convolve these masks with a `Gaussian2DKernel` with a width of $2\sigma$ (`GAUSS_WIDTH` = 2) to smooth the jagged outer edges of the masks caused by noise contamination. We create an additional smoothed mask for the target galaxy (panel 1C) to omit it from the estimation of background levels.

Following the source masking, we subtract the background using `GalPRIME`'s 2D background subtraction method.

[3] Documentation on `GalPRIME` can be found at: https://galprime.readthedocs.io/en/latest/

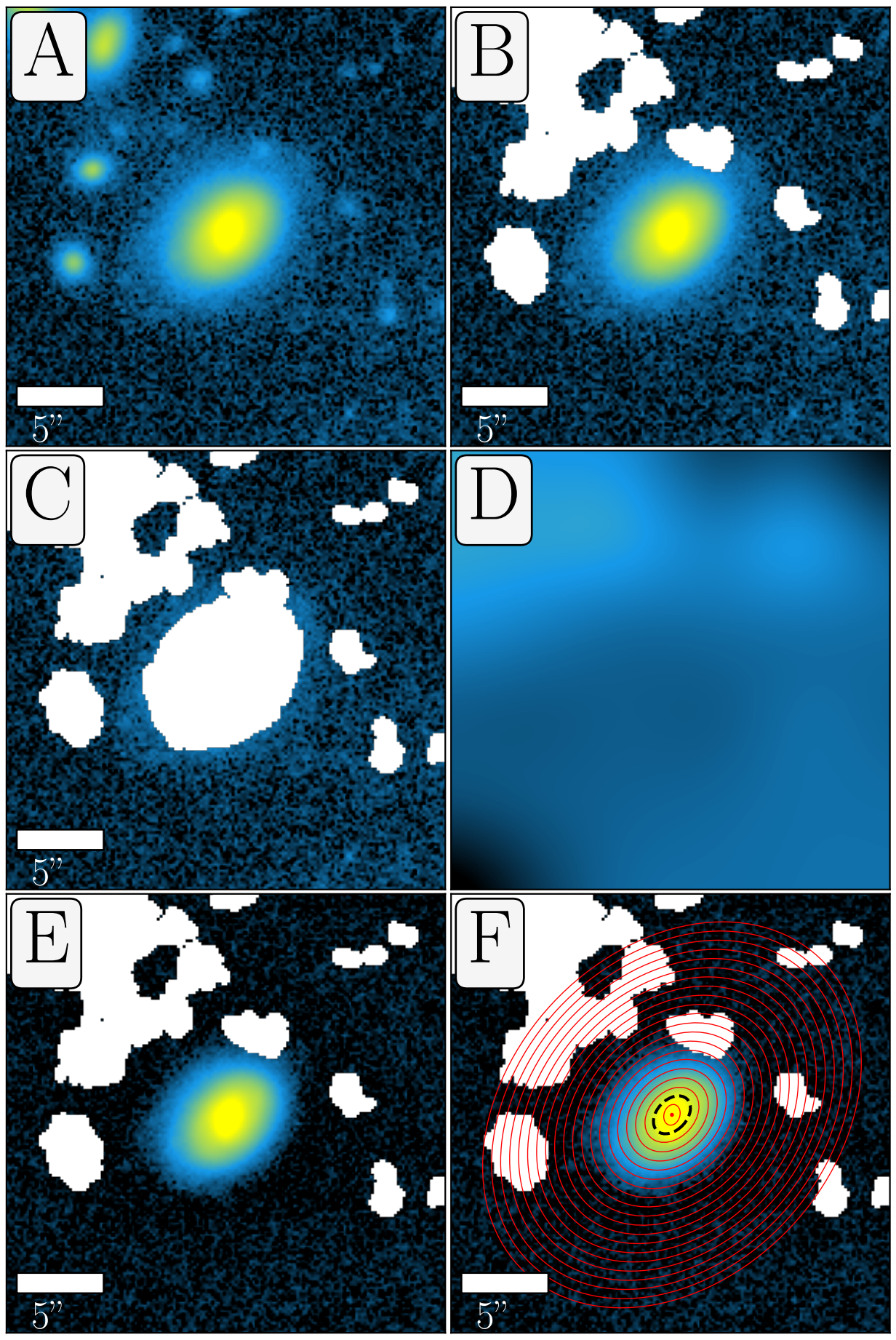

**Figure 1.** Illustration of the source masking and 2D background (BG) subtraction procedures in `GalPRIME`. Panel A: $i$-band image ($25'' \times 25''$) of a SFG with $M_\star = 10^{10.9} M_\odot$ at $z = 0.23$. Panel B: Masks (white pixels) created for nearby sources. Panel C: Mask created for the target galaxy to omit from BG calculations. Panel D: 2D BG levels estimated from the masked image. Panel E: BG-subtracted and masked galaxy cutout. Panel F: Final corrected cutout with fitted isophotes (red) used for $\mu$ profile extraction. The galaxy's size ($R_e$ = 6.7 kpc, Sec. 4.3) is represented by the black dashed isophote. Image intensity levels have been adjusted to better highlight the variations in the 2D background and to increase the visibility of the of background subtraction effects.

This 2D subtraction technique performs well when background emission levels vary across an image, as they typically do in real galaxy images (Astropy Collaboration et al. 2013). In this method, a mesh grid is generated over the original galaxy image (omitting regions within the source masks) and median background intensity levels are computed within each box of the grid (`BOX_SIZE` $= 41$) using the `sigma_clipped_stats` package ($\sigma = 3$ and 5 iterations) from `Astropy`. The 2D image of varying background levels that is created is then median filtered using a filter window size of 6 pixels (`FILTER_SIZE` $= 6$) to help suppress local fluctuations due to overly bright sources in a particular box. We subtract this 2D background image (panel 1D) from the masked galaxy cutout (panel 1B) to produce the final masked and background-corrected galaxy image (panel 1E) that is used later during $\mu$ profile extraction (panel 1F, procedure described in Sec. 3.2).

Both the source masking and 2D background subtraction routines in `GalPRIME` can be adjusted by tuning different input parameters discussed here (i.e., `NSIGMA`, `NPIX`, `GAUSS_WIDTH`, `BOX_SIZE`, and `FILTER_SIZE`). Our chosen parameter values are based on tests of `GalPRIME`'s performance using simulated galaxies inserted into CLAUDS+HSC-SSP images (Souchereau et al. in prep.). The additional analysis we describe in Appendix A demonstrates that the variations in these chosen parameters have negligible effects on the integrated stellar halo luminosity of galaxies in our sample (explored in Sec. 5.4).

#### 3.1.2. *Correcting for the PSF*

Before extracting galaxy $\mu$ profiles (Sec. 3.2) we correct for the effect of filter-specific point-spread functions (PSF). This correction is needed as the PSF will suppress central $\mu$ levels in galaxies and redistribute this light out to larger radii, increasing the $\mu$ levels measured in the outer portions of $\mu$ profiles (e.g., ~0.5-1 mag/arcsec$^2$; Sandin 2014, 2015; Wang et al. 2019). The contribution from this effect depends both on the photometric filter used for observation and the specific properties of individual galaxies (e.g., Trujillo et al. 2001; de Jong 2008; Szomoru et al. 2012; Borlaff et al. 2017; Gilhuly et al. 2022).

To correct for PSF-related effects, we adopt the procedure of Borlaff et al. (2017), where a PSF-convolved galaxy model is fit to a raw galaxy image to isolate and subtract the effect of the PSF. This procedure has been used in previous observational studies of galaxy stellar haloes and $\mu$ profiles (e.g., Szomoru et al. 2010, 2012; Trujillo & Bakos 2013b; Trujillo & Fliri 2016; Gilhuly et al. 2022). This subsection gives a brief overview of our computational implementation of the PSF correction procedure. In Appendix B we describe tests of our procedure using simulated galaxies and discuss how results depend on different filters and galaxy parameters.

To accurately measure the contribution from the PSF in galaxy outskirts, we require PSF models that extend to similar sizes as galaxy images (de Jong 2008; Sandin 2014; Trujillo & Fliri 2016; Gilhuly et al. 2022). We obtain raw PSF models for the $grizy$ filters from the HSC-SSP PDR3 (Aihara et al. 2022) PSF Picker[4] tool. We follow the same methodology as George et al. (2024) and segment each HSC-SSP image into 36 equal regions, using one PSF model to represent each region. We extend these initial HSC-SSP PSFs to match a galaxy cutout size by fitting a three-component `Astropy` model to the raw PSFs. In Appendix B we discuss the components of this model and describe tests of the fit in different $grizy$ filters.

[4] PSF Picker can be found at the HSC-SSP PDR3 (Aihara et al. 2022) website: https://hsc-release.mtk.nao.ac.jp/doc/index.php/data-access__pdr3/.

For each galaxy in our sample, we use coordinates from the CLAUDS+HSC-SSP catalogues (Sec. 2.2) to find the closest matching PSF model, as the smearing effect induced by PSFs can vary with position. We create a simulated galaxy model that is a combination of two Sérsic (Sérsic 1963) components (using `Sersic2D` from `Astropy`), as two-component models outperformed one- and three-component models in tests of our procedure (Appendix B). We convolve this simulated model with the extended $grizy$ PSF selected for the galaxy and fit it to the original input image using the `fit_model` function from `PetroFit`. This function improves the fit between the PSF-convolved model and original galaxy image through iterations by adjusting `Sersic2D` parameters for both components until incremental improvements fall below the threshold level[5] of $10^{-9}$.

We allow `fit_model` to select any combination of `Sersic2D` parameters for the two Sérsic components that produce the best fit, even if the parameter combinations do not correspond to typical galaxy components seen in 2D decompositions of galaxies (e.g., a pure disk or bulge component; Simard et al. 2011). Once the procedure converges to the best-fit solution we subtract the PSF-convolved simulated model from the original galaxy image to produce residuals which capture features absent from the parametric fits (e.g., irregular morphologies, low surface brightness features). We add these residuals to the underlying non-convolved galaxy model to produce the final PSF-corrected galaxy image we use to extract radial $\mu$ profiles (Sec. 3.2). The steps of this PSF correction procedure are summarized in Eq. 5 and 6 of Borlaff et al. (2017).

We test our PSF correction procedure using 15000 simulated galaxies in Appendix B. The main conclusion from these tests is that our procedure achieves similar success in all $grizy$ filters within uncertainties, quantified by the root-mean-squared error (RMSE) between the initial simulated galaxy $\mu$ profile and the final PSF-corrected $\mu$ profile. We obtain a median RMSE of $\sim 0.10^{+0.06}_{-0.04}$ mag/arcsec$^2$ across all tests. Additionally, we identify no strong correlations between RMSE and any input galaxy parameters. We conclude that our procedure introduces no bias across different types of galaxies in our sample.

### 3.2. *The Extraction of Galaxy Light Profiles*

We extract galaxy $\mu$ profiles using `GalPRIME`'s wrapper for the `photutils` implementation of the elliptical isophote analysis method of Jedrzejewski (1987). This method, illustrated in Fig. 2, fits a series of elliptical isophotes to an image (red ellipses, right panel) to represent the observed $\mu$ distribution of a galaxy as a function of distance from its center.

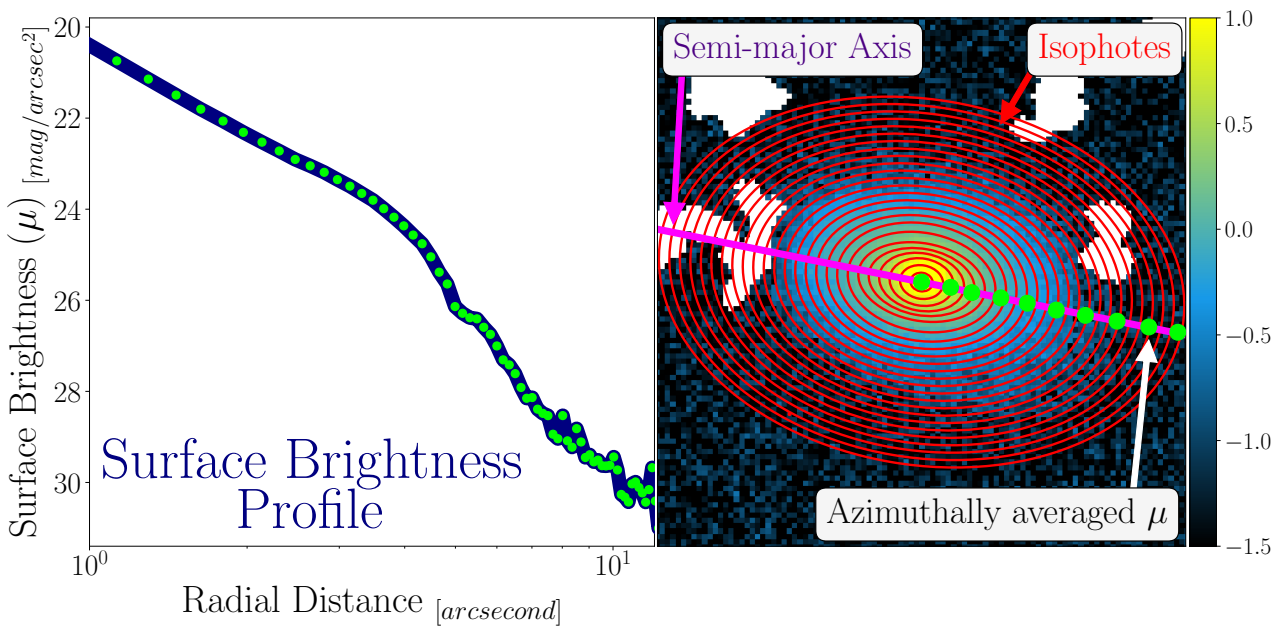

**Figure 2.** Illustration of elliptical isophote analysis (Jedrzejewski 1987). Left: Surface brightness profile (blue) extracted from the galaxy image. Right: HSC-SSP $i$-band image of a galaxy fit with isophotes (red). Each point in the radial profile (green dots, left panel) represents the mean value along an isophote (i.e. the azimuthally averaged value) at a given semi-major axis length (purple line, right panel). The blue shaded regions (left panel) represent uncertainties on the azimuthally averaged values and are sometimes smaller than the width of the profile shown. For visual clarity, the number of isophotes shown has been reduced. In practice, isophotes are more numerous and finely spaced.

The fitting procedure allows for variations in the ellipticities and position angles of isophotes with increasing radius along the major axis. Each point in the radial $\mu$ profile (green points, left panel) represents the mean $\mu$ value along an individual isophote (i.e. the *azimuthally averaged* value) at that semi-major axis distance (purple line, right panel), which increases the signal-to-noise ratio (Jedrzejewski 1987). Uncertainties on the averaged $\mu$ values grow larger with increasing radial distance (blue shaded region, left panel), typically ranging from $\sim 0.05$ mag/arcsecond$^2$ in the inner regions of profiles to $\sim 1$ mag/arcsecond$^2$ in the outermost regions.

We extract $\mu$ profiles from individual galaxy cutouts ($grizy$ images) that are $250 \times 250$ pixels ($42'' \times 42''$) in size for the majority of our sample. For lower redshift ($0.2 \leq z < 0.35$) galaxies, we use larger cutouts of $300 \times 300$ pixels ($50.4'' \times 50.4''$) to account for their larger apparent sizes (e.g., Li et al. 2021). A small fraction ($\sim$2%) of $\mu$ profile extractions failed for individual galaxies. These failures are caused by artifacts in the images not caught by the quality cuts applied to our sample (Table 1), or issues from the source masking procedure implemented in `GalPRIME` (Sec. 3.1.1). The number of galaxies removed from our sample due to these failures is included in the initial sample number in Table 1.

Extracted $\mu$ values in profiles are converted to AB magnitudes/arcsecond$^2$ using the zero-point offset of HSC-SSP (27 mag) and corrected for cosmological surface bright-ness dimming ($\mu \propto (1+z)^{-3}$ when using AB magnitudes; Whitney et al. 2020). Additionally, we use absolute solar magnitudes in $grizy$ bands obtained from Willmer (2018) to convert $\mu$ values into $L_{\odot}$/pc$^2$ units. We also convert the major-axis profiles that result from `GalPRIME` into circularized profiles via $R = \sqrt{ab}$ (also called *geometric mean* profile; Graham & Driver 2005), where $a$ and $b$ represent the major and minor axes, respectively.

[5] $acc$ parameter from `Astropy`'s LevMarLSQFitter: https://docs.astropy.org/en/stable/api/astropy.modeling.fitting.LevMarLSQFitter.html

## 4. MEDIAN LIGHT PROFILES OF DIFFERENT GALAXY SUBPOPULATIONS

In this section, we describe our procedure for computing median rest-frame $g$-band $\mu$ profiles of different galaxy subpopulations. In Sec. 4.1 we divide our galaxy sample into smaller bins based on stellar mass ($M_{\star}$), photometric redshift ($z$), and star-formation activity (i.e. QG vs. SFG), and outline our method of tracing rest-frame $g$-band emission across the full redshift range ($0.2 \leq z \leq 1.1$). In Sec. 4.2 we discuss $\mu$ limits and compute median profiles for the various $M_{\star}$ + redshift bin combinations (i.e. subpopulations) in the SFG and QG samples. Lastly, in Sec. 4.3 we summarize our methods of quantifying and comparing the evolution in median profiles between different galaxy subpopulations.

### 4.1. *Tracing The Rest-Frame $g$-Band Light Profiles of $0.2 \leq z \leq 1.1$ Galaxies*

To study trends in stellar mass assembly for different galaxy subpopulations we divide our SFG and QG samples into smaller bins based on $M_{\star}$ and redshift (Table 2). We create our redshift bins to trace the rest-frame $g$-band emission in galaxies throughout the full redshift interval ($0.2 \leq z \leq 1.1$). Emission at these wavelengths ($\sim$ 4000-5500Å) traces the long-lived lower-mass stars that form the bulk of a galaxy's stellar mass. Previous studies have also targeted a similar rest-frame wavelength range to study galaxy stellar haloes and $\mu$ profiles (e.g., Spavone et al. 2017, 2021; Huang et al. 2018; Gilhuly et al. 2022).

We use different observed broadband filters (i.e. $grizy$, see Fig. 2 in Aihara et al. 2018) to trace the same approximate wavelength range (centred at $\sim$5000Å) in each bin following the relation $\lambda_g = \lambda_{obs} \cdot (1+z)^{-1}$, where $\lambda_g$ and $\lambda_{obs}$ represent rest-frame $g$-band and observed-band wavelengths, respectively, and $z$ represents galaxy photometric redshift. The four specific filters used for this process are $r, i, z$, and $y$-band for low to high redshift bins (top two rows in Table 2). Throughout the rest of this text, we use $\mu_g$ to refer to galaxy surface brightness in the rest-frame $g$-band. We note that we do not make corrections to extracted $\mu_g$ values in profiles to account for differences in $grizy$ filter bandpasses (i.e. $K$ corrections; Hogg et al. 2002), but discuss their potential impact on our results in Sec. 6.4.

We divide galaxies in our mass-complete sample ($M_{\star} \geq 10^{9.5} M_{\odot}$) into four separate $M_{\star}$ bins (first column in Table 2). An influential factor that determines our $M_{\star}$ bins is the *pivot mass* ($M_p \sim 10^{10.5\pm0.4} M_{\odot}$) of observed galaxy size-stellar mass relations which marks the transition into a steeper size-stellar mass relation slope for more massive galaxies (e.g., Lange et al. 2015; Mowla et al. 2019a; Kawinwanichakij et al. 2021; Damjanov et al. 2022; George et al. 2024). This change in slope has been interpreted as due to an increased influence of merger-driven accretion in more massive galaxies, based on predicted ex-situ fractions of galaxies in cosmological simulations (e.g., Rodriguez-Gomez et al. 2016; Tacchella et al. 2019; Huško et al. 2022; Davison et al. 2020).

We construct two $M_{\star}$ bins above and two below the pivot mass to study the different mechanisms that drive galaxy assembly in the two mass regimes. Throughout this text, we refer to the two upper $M_{\star}$ bins collectively as the *high-mass* ($M_{\star} \geq 10^{10.5} M_{\odot}$) galaxy sample, and the two lower $M_{\star}$ bins as the *low-mass* ($10^{9.5} M_{\odot} \leq M_{\star} < 10^{10.5} M_{\odot}$) galaxy sample.

### 4.2. *Computing Median Light Profiles*

We compute median $\mu_g$ profiles of galaxies in each $M_{\star}$ and redshift bin combination of our SFG and QG samples (Table 2), using a procedure illustrated in Fig. 3. Median light profiles (dark blue in Fig. 3) are representative of the majority of individual light profiles (grey in Fig. 3) of galaxies in that bin. The scatter between individual profiles provides us with information on the range of assembly histories of galaxies in that subpopulation. We choose to use median profiles rather than mean profiles as they are less affected by outliers (i.e. the brightest or faintest grey profiles in Fig. 3), which is especially important at radii where some individual profiles drop below our surface brightness limit ($\mu_{lim}$, brown horizontal line in Fig. 3).

We compute bootstrapped uncertainties on median $\mu_g$ profiles via `sklearn`'s `resample` package. This bootstrap method refers to sampling with replacement, where 10000 bootstrapped samples of median profiles are computed and the standard deviation of that distribution is used as the $1\sigma$ bootstrapped error on the median profile. Bootstrapped errors on the median are relatively small ($1\sigma \lesssim 0.1$ mag/arcsec$^2$) across all subpopulations. We also compute 25 and 75 percentiles (cyan dashed profiles in Fig. 3), which represent the intrinsic scatter in the distribution of individual galaxy profiles from a given subpopulation.

For each median $\mu_g$ profile, we compute a limiting radius ($R_{lim}$, vertical brown line in Fig. 3) beyond which any emission is likely dominated by residual background noise and is omitted from analysis. We determine this limit by finding the radius at which $\mu$ levels in the median profile drop below the median background level ($\mu_{lim}$). Median background levels are calculated from individual galaxy background lev-els which are computed via `sigma_clipped_stats` after we apply source masking and 2D background subtraction to images (Sec. 3.1.1). For our high to low redshift bins median $\mu_g$ profiles are limited to $\mu_{lim}$ = 30.15, 30.60, 30.85, and 30.95 mag/arcsecond$^2$. Bootstrapped errors on median $\mu_{lim}$ values are very small ($1\sigma \approx$ 0.002-0.004 mag/arcsec$^2$).

**Table 2.** The number of SFGs and QGs in the stellar mass and photometric redshift bins we define. The full CLAUDS+HSC-SSP sample spans $0.2 \leq z \leq 1.1$ and $M_\star \geq 10^{9.5} M_\odot$ (Sec. 2). We use different observed-band filters (second row) to trace a similar rest-frame $g$-band wavelength range in each redshift bin.

| **Redshift Bins:** | **0.2 $\leq z <$ 0.35** | **0.35 $\leq z <$ 0.7** | **0.7 $\leq z <$ 0.9** | **0.9 $\leq z \leq$ 1.1** |
|---|---|---|---|---|
| Observed-band filter used | HSC-$r$ | HSC-$i$ | HSC-$z$ | HSC-$y$ |
| **Stellar Mass Bins** | **Star-forming Galaxies (SFG)** | | | |
| $10^{9.5} M_\odot \leq M_\star < 10^{10} M_\odot$ | 8340 | 41867 | 42762 | 41671 |
| $10^{10} M_\odot \leq M_\star < 10^{10.5} M_\odot$ | 4409 | 23762 | 22638 | 20097 |
| $10^{10.5} M_\odot \leq M_\star < 10^{11} M_\odot$ | 1867 | 10546 | 9433 | 8497 |
| $M_\star \geq 10^{11} M_\odot$ | 395 | 2391 | 2037 | 1744 |
| | **Quiescent Galaxies (QG)** | | | |
| $10^{9.5} M_\odot \leq M_\star < 10^{10} M_\odot$ | 1546 | 6095 | 2660 | 979 |
| $10^{10} M_\odot \leq M_\star < 10^{10.5} M_\odot$ | 2087 | 11064 | 7728 | 5859 |
| $10^{10.5} M_\odot \leq M_\star < 10^{11} M_\odot$ | 2260 | 12596 | 10374 | 8538 |
| $M_\star \geq 10^{11} M_\odot$ | 1044 | 6060 | 4950 | 4581 |

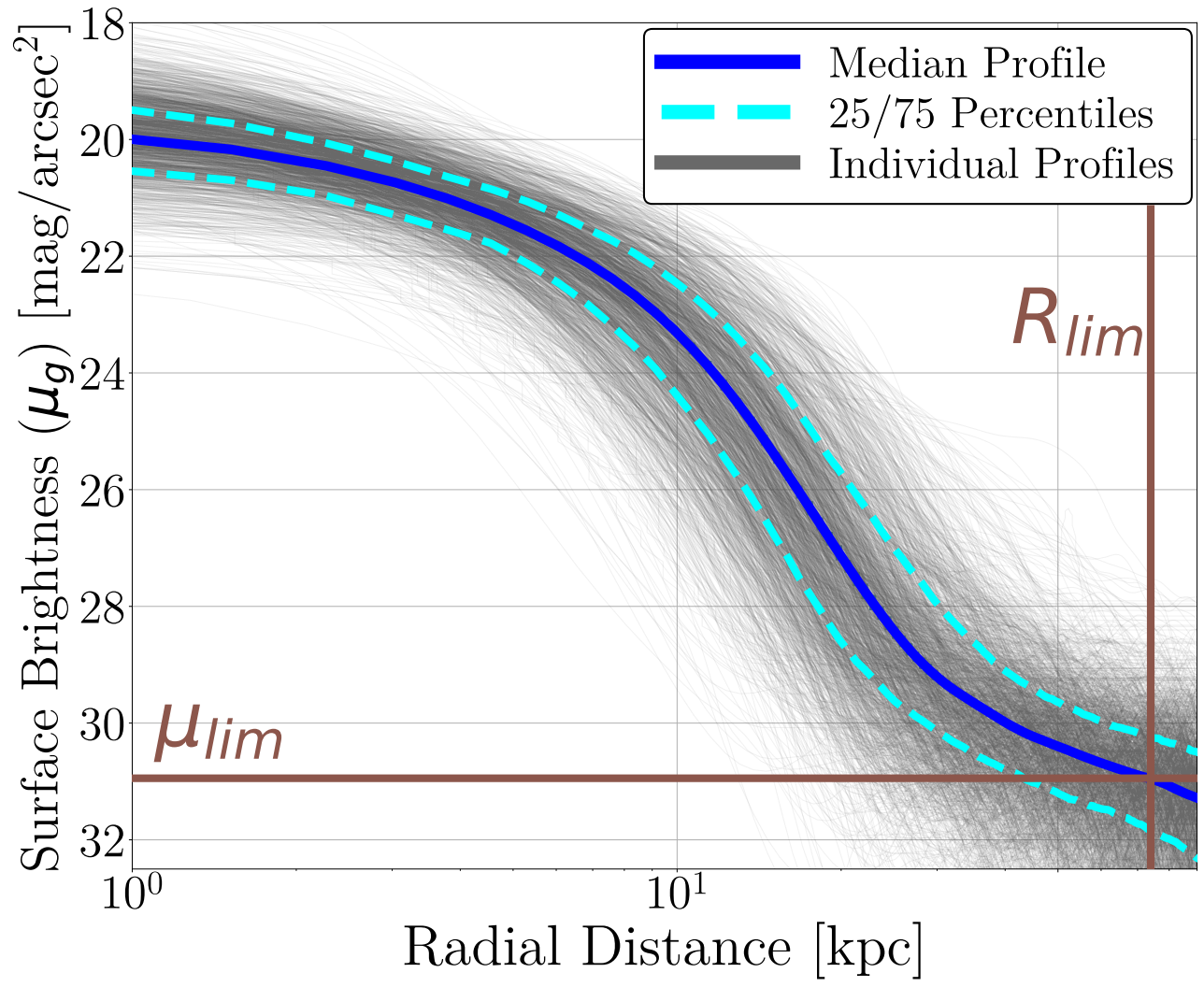

**Figure 3.** Example of a median rest-frame $g$-band surface brightness profile (blue) calculated from the individual profiles (grey) of a sample of SFGs ($10^{10} M_\odot \leq M_\star < 10^{10.5} M_\odot$, $0.2 \leq z < 0.35$). Dashed cyan lines represent the 25/75 percentile range. Bootstrapped errors on the median are roughly the width of the blue median profile shown. The brown horizontal line represents the surface brightness limit ($\mu_{lim}$). The vertical brown line indicates the radius ($R_{lim}$) where surface brightness values in the median profile have reached $\mu_{lim}$.

Differences in $\mu_{lim}$ across our four redshift bins arise from variations in the background level measured from images in different $grizy$ filters, due to the different transmission functions of each filter and the depths reached in their images (Aihara et al. 2022). The varying limits reached at different redshifts do not affect the results presented in our study, as we use $10R_e$ (size measurements described in Sec. 4.3) as the largest radius to which all median profiles extend. We choose $10R_e$ as it lies within $R_{lim}$ for all subpopulations. Truncating median $\mu_g$ profiles at $10R_e$ results in $\mu$ levels in profile outskirts being $> 5\sigma$ above the background for all subpopulations in our sample except for $M_\star \geq 10^{11} M_\odot$ galaxies in the highest and lowest redshift bins ($0.2 \leq z < 0.35$ and $0.9 \leq z \leq 1.1$ ), where $\mu$ levels in profile outskirts are $\gtrsim 2.5\sigma$ above the background.

To estimate the fraction of light in galaxies we are omitting by truncating median $\mu_g$ profiles at $10R_e$, we create different simulated galaxy $\mu$ profiles and measure the percentage of integrated light that extends beyond $10R_e$. Regardless of $\mu$ profile type (e.g., 1- vs. 2-component, or bulge- vs. disk-dominated), the fraction of light beyond $10R_e$ is $\lesssim$ 2-3% even when the radial distance we reach is $\sim$ 40-50$R_e$.

### 4.3. *Analyzing Evolution in Galaxy Light Profiles*

To study the evolution in median $\mu_g$ profiles of galaxies as a function of redshift we analyze profiles within radial regions that correspond to galaxy physical components (e.g., bulges, stellar haloes). We choose to define different regions based on the median $R_e$ within a given galaxy subpopulation (bins in Table 2) to facilitate testing theoretical predictions with our observed results. In their study of galaxy stellar halo assembly in the Illustris simulation, Cook et al. (2016) define the stellar halo region as 2-4$R_e$ (or $R > 2R_e$ if observations are sufficiently deep to surpass $4R_e$). Hirschmann et al. (2015) define the stellar halo region as 2-6$R_e$ in their analysis of stellar population gradients in simulated galaxies. Merritt et al. (2020) propose the fraction of stellar mass beyond $2R_e$ as a proxy for the stellar halo mass based on their study of galaxy stellar haloes in IllustrisTNG and the Dragonfly Nearby Galaxies Survey (Merritt et al. 2016b).

Based on these previous definitions, we adopt $R \geq 2R_e$ as the *stellar halo* region. We briefly discuss caveats of this chosen definition in Sec. 6.4; a comprehensive investigation of other suitable definitions for the stellar halo region is beyond the scope of this study. We use $10R_e$ as the limiting radius of our median $\mu_g$ profiles (Sec. 4.2) and define the stellar halo as the region between 2 and $10R_e$ from the galactic center. In addition, we define $R < 2R_e$ simply as the *inner galaxy* region which may contain different galaxy components (e.g., bulges, bars, or disks) depending on the type of galaxy (e.g., QG or SFG).

We obtain effective radii from the $\mu_g$ profiles of individual galaxies based on a curve of growth procedure (Eq. 1). The total luminosity within some radius ($R$) is defined as

$$L(< R) = \int_0^R \mu_g(R')2\pi R' dR', \quad (1)$$

where $\mu_g$ represents the rest-frame $g$-band surface brightness of the galaxy (in $L_\odot$/pc$^2$ units), $R$ represents the radial distance from the galaxy's center (in parsecs), and the $2\pi R dR$ factor is the surface area element. The $R_e$ of a galaxy is defined as the radius where the integrated area under the light profile has reached half of the total light (i.e. $L(R_e)$ = $0.5L_{tot}$). We note again that we use geometric mean profiles (where $R = \sqrt{ab}$, Sec. 3.2), and thus the effective radii we obtain are circularized radii represented as $R_e$ throughout this text.

We compute median $R_e$ (rest-frame $g$-band) for each $M_\star$ and redshift bin in our SFG and QG samples and use these to separate the different radial regions (i.e. inner galaxy vs. stellar halo) where we analyze median $\mu_g$ profile evolution. For an in-depth discussion of galaxy size evolution and size-mass relations in the CLAUDS+HSC-SSP datasets, see George et al. (2024). In Appendix C we compare our median sizes with others from the literature. We find our sizes are in good agreement ($\lesssim$ 1-2$\sigma$) with those from previous studies (e.g., van der Wel et al. 2014; Roy et al. 2018; George et al. 2024) and follow similar trends with decreasing redshift.

To quantify the evolution in median $\mu_g$ profiles within the defined radial regions, we measure changes in the integrated luminosity (Eq. 1) contained within a given region. Additionally, we calculate profile gradients, which enables us to connect to predictions from simulations that suggest different galaxy assembly processes influence $\mu$ gradients in different ways (e.g., Hopkins et al. 2010; Hilz et al. 2013; Pillepich et al. 2014; Hirschmann et al. 2015; Cook et al. 2016). The $\mu$ gradients we compute are defined as

$$\nabla\mu_g = \frac{d\log\mu_g}{d\log R}, \quad (2)$$

where $R$ and $\mu_g$ represent the same quantities as in Eq. 1. For each median profile and galactocentric region within it, we calculate an array of gradients using `NumPy`'s `gradient` function and use its mean value for the subsequent analysis.

## 5. RESULTS

### 5.1. *Median Rest-Frame $g$-band Light Profiles of SFGs and QGs*

Figure 4 and Figure 5 show the median $\mu_g$ profiles of our SFG and QG samples, respectively. Each panel shows a median profile of a different $M_\star$ and redshift bin combination ($M_\star$ increases downward, $z$ decreases rightward). The x-axis in each panel is displayed in units of median $R_e$ (Sec. 4.3) so that all median profiles span a common normalized radial range (limited to $10R_e$, Sec. 4.2). Dashed vertical lines in each panel represent 1, 2, 4, and 10 $R_e$ for a given bin and highlight the smaller radial regions where we analyze profile evolution.

All galaxy subpopulations in the SFG and QG samples (rows in Fig. 4 and Fig. 5) display growth (i.e. an increase in rest-frame $g$-band luminosity) in their median $\mu_g$ profiles over a period of $\sim$4.5 Gyr (from the centres of our lowest and highest redshift bins, $z \sim 0.275$ and $z \sim 1$). To quantify this growth at a given stellar mass, we integrate the median $\mu_g$ profiles via Eq. 1 to obtain the change in total rest-frame $g$-band luminosity ($L_{g,\,tot}$) between the highest and lowest redshift bins (i.e. $0.2 \leq z < 0.35$ and $0.9 \leq z \leq 1.1$).

In the low to high $M_\star$ bins, the QG sample increases in total luminosity, $L_{g,\,tot}$, by a factor of 2.66$\pm$0.04, 2.90$\pm$0.04, 3.17$\pm$0.06, and 3.27$\pm$0.08, while the SFG sample increases $L_{g,\,tot}$ by a factor of 2.38$\pm$0.03, 2.46$\pm$0.04, 2.86$\pm$0.05, and 3.41$\pm$0.07. Thus, more massive galaxies (lower rows in Fig. 4 and Fig. 5) exhibit more profile evolution over $0.2 \leq z \leq 1.1$ (left to right columns). At fixed stellar mass, QGs exhibit more evolution in their median $\mu_g$ profiles than SFGs except at the highest stellar masses ($M_\star \geq 10^{11} M_\odot$).

More profile evolution occurs at larger radii ($R \gtrsim 2R_e$) over $0.2 \leq z \leq 1.1$ than at smaller $R$. This trend in radial distance is independent of stellar mass or population type (i.e. SFG or QG). The innermost regions of median $\mu_g$ profiles (e.g., $R < 0.5R_e$) show little to no change in $\mu$ levels over the full redshift range. In Sec. 5.3 we quantify and analyze the growth in different regions of the median $\mu_g$ profiles further.

Our results confirm those found in small samples of massive ($M_\star \geq 10^{10.7} M_\odot$) galaxies in the redshift range $0 < z < 2$ by van Dokkum et al. (2010) and Patel et al. (2013). Both these studies find the majority of growth in galaxy surface mass density profiles since $z = 2$ occurs in the outer regions of the profiles. Li et al. (2021) used HSC-SSP observations to compute median $r$-band $\mu$ profiles for massive ETGs ($M_\star \sim 10^{11.6-12.1} M_\odot$) over a smaller redshift range of $z$ = 0.19-0.50. Galaxy $\mu$ profiles in that study exhibit growth with decreasing redshift which is most pronounced in profile outskirts (see their Fig. 6). Our results at low redshift ($0.2 \leq z < 0.35$, last column in Fig. 4 and Fig. 5) align with findings from D'Souza et al. (2014) who studied $M_\star > 10^{10} M_\odot$ galaxies in lower redshift ($0.06 \leq z \leq 0.1$) SDSS data. Similar to our results, in that study more massive galaxies exhibit an excess of $\mu$ in the outskirts of their $r$-band $\mu$ profiles compared to the profiles of lower $M_\star$ galaxies.

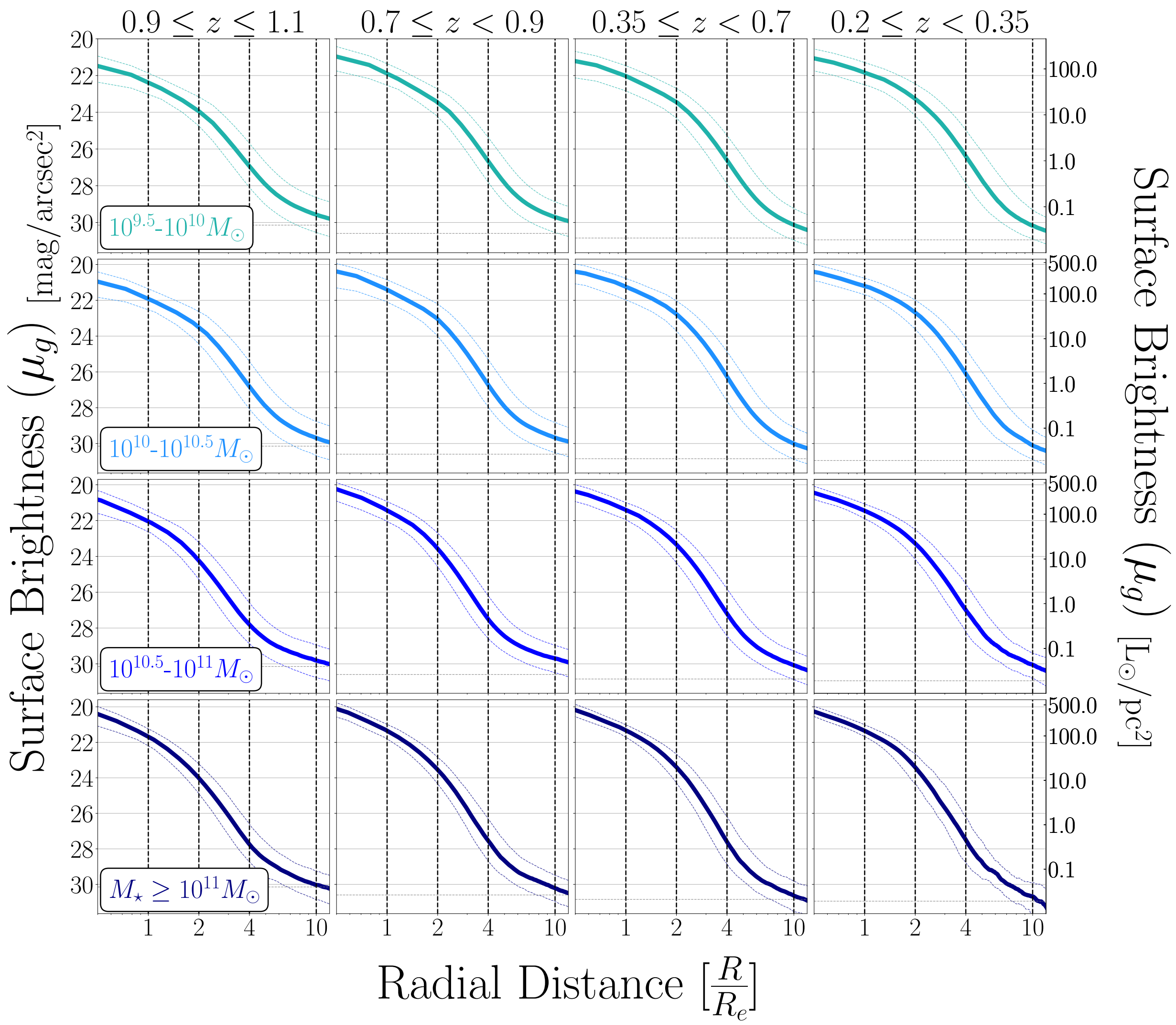

**Figure 4.** Median rest-frame $g$-band surface brightness profiles of our *star-forming galaxy (SFG)* sample. Rows and colours represent different $M_\star$ bins (text boxes, first column), and columns represent redshift bins, as labelled. Solid coloured lines in each panel represent median profiles. Widths of median profiles represent bootstrapped errors on the median, while coloured shaded regions show the 25 and 75 percentiles. The x-axis in each panel is shown in units of median $R_e$ for a given subpopulation ($M_\star$ and redshift bin combination). Grey-shaded regions indicate the surface brightness limit in a given redshift range ($\mu_{lim}$, Sec. 4.2). Vertical dashed black lines indicate 1, 2, 4, and 10 $R_e$.

In summary, we find that over $0.2 \leq z \leq 1.1$, more massive galaxies in both our SFG and QG samples display more evolution in their median $\mu_g$ profiles (i.e. larger increases to $L_{g,\,tot}$) than do lower $M_\star$ galaxies. QGs exhibit more profile evolution than SFGs at fixed $M_\star$, except in the highest stellar mass range we study ($M_\star \geq 10^{11} M_\odot$). Most of the evolution in median $\mu_g$ profiles occurs throughout the outer regions of galaxies ($R \geq 2R_e$). Our results are in agreement with previous studies but cover a larger range of redshift and stellar mass than previously studied.

### 5.2. *Evolution in Surface Brightness Gradients*

Based on predictions from hydrodynamical cosmological simulations, $\mu$ gradients exhibit different changes depending on the processes driving the assembly of stellar material in galaxies (e.g., Hopkins et al. 2010; Hilz et al. 2013; Pillepich et al. 2014; Hirschmann et al. 2015). In the Illustris simulation, flatter $\mu$ gradients within 1-2$R_e$ and 2-4$R_e$ are driven by stellar mass growth through accretion, while steeper gradients arise from in-situ growth via star formation (Cook et al. 2016).

To further analyze the evolution in median $\mu_g$ profiles shown in Fig. 4 and Fig. 5, and to connect our results with predictions, we measure gradients ($\nabla\mu_g$, Sec. 4.3) within an inner galaxy (1-2$R_e$) and stellar halo (2-4$R_e$) region. We also measure gradients in profile outskirts (4-10$R_e$, not shown), but are unable to use their trends with redshift to distinguish between assembly mechanisms in different galaxy subpopulations. This is because the outer gradients of all subpopulations grow steeper over time simply due to very little luminosity being added in the distant outskirts of profiles ($R \sim 10R_e$) compared to more inner regions ($R \sim 4R_e$). This similarity in the shape of $\mu$ profile outskirts between different galaxy subpopulations is also seen in galaxies of $10^{9.2} \leq M_\star \leq 10^{11.4} M_\odot$ at lower redshifts ($0 \leq z < 0.28$) in HSC-SSP data (Wang et al. 2019).

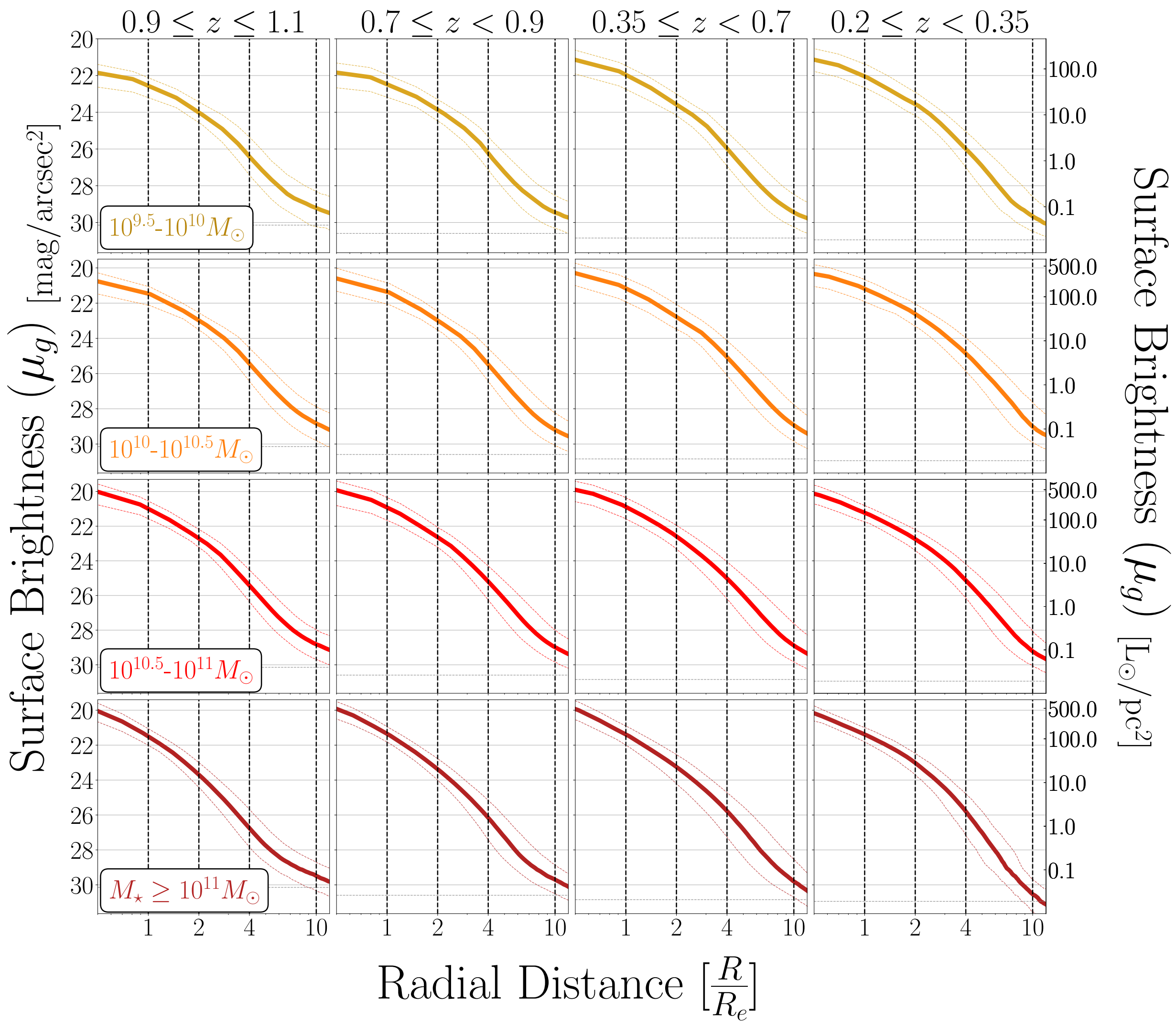

**Figure 5.** Same as Fig. 4, but for the *quiescent galaxy (QG)* sample.

Figure 6 shows the median inner galaxy (top row) and stellar halo (bottom row) $\mu_g$ gradients of our SFG and QG samples as a function of redshift, with colours corresponding to the same $M_\star$ bins as the median $\mu_g$ profiles in Figs. 4 and 5. Open symbols in each panel represent low redshift ($0.03 \leq z \leq 0.34$) counterparts from the literature and are colour-coded to match the comparable $M_\star$ bin in our sample.

In the inner galaxy regions of SFGs (panel 6A), more massive galaxies (darker colours) have steeper gradients at all redshifts. Over $0.2 \leq z \leq 1.1$ the gradients of low-mass SFGs (two lighter colours) become steeper over time by a factor of ∼1.44 ($10^{9.5} M_\odot \leq M_\star < 10^{10} M_\odot$ bin) and ∼1.26 ($10^{10} M_\odot \leq M_\star < 10^{10.5} M_\odot$ bin). The gradients of high-mass SFGs (two darker colours) grow only slightly steeper over the full redshift range, changing by a factor of ∼1.09 ($10^{10.5} M_\odot \leq M_\star < 10^{11} M_\odot$ bin) and ∼1.06 ($M_\star \geq 10^{11} M_\odot$ bin). However, within 25/75 percentiles on median gradients (∼0.08-0.1), the trends in the inner regions of high-mass SFGs can be considered flat (i.e., no change with redshift).

The gradients within inner galaxy regions of QGs (panel 6B) follow different trends with redshift than SFGs. At higher redshifts ($z \sim 0.7 - 1.1$), low-mass QGs (two lighter colours) display flatter gradients than high-mass QGs (two darker colours), similar to the SFG sample. At lower red-shifts ($z \sim 0.2 - 0.7$) however, this trend in stellar mass reverses and the gradients of low-mass QGs are steeper than those of high-mass QGs. Over $0.2 \leq z \leq 1.1$, gradients of low-mass QGs grow steeper by a factor of $\sim$1.99 ($10^{9.5} M_\odot \leq M_\star < 10^{10} M_\odot$ bin) and $\sim$1.34 ($10^{10} M_\odot \leq M_\star < 10^{10.5} M_\odot$ bin). In contrast, gradients of high-mass QGs grow flatter by a factor of $\sim$1.16 ($10^{10.5} M_\odot \leq M_\star < 10^{11} M_\odot$ bin) and $\sim$1.33 ($M_\star \geq 10^{11} M_\odot$ bin).

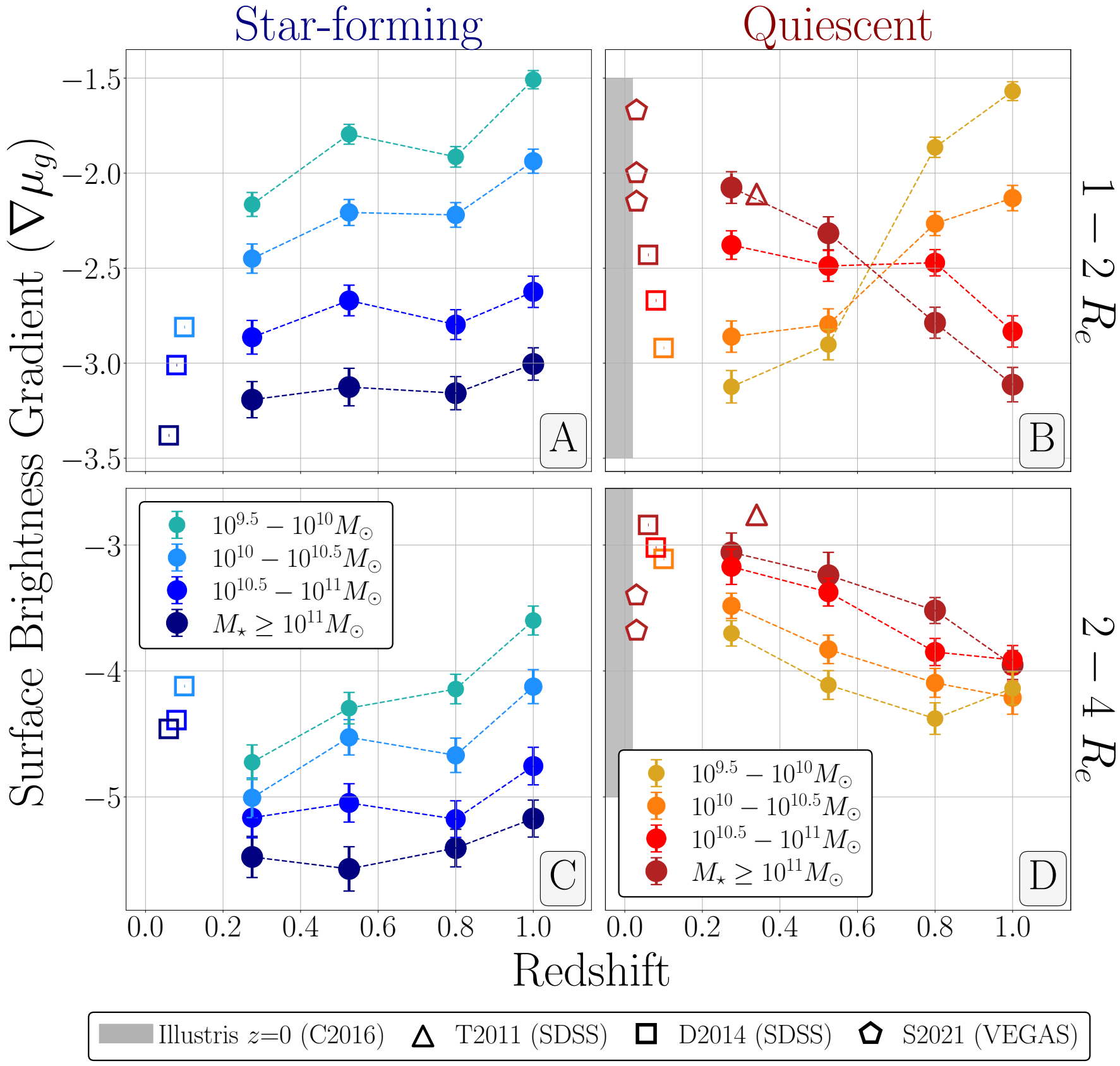

**Figure 6.** Median surface brightness gradients ($\nabla \mu_g$) as a function of redshift for our SFG (left column) and QG samples (right column). Different rows show different radial regions. Colours represent different $M_\star$ bins, with higher $M_\star$ galaxies represented by larger circles and darker colours. Different open symbols represent comparisons from the literature (triangles: Tal & van Dokkum (2011), squares: D'Souza et al. (2014), pentagons: Spavone et al. 2021) and are colour-coded to match the comparable $M_\star$ bin. The grey band in panels B and D represents predictions of gradients in ETGs from the Illustris simulation at $z = 0$ (Cook et al. 2016). Error bars on gradients represent 25/75 percentiles on median $\mu_g$ profiles.

From comparing the evolution in inner galaxy gradients between the SFG and QG populations (panels 6A and 6B), high-mass ($M_\star \geq 10^{10.5} M_\odot$) QGs exhibit flatter gradients than high-mass SFGs at ($z \sim 0.2 - 0.9$) but both populations have similar gradients in the $0.9 \leq z \leq 1.1$ bin. Low-mass ($10^{9.5} M_\odot \leq M_\star < 10^{10.5} M_\odot$) QGs display steeper gradients than low-mass SFGs at lower redshifts ($z \sim 0.2 - 0.7$), but at higher redshifts ($z \sim 0.7 - 1.1$) the two low-mass populations have comparable gradients.

In the stellar halo region (bottom row in Fig. 6), more massive SFGs (darker colours, panel 6C) exhibit steeper gradients than less massive SFGs at all redshifts, similar to inner galaxy regions. The stellar halo gradients of SFGs grow steeper by a factor of $\sim$1.31, $\sim$1.21, $\sim$1.09, and $\sim$1.06 (low to high $M_\star$ bins) over $0.2 \leq z \leq 1.1$. Within the 25/75 interquartile range on median gradients the trends in high-mass SFGs are again consistent with being flat.

In the QG sample, more massive QGs (darker colours, panel 6D) display flatter stellar halo gradients than less massive QGs at all redshifts, opposite to the trend in SFGs. The stellar halo gradients of QGs grow flatter over $0.2 \leq z \leq 1.1$ by a factor of $\sim$1.11, $\sim$1.17, $\sim$1.19, and $\sim$1.23 (low to high $M_\star$ bins). In comparing the stellar halo gradients between QGs and SFGs across the full redshift range, high-mass QGs always exhibit flatter gradients than high-mass SFGs at fixed $M_\star$. In the low-mass sample, this trend is only true at lower redshifts ($z \sim 0.2 - 0.7$), while at higher redshifts ($z \sim 0.7 - 1.1$) the low-mass QG and SFG samples have comparable stellar halo gradients.

The spread in gradient values between galaxies of different $M_\star$ ranges within the same population (i.e., SFG or QG) is much smaller in the stellar halo region (panels 6C and 6D) than in the inner galaxy region (panels 6A and 6B). This result is also seen in the gradients of D'Souza et al.

2014 (open squares in Fig. 6) which represent galaxies of $M_\star > 10^{10} M_\odot$ at $0.06 \leq z \leq 0.1$ in SDSS data. The low redshift gradients of our QG sample ($z = 0.2$ data points, panel 6B and 6D) agree within $\lesssim 2\sigma$ with galaxies of comparable $M_\star$ ranges in their study. The low redshift gradients of our SFG sample agree ($\lesssim 2\sigma$) with those of D'Souza et al. (2014) in the inner galaxy region (1-2$R_e$, panel 6A), but are steeper by a factor of ∼1.2 in the stellar halo region (2-4$R_e$, panel 6C).

The gradients of our QG sample fall within the predicted range of values in ETGs at $z = 0$ from Illustris (Cook et al. 2016, grey bands in panel 6B and 6D). Additionally, gradients of our massive ($M_\star \geq 10^{11} M_\odot$) QG sample (dark red) agree with those found in small samples of similarly massive ETGs at $z \sim 0.03$ (Spavone et al. 2021, open pentagons in panel 6B and 6D) and $z \sim 0.34$ (Tal & van Dokkum 2011, open triangles in panel 6B and 6D).

In conclusion, based on our analysis of $\mu_g$ gradients in our large mass-complete ($M_\star \geq 10^{9.5} M_\odot$) sample of galaxies over a much wider range of redshift ($0.2 \leq z \leq 1.1$) than previously studied, we find the gradients of QGs and SFGs follow different evolutionary trends (i.e., grow steeper or flatter) with decreasing redshift throughout their inner galaxy (1-2$R_e$) and stellar halo (2-4$R_e$) regions. In the QG sample, more massive galaxies exhibit flatter gradients by $z = 0.2$ than less massive galaxies, while the opposite is true in the SFG sample. In comparing the two populations, by $z \sim 0.2$ QGs (of $M_\star \geq 10^{9.5} M_\odot$) exhibit flatter stellar halo gradients than SFGs at fixed $M_\star$. In the inner galaxy regions, this trend is only true for high-mass ($M_\star \geq 10^{10.5} M_\odot$) galaxies. Based on predictions (e.g., Hopkins et al. 2010; Cook et al. 2016), these results may indicate that different assembly mechanisms (i.e., in-situ vs. ex-situ processes, Sec. 1) are driving the majority of stellar mass growth in the two galaxy populations. In Sec. 6.1 we discuss the impact of different assembly processes on the evolution of gradients shown in Fig. 6.

### 5.3. *Inner vs. Outer Light Profile Growth*

Figure 4 and Figure 5 show that over $0.2 \leq z \leq 1.1$ the median $\mu_g$ profiles of galaxies exhibit more growth (i.e. an increase in luminosity) at larger radii (i.e. towards the $2R_e$ and $4R_e$ dashed lines) than at smaller radii (e.g., $R <$ 1-2$R_e$). In this section, we use these profiles to quantify how much of this total luminosity increase over the full redshift range (i.e. $\Delta L_{g,\,tot}$) occurs within the inner ($R < 2R_e$) and stellar halo (2-10$R_e$) regions of galaxies. We split the stellar halo region into 2-4$R_e$ and 4-10$R_e$ to analyze where in the outskirts the majority of growth is occurring in different galaxy subpopulations (bins in Table 2). As discussed in Sec. 4.2, the fraction of light we are potentially missing by truncating profiles at $10R_e$ is negligible.

In Fig. 7 we show the fraction of $\Delta L_{g,\,tot}$ that occurs within the different radial regions (separate panels in Fig. 7) of our SFG (blue stars) and QG (red circles) subpopulations. To calculate these fractions we integrate the median $\mu_g$ profiles of the highest and lowest redshift bins via Eq. 1 (using integration limits corresponding to the different radial regions) to obtain the relative change in luminosity within a particular region over $0.2 \leq z \leq 1.1$.

In the QG sample, $\sim$29% of $\Delta L_{g,\,tot}$ (mean across all $M_\star$ bins) occurs throughout their inner galaxy regions (panel 7A). There is a negative trend in stellar mass where less massive QGs exhibit larger fractions of their total $\mu_g$ profile growth within these inner regions (e.g., ∼0.41 in $10^{9.5} M_\odot \leq M_\star < 10^{10} M_\odot$ galaxies vs. ∼0.21 in $M_\star \geq 10^{11} M_\odot$ galaxies, factor of ∼1.95 larger). In comparison, $\sim$36% of $\Delta L_{g,\,tot}$ in SFGs (mean across all $M_\star$ bins) occurs throughout their inner regions and the same trend in stellar mass is present (∼0.50 in $10^{9.5} M_\odot \leq M_\star < 10^{10} M_\odot$ galaxies vs. ∼0.32 in $M_\star \geq 10^{11} M_\odot$ galaxies) albeit somewhat weaker than in the QG sample (factor of ∼1.56 vs. ∼1.95 in SFGs and QGs, respectively).

At larger radii, $\sim$71% of the luminosity increase in QGs and $\sim$64% of the increase in SFGs (mean across all $M_\star$ bins, solid red and blue lines in panel 7B and 7C) occurs throughout their full stellar halo regions (2-10$R_e$). There is a positive trend in stellar mass within both the SFG and QG samples (the reverse of the inner region trend), which indicates that more massive galaxies exhibit more of their $\mu_g$ profile evolution with decreasing redshift throughout these outer regions.

These results confirm previous findings based on a small sample of massive ($M_\star \geq 10^{11} M_\odot$) galaxies at $0 < z < 2$ (van Dokkum et al. 2010). In their study, the majority of growth in the average surface mass density profiles of massive galaxies over this period occurs in their outer regions ($R > 5$ kpc). The positive trend with stellar mass seen in our QG sample agrees with results from D'Souza et al. (2014) who studied galaxies of $M_\star \geq 10^{10} M_\odot$ at $0.06 \leq z \leq 0.1$ in SDSS data. The authors measure the fraction of light contained in the outer Sérsic component of their fits to galaxy $\mu_r$ profiles and find a positive trend in stellar mass within their high-concentration galaxy sample (comparable to QGs).

When we divide the stellar halo regions into two (panels 7B and 7C), we see a stark difference in the fraction of $\mu_g$ profile growth that occurs in SFGs and QGs. SFGs exhibit much larger fractions of their total luminosity increase throughout the 2-4$R_e$ region than their extended outskirts (4-10$R_e$), with $\sim$50% and $\sim$14% of $\Delta L_{g,\,tot}$ occurring in these regions, respectively (dashed blue lines in panels 7B and 7C). In contrast to SFGs, more comparable fractions of total luminosity increase are seen in the subdivided stellar halo regions of QGs, with $\sim$40% and $\sim$31% of $\Delta L_{g,\,tot}$ occurring within 2-4$R_e$ and 4-10$R_e$, respectively (dashed red lines in panels 7B and 7C).

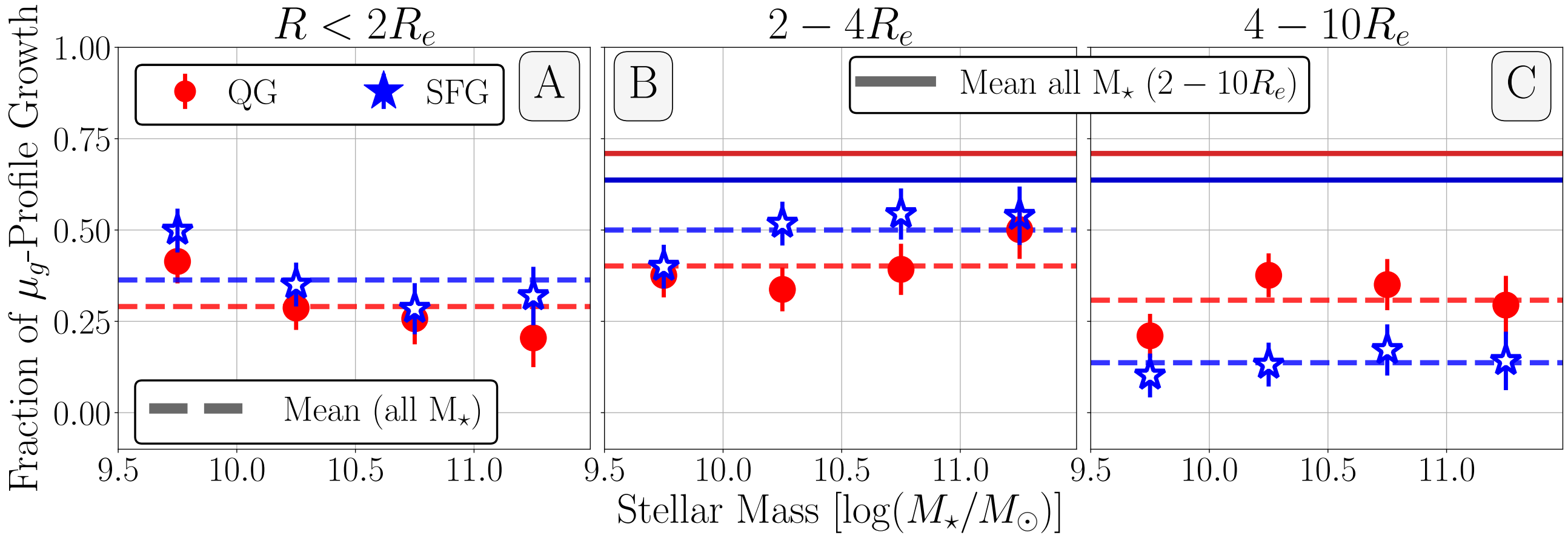

**Figure 7.** The fraction of total median $\mu_g$ profile growth over $0.2 \leq z \leq 1.1$ as a function of stellar mass. Different panels show different radial regions (top labels). Red circles and blue stars represent the QG and SFG samples, with symbols placed at the center of stellar mass bins that are 0.5 wide. In each panel, dashed red and blue horizontal lines represent the mean fraction across all $M_\star$ bins for that radial region. The solid red and blue lines in the second and third panels represent the mean fraction across all $M_\star$ bins for both stellar halo regions combined (i.e. $2 - 10R_e$). Error bars on fractions shown come from 25/75 percentiles on median $\mu_g$ profiles. Bootstrapped errors on median $\mu_g$ profiles produce error bars smaller than the size of symbols shown.

A possible cause of this difference in growth in galaxy outskirts between the SFG and QG samples is a difference in stellar mass assembly mechanisms. If QGs are building stellar mass via minor mergers, the accreted stellar material gets distributed throughout the outer regions of the host galaxies and fuels the assembly of their extended stellar haloes. The comparable amount of $\mu_g$ profile growth we see in QGs within the two subdivided stellar halo regions (i.e. 2-4$R_e$ and 4-10$R_e$, panel 7B and 7C) supports this evolutionary scenario.

On the other hand, the stellar mass assembly of SFGs must be primarily driven by in-situ star formation, although there is a possibility that high-mass SFGs complement this with growth via accretion (investigated in Sec. 6.1). Buildup through star formation would occur more at smaller radii in the halo than in the extended outskirts due to much lower SFRs (e.g., Chamba et al. 2022; Trujillo et al. 2020). This reduction in SFR-related growth at large radial distances could explain why SFGs exhibit much more $\mu_g$ profile growth throughout the 2-4$R_e$ region (panel 7B) than 4-10$R_e$ (panel 7C). We also note that based on recent studies of massive disk galaxies at $z < 0.09$ and $z = 1$ which define $R_{edge}$ as the outer edge of the star-forming disk, the 2–4$R_e$ region in our high-mass SFGs could include some light from both their stellar disks and haloes (e.g., Trujillo et al. 2020; Chamba et al. 2022; Buitrago & Trujillo 2024).

In summary, the majority of evolution in the median $\mu_g$ profiles of $M_\star \geq 10^{9.5} M_\odot$ galaxies over $0.2 \leq z \leq 1.1$ occurs throughout their extended stellar halo regions (2-10$R_e$). On average across our full stellar mass range ($M_\star \geq 10^{9.5} M_\odot$), $\sim$71% of the luminosity increase in QGs and $\sim$64% of the increase in SFGs occurs throughout these outer regions, with a small positive trend with stellar mass in both populations. Furthermore, most of the growth observed throughout the stellar halo regions of SFGs occurs within inner halo regions (2-4$R_e$) than in outer halo regions (4-10$R_e$). In contrast, the QG sample exhibits a similar amount of growth throughout both subdivided stellar halo regions.

### 5.4. *Assembly of Stellar Halo Material*

Results from the previous subsection demonstrate that the majority of $\mu_g$ profile growth (and underlying stellar mass assembly) in galaxies of $M_\star \geq 10^{9.5} M_\odot$ over $0.2 \leq z \leq 1.1$ occurs throughout their stellar halo regions (2-10$R_e$), with important distinctions between how QGs and SFGs assemble this material. In this subsection, we analyze the cumulative buildup of this stellar halo material in different galaxy subpopulations by integrating the median $\mu_g$ profiles (Fig. 4 and Fig. 5) over the stellar halo regions (i.e. via Eq. 1 using 2-10$R_e$ as limits) to obtain total stellar halo luminosity ($L_{halo}$). As mentioned in Sec. 4.2, the fraction of light (and thus stellar halo mass) we are omitting by truncating profiles at 10$R_e$ is negligible.

In Fig. 8 we show how $L_{halo}$ evolves with redshift in the different SFG (top panel) and QG (bottom panel) $M_\star$ bins (different colours). Values in Fig. 8 are normalized by the stellar halo luminosity contained in the median $\mu_g$ profile from the highest redshift bin (i.e. $0.9 \leq z \leq 1.1$, or $L_{halo,\, z\sim1}$) to more easily compare trends with redshift between the different subpopulations.

More massive galaxies (larger and darker coloured symbols in Fig. 8) build up fractionally more stellar halo material than less massive galaxies over $0.2 \leq z \leq 1.1$. In the SFG sample, galaxies increase $L_{halo}$ by a factor of 1.99±0.04, 2.73±0.04, 3.74±0.09, and 4.24±0.13 (low to high $M_\star$ bins). In comparison, the same $M_\star$ bins in the QG sample increase $L_{halo}$ by a factor of 2.39±0.04, 3.18±0.05, 3.94±0.08, and 5.42±0.14 over the full redshift range. These results imply that in galaxies of $M_\star \geq 10^{9.5}M_\odot$, QGs assemble fractionally more stellar halo material than SFGs at fixed $M_\star$ over $0.2 \leq z \leq 1.1$.

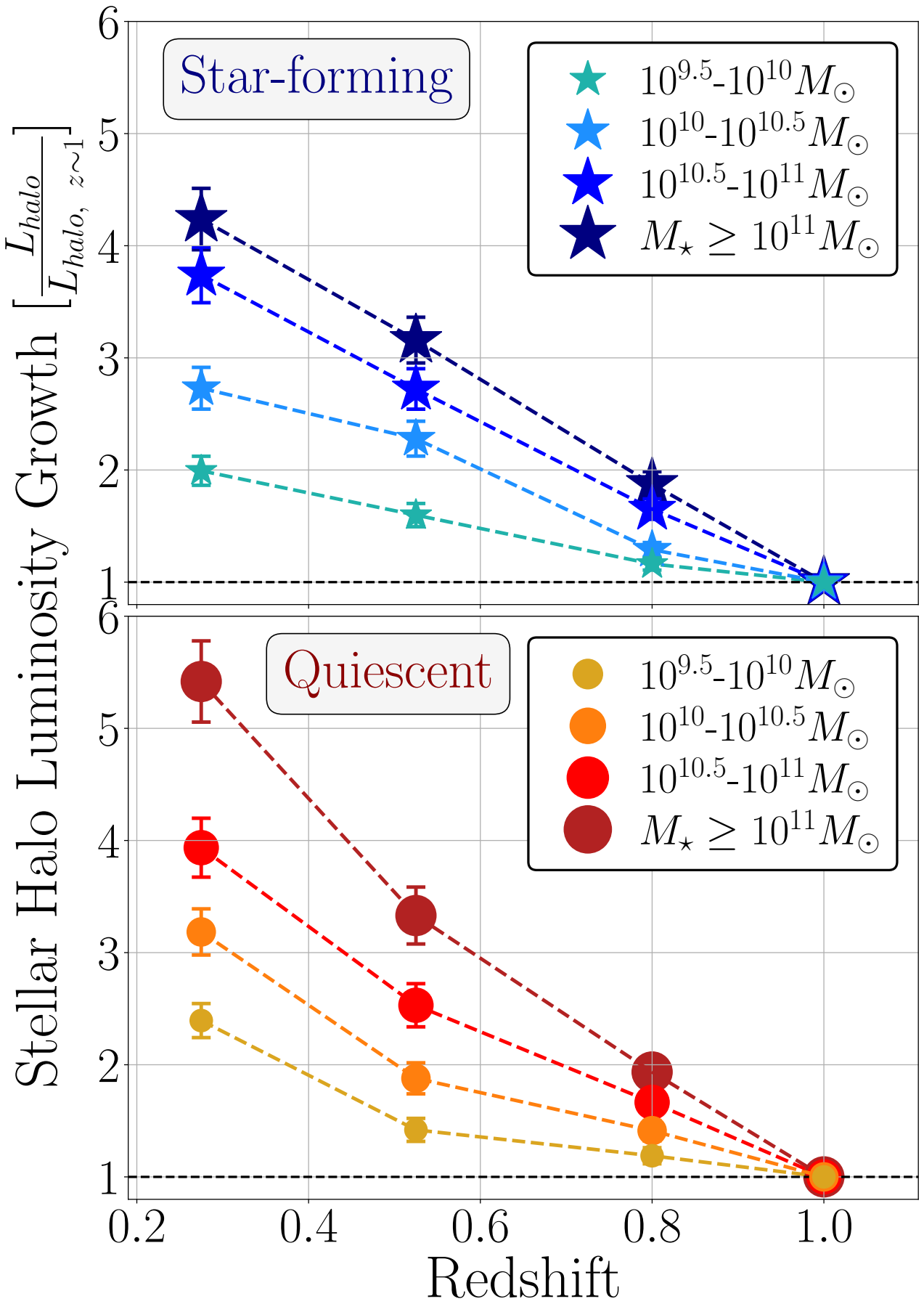

**Figure 8.** Cumulative increases in the integrated stellar halo light ($L_{halo}$) as a function of redshift for different SFG (stars, top panel) and QG (circles, bottom panel) subpopulations. Different $M_\star$ bins are represented by different colours, with more massive galaxies shown in darker colours and larger symbols. Data points represent median $L_{halo}$ values and are placed at the centers of our four redshift bins (Table 2). Values are shown in units of $L_{halo}$ at high redshift ($0.9 \leq z \leq 1.1$ bin, or $L_{halo,\ z\sim1}$). Error bars represent 25/75 percentiles on median $\mu_g$ profiles, and in some cases are smaller than the size of symbols shown.

A similar trend in stellar mass is seen in the stellar haloes of $M_\star \sim 10^{9.7-10.9}M_\odot$ galaxies in the local universe ($D \lesssim$ 24 Mpc; Gilhuly et al. 2022). In that study, more massive galaxies exhibit larger stellar halo mass fractions. Huang et al. (2018) integrated the surface mass density profiles of massive ($M_\star > 10^{11.4}M_\odot$) galaxies at higher redshifts ($0.3 < z < 0.5$) in HSC-SSP data. The authors find that the stellar mass fraction within 10-100 kpc also scaled with increasing stellar mass. The growth of $L_{halo}$ with decreasing redshift (Fig. 8) agrees with the trend found in a small sample of massive galaxies ($M_\star \geq 10^{11}M_\odot$) at $0 < z < 2$ by van Dokkum et al. (2010). The authors find that the stellar mass fraction contained at $R > 5$ kpc grows with decreasing redshift.

Our results also agree with predictions of stellar halo assembly based on the Illustris and IllustrisTNG100 simulations (Elias et al. 2018; Merritt et al. 2020). While stellar halo region definitions differ slightly, the fraction of stellar mass contained within the stellar halo regions of their simulated galaxies is larger in more massive galaxies and in galaxies with early-type morphologies rather than disk-dominated ones (see Fig. 7 in Elias et al. 2018).

In conclusion, results from our analysis of stellar halo luminosity growth in our large CLAUDS+HSC-SSP sample indicate that more massive galaxies in both populations (QGs and SFGs) build up their stellar halo material more rapidly over $0.2 \leq z \leq 1.1$ (i.e. larger fractional increases to $L_{halo}$). For galaxies of $M_\star \geq 10^{9.5}M_\odot$, QGs build up fractionally more stellar halo material than do SFGs at fixed $M_\star$.

## 6. DISCUSSION

The observed change in the median $\mu_g$ profiles of galaxies (Sec. 5) in the mass-complete sample ($M_\star \geq 10^{9.5}M_\odot$) that spans $0.2 \leq z \leq 1.1$ confirms that both SFGs and QGs grow their stellar haloes over this period. In this section, we investigate the impact of different processes on the evolution of galaxy $\mu_g$ profiles. In Sec. 6.1 we estimate the contribution to galaxy stellar halo growth from in-situ star formation over $0.2 \leq z \leq 1.1$. In Sec. 6.2, we investigate the role of merger-driven accretion in fueling stellar halo assembly by testing predictions about ex-situ fractions from cosmological simulations. In Sec. 6.3 we discuss additional physical processes that may impact the evolution in galaxy $\mu_g$ profiles and the buildup of galaxy stellar haloes. Lastly, in Sec. 6.4 we summarize some limitations and caveats of our analysis.

### 6.1. *In-situ Contributions to Stellar Halo Growth*

#### 6.1.1. *Star Formation in SFGs*

Simulations suggest that stellar mass growth throughout the stellar halo regions of SFGs could be driven by star formation despite lower SFRs in galaxy outskirts (e.g., Trujillo et al. 2020; Chamba et al. 2022). Based on the zoomed-in Eris simulation (Guedes et al. 2011), Pillepich et al. (2015) find that simulated Milky Way analogues can contain non-zero in-situ fractions throughout their inner (∼0.25) and outer (∼0.03) stellar halo regions. Galaxies in the Illustris simulation exhibit non-negligible in-situ fractions ($\lesssim 0.3$) at larger radii (e.g., $R \geq 4R_e$) that follow a negative trend with stellar mass over $M_\star \sim 10^{10-12}M_\odot$ and become smaller with increasing radii at fixed $M_\star$ (Rodriguez-Gomez et al. 2016).

To estimate the contribution from star formation to the median $\mu_g$ profile growth (i.e. the luminosity increase over 0.2 $\leq z \leq 1.1$) in our SFG sample, we compute the total possible amount of stellar mass that can be formed via star formation over our redshift range ($\Delta M_{\star,\ form}$). We use SFRs obtained via `LePhare` SED fitting performed by Chen et al. in prep. (Sec. 2.2), and calculate median SFRs for each $M_\star$ and redshift bin combination (Table 2). For a given SFG stellar mass bin, $\Delta M_{\star,\ form}$ is then calculated as

$$\Delta M_{\star,\ form} = \sum (SFR_{zi} \cdot T_{zi}), \tag{3}$$

where $SFR_{zi}$ refers to the median SFR of the $i$-th redshift ($z$) bin (four $z$ bins in total, Table 2), and $T_{zi}$ represents the time in Gyr covered by the redshift interval of the bin. For the highest ($0.9 \leq z \leq 1.1$) and lowest ($0.2 \leq z < 0.35$) redshift bins we limit the time period to the median redshift value within the bins (i.e. the total period extends from $z = 0.275$ to $z = 1$), to reflect the fact that we are studying median profiles within those redshift ranges. We assume that median SFRs are constant throughout a redshift interval and thus the values we obtain for $\Delta M_{\star,\ form}$ are considered upper-limit estimates of the total growth through star formation.

To convert $\Delta M_{\star,\ form}$ into luminosity (i.e. $\Delta L_{g,\ form}$), we use the stellar mass-to-light ratio in the rest-frame $g$-band ($M_\star/L_g$). For each SFG $M_\star$ bin, we obtain a $M_\star/L_g$ ratio based on the median $(U-g)$ colour and the relation between $(U-g)$ colour and $M_\star/L_g$ ratio from Szomoru et al. (2013, Fig. 1). We compute global rest-frame $(U-g)$ colours for galaxies from their absolute $U$- and $g$-band magnitudes (Sec. 2.2).

To compare this change in luminosity expected from star formation ($\Delta L_{g,\ form}$) to the actual profile growth measured in our sample, we integrate (via Eq. 1) the median $\mu_g$ profiles of each SFG $M_\star$ bin to calculate the change in total luminosity ($\Delta L_{g,\ tot}$) from the highest redshift profile ($0.9 \leq z \leq 1.1$) to the lowest redshift profile ($0.2 \leq z < 0.35$). By dividing $\Delta L_{g,\ form}$ by $\Delta L_{g,\ tot}$, we obtain a fractional increase in total luminosity that could arise from star formation over our full redshift range.

The left panel of Fig. 9 shows the contribution to stellar halo growth (i.e. the increase in $L_{halo}$ over $0.2 \leq z \leq 1.1$) that can arise from star formation (blue) for our four SFG $M_\star$ bins (separate bars in Fig. 9). The remaining fractions of assembled stellar halo material in each subpopulation are shown as cyan hashed bars.

Low-mass SFGs ($10^{9.5} M_\odot \leq M_\star < 10^{10.5} M_\odot$) can potentially build up their stellar halo material solely via star formation over $0.2 \leq z \leq 1.1$ (two full blue bars, Fig. 9). This stellar material likely gets rearranged into a stellar halo component through secular processes (e.g., bar dynamics; Kormendy & Kennicutt 2004; Athanassoula 2005) or from the close interactions and tidal forces these low-mass SFGs experience as they fall into denser environments of the cosmic web (e.g., Hopkins et al. 2010; Tissera et al. 2013; Boselli et al. 2022; Zhu et al. 2022). As our calculations for the contribution from star formation are upper-limit estimates, these low-mass SFGs may still accumulate small amounts of stellar mass via accretion events (e.g., Sestito et al. 2023; Jensen et al. 2024).

For high-mass SFGs ($M_\star \geq 10^{10.5} M_\odot$), at most 52%±5 ($10^{10.5} M_\odot \leq M_\star < 10^{11} M_\odot$ bin) and 23%±3 ($M_\star \geq 10^{11} M_\odot$ bin) of stellar mass growth can be due to in-situ star formation. The remaining fractions of assembled stellar halo material (48% and 77%, hashed cyan bars in Fig. 9) strongly imply that high-mass SFGs are growing their stellar haloes through a combination of star formation and merger-driven accretion over $0.2 \leq z \leq 1.1$. Major mergers are expected to destroy disks and induce transformations into elliptical morphologies, eventually leading to the galaxy being quenched (Jackson et al. 2020, 2022). Because of this, the ex-situ contributions in our high-mass SFG sample are likely a result of accretion via minor mergers over $0.2 \leq z \leq 1.1$. This agrees with several previous studies showing evidence for a minor merger-driven period of growth of massive galaxies over similar redshift ranges (e.g., $z \lesssim 1-2$; Bundy et al. 2009; Trujillo et al. 2011; Ownsworth et al. 2014; Matharu 2019; Jackson et al. 2022).

The evolution in $\mu_g$ gradients (Fig. 6) in our SFG sample supports the interpretation that the stellar haloes of high-mass SFGs grow through minor mergers. Based on the Illustris simulation, steeper $\mu$ profile gradients within 1-2$R_e$ and 2-4$R_e$ are predicted to be caused by in-situ stellar mass assembly. In comparison, growth through accretion is expected to induce flatter gradients in these regions (Cook et al. 2016). In Fig. 6 (panels A and C) our results show that while all $M_\star$ bins in our SFG sample exhibit steeper gradients over time in these regions, the trends of more massive SFGs are smaller (i.e. closer to being a flattening trend). This result would be expected if high-mass SFGs assemble stellar mass through both in-situ and ex-situ mechanisms over our redshift range. Accretion via major mergers would induce much stronger flattening trends in the gradients of high-mass SFGs (due to the increase in accretion). Therefore growth via minor mergers is the most likely scenario.

In summary, in-situ star formation can explain all of the evolution observed in the median $\mu_g$ profiles of low-mass SFGs ($10^{9.5} M_\odot \leq M_\star < 10^{10.5} M_\odot$) over $0.2 \leq z \leq 1.1$. In contrast, high-mass SFGs ($M_\star \geq 10^{10.5} M_\odot$) require an additional assembly mechanism beyond star formation to explain the buildup of their stellar halo material (i.e. increase to $L_{halo}$) over our redshift range.

#### 6.1.2. *Effect of Newcomers in QG Populations*

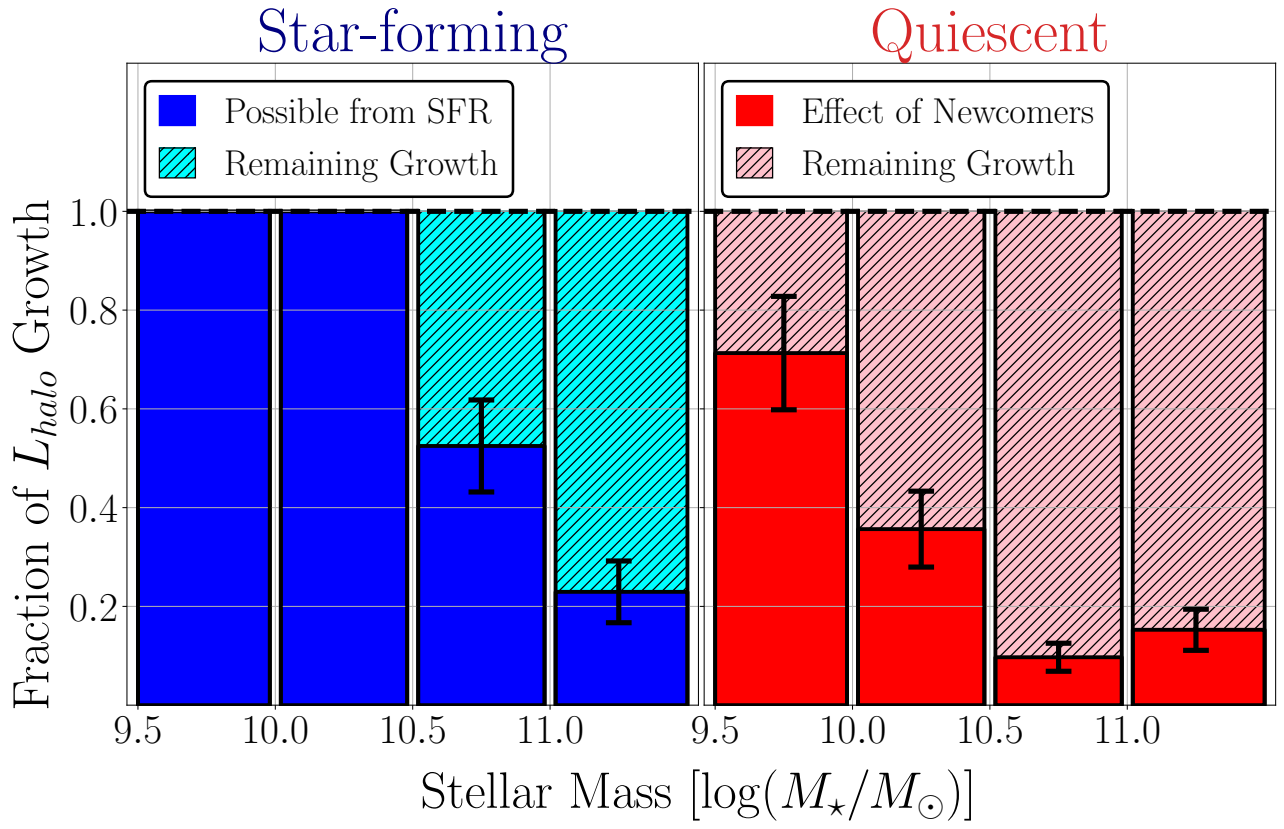

**Figure 9.** The contribution to stellar halo growth over $0.2 \leq z \leq 1.1$ from star formation (SF) in SFGs (left panel) and the effect of newcomers (NC) in QGs (right panel) as a function of stellar mass. Different bars represent $M_\star$ bins. Blue and red portions of bars show contributions from SF and NC respectively, while cyan and pink hashed bars show the fraction of remaining stellar halo material assembled over $0.2 \leq z \leq 1.1$. Error bars shown represent 25/75 interquartile range on median $\mu_g$ profiles.

For QGs, we estimate the contribution to median $\mu_g$ profile growth that can be attributed to the effect of recently quenched galaxies (i.e. *newcomers*), also referred to as progenitor bias (van Dokkum & Franx 1996, 2001; Carollo et al. 2013; Saglia et al. 2016; Damjanov et al. 2019). When comparing samples of QGs at different redshift ranges, we must take into account that some low redshift QGs were, in fact, SFGs at higher redshifts which have since been quenched. This newcomer effect has been shown to account for some of the observed size ($R_e$) growth with decreasing redshift in the QG population (e.g., Damjanov et al. 2022; George et al. 2024). We use the contribution from newcomers as a proxy for the "in-situ" contribution to average stellar halo growth in QGs over $0.2 \leq z \leq 1.1$, as it reflects stellar mass assembly that occurred when galaxies were still star-forming.

To calculate the contribution from newcomers we use the fractional change in QG number densities ($\Phi$) over $0.2 \leq z \leq 1.1$ based on galaxy stellar mass functions from Weaver et al. (2023). We use the QG stellar mass functions from the $0.2 < z < 0.5$ (low $z$) and $0.8 < z < 1.1$ (high $z$) redshift bins from that study, and calculate a fractional change (i.e. $\frac{\Phi_{low\ z} - \Phi_{high\ z}}{\Phi_{high\ z}}$) using the number density values at the centres of our four stellar mass bins (Table 2). Over our full redshift range, we estimate the QG number density increases by 58%, 27%, 8%, and 10% (low to high $M_\star$ bins). These percentages serve as fractions of QGs at low redshift ($z \sim 0.2$) that are likely SFGs which quenched over $0.2 \leq z \leq 1.1$.

We remove the corresponding number of individual galaxies from all mass bins in the low-redshift regime ($0.2 \leq z < 0.35$) and recalculate median $\mu_g$ profiles (Sec. 4.2) and median sizes ($R_e$, Sec. 4.3). We specifically remove individual galaxies with the largest sizes as SFGs, which are the source of the newcomer population, have larger sizes than QGs at fixed $M_\star$ (e.g., van der Wel et al. 2014; Mowla et al. 2019b; Kawinwanichakij et al. 2021; George et al. 2024). Using new median $\mu_g$ profiles adjusted for the effect of newcomers we recalculate the fractional increase in $L_{halo}$ over $0.2 \leq z \leq 1.1$ (Sec. 5.4) for each QG $M_\star$ bin. We take the difference between the original increase to $L_{halo}$ and the increase using the adjusted median $\mu_g$ profiles to be the contribution to the stellar halo growth due to the influx of newcomers.

The right panel of Fig. 9 shows the contribution to stellar halo growth from newcomers in red and the remaining fraction of assembled stellar halo material as pink hashed bars for our four QG $M_\star$ bins (separate bars). In our low to high $M_\star$ bins, newcomers can account for 71%±6, 36%±3, 10%±1, and 15%±2 of the luminosity increase observed in the median $\mu_g$ profiles of QGs over $0.2 \leq z \leq 1.1$ (red in Fig. 9). This implies that ~29%, ~64%, ~90%, and ~85% (low to high $M_\star$ bins) of the stellar halo growth observed in our QG sample over the full redshift range cannot be explained by in-situ stellar mass assembly occurring in newcomers prior to quenching.

The most likely explanation for the remaining fractions of assembled stellar halo material in QGs (pink hashed bars in Fig. 9) is that minor merger-driven accretion is fueling this growth. Our results in Fig. 8 show that more massive QGs assemble fractionally more stellar halo material than less-massive QGs over $0.2 \leq z \leq 1.1$, which mimics the trend with stellar mass seen in predicted ex-situ fractions of simulated galaxies (e.g., Rodriguez-Gomez et al. 2016; Davison et al. 2020; Huško et al. 2022). Figure 7 shows the majority of growth in QGs over our redshift range occurs throughout their extended outskirts ($R \geq 2R_e$) where accreted stellar material from minor interactions is expected to be deposited (e.g., Trujillo et al. 2011; Lambas et al. 2012; Ownsworth et al. 2014; Montenegro-Taborda et al. 2023).

Furthermore, the evolution in $\mu_g$ gradients in our QG sample (Fig. 6) supports this scenario of QG stellar halo growth being driven by accretion. Based on predictions from Illustris, flatter $\mu$ profile gradients within 1-2$R_e$ and 2-4$R_e$ are driven by accretion (Cook et al. 2016). In both regions (panels B and D in Fig. 6), we find that more massive QGs have flatter gradients by $z = 0.2$ which would imply a larger influence from accretion. While the gradients of high-mass QGs ($M_\star \geq 10^{10.5} M_\odot$) grow flatter with decreasing redshift in both regions, low-mass QGs ($10^{9.5} M_\odot \leq M_\star < 10^{10.5} M_\odot$) only exhibit a flattening trend with redshift in the 2-4$R_e$ region. This difference in trends may reflect the fact that more massive galaxies (with their deeper gravitational potential wells) are more efficient at funnelling accreted material deeper into their inner regions (e.g., Boylan-Kolchin et al. 2008; Gan et al. 2010).

In conclusion, the effect of newcomers can account for a portion of the growth observed in the median $\mu_g$ profiles of QGs over $0.2 \leq z \leq 1.1$. The contribution from newcomers is larger in less massive QGs, with an average of ∼54% in low-mass QGs ($10^{9.5} M_\odot \leq M_\star < 10^{10.5} M_\odot$) vs. ∼13% in high-mass QGs ($M_\star \geq 10^{10.5} M_\odot$). The remaining fractions of assembled stellar material we measure strongly suggest that, over the $0.2 \leq z \leq 1.1$ redshift interval, QGs of $M_\star \geq 10^{9.5} M_\odot$ build up their stellar haloes primarily through accretion via minor mergers.

### 6.2. *Accretion and Merger-Driven Growth*

The buildup of galaxy stellar haloes is predicted to be primarily fueled by the accretion of stellar material via the hierarchical merging of galaxies over time (e.g., Oser et al. 2010; Cooper et al. 2010; Cook et al. 2010, 2016; Pillepich et al. 2014; Huško et al. 2022). Our observational results confirm that high-mass ($M_\star \geq 10^{10.5} M_\odot$) SFGs and QGs of $M_\star \geq 10^{9.5} M_\odot$ require an additional mechanism beyond in-situ star formation to explain the majority of their stellar halo growth over $0.2 \leq z \leq 1.1$ (Sec. 6.1). Here, we test whether the additional stellar halo growth is consistent with stellar mass growth via accretion. In Sec. 6.2.1, we compare our results with predictions from cosmological simulations regarding the evolution of ex-situ fractions in galaxies over a similar redshift range. In Sec. 6.2.2, we discuss the change in observed ex-situ fractions with stellar mass for SFGs and QGs in our sample.

#### 6.2.1. *Testing Ex-situ Fraction Predictions*

In observational studies of stellar haloes of massive ETGs, both Huang et al. 2018 (HSC-SSP, $0.3 < z < 0.5$) and Buitrago et al. 2017 (HST, $z \sim 0.65$) proposed the fraction of stellar mass contained beyond 10 kpc as an observational proxy for galaxy ex-situ fraction (i.e., $[M_{\star,\,(>10\ kpc)}/M_{\star,\,tot}] \approx M_{ex}$). This definition is motivated by the two-phase galaxy formation scenario that predicts the early formation (e.g., $z \gtrsim 2$) of galaxy cores through in-situ star formation, and the assembly of outer stellar envelopes via accretion at later times ($z < 2$; Oser et al. 2010; Cook et al. 2010). Additionally, both simulations (e.g., Rodriguez-Gomez et al. 2016) and observations (e.g., van Dokkum et al. 2010) suggest the in-situ component dominates the stellar mass contained at small radii. Here we expand on the analysis of Buitrago et al. (2017) and Huang et al. (2018) and investigate this ex-situ fraction proxy over a broad range of stellar mass (i.e. $M_\star \geq 10^{9.5} M_\odot$ ) and redshift ($0.2 \leq z \leq 1.1$) for SFGs and QGs.

Fig. 10 shows the fraction of total stellar mass present at galactocentric distances of $R \geq 10$ kpc for all galaxies in our sample (i.e., SFGs and QGs combined) separated into four redshift bins (black squares in all four figure panels). These median stellar mass fractions are proportional to the median luminosity fractions we calculate by integrating median $\mu_g$ profiles via Eq. 1 (with adjusted integration limits). Under the assumption of constant $M_\star/L$ ratios throughout galaxy $\mu_g$ profiles, the conversion between luminosity fractions and stellar mass fractions is not required (i.e. if $M_\star/L(R) \approx$ constant $\Longrightarrow L_{>10kpc}/L_{tot} \approx M_{\star,\,>10kpc}/M_{\star,\,tot}$).

Low-mass galaxies ($10^{9.5} M_\odot \leq M_\star < 10^{10.5} M_\odot$, first two data points in each panel) exhibit small ex-situ fractions of ∼ 4-12% over $0.2 \leq z \leq 1.1$, increasing by a minor factor of $\sim 1.1 - 1.3$ from the highest to lowest redshift bins ($0.9 \leq z \leq 1.1$ to $0.2 \leq z < 0.35$). High-mass galaxies ($M_\star \geq 10^{10.5} M_\odot$, last two data points in each panel) exhibit much larger ex-situ fractions (∼11-46%) that show considerable growth over the full redshift range (factor of ∼2.3-2.6). We find a good agreement (at $< 2\sigma$ level) between the ex-situ fractions of our high-mass ($M_\star \geq 10^{10.5} M_\odot$) QG sample and those of similarly massive ETGs from the observational samples of Buitrago et al. 2017 (purple squares, top-right and bottom-right panels in Fig. 10) and Huang et al. 2018 (purple contour, top right panel).

Fig. 10 also shows ex-situ fraction predictions based on samples of both spiral and elliptical galaxy models. We include predictions from several simulations as ex-situ fractions differ between theoretical models (e.g., see Huško et al. 2022, their Fig. 2). Differences in predicted ex-situ fractions can arise from several factors, such as mass resolution and the ability to trace lower-mass mergers (e.g., *mini mergers*; Bottrell et al. 2024), as well as methods for constructing merger trees and defining ex-situ mass (Rodriguez-Gomez et al. 2016; Tacchella et al. 2019; Qu et al. 2017; Davison et al. 2020). The strength of galaxy feedback (e.g., from AGN or supernovae) can also impact ex-situ fractions, particularly in galaxies of $M_\star \geq 10^{10.5} M_\odot$ based on predictions from the EAGLE simulation (Qu et al. 2017, their Fig. 10).

We find good agreement ($\lesssim$ 1-2$\sigma$) between our computed ex-situ fractions and predicted ranges (coloured bands) across our four redshift intervals (separate panels). This agreement suggests that simulation differences produce only subtle variations in ex-situ fractions and all models are consistent with observations within uncertainties. One exception is in the highest redshift bin ($0.9 \leq z \leq 1.1$, bottom-right panel), where our results at larger stellar masses ($M_\star > 10^{10.5} M_\odot$) agree more with Illustris than IllustrisTNG. This may indicate that IllustrisTNG is over-predicting the amount of accreted material in massive galaxies at higher redshifts (e.g., $z \gtrsim 1$). This could be due to differences in how both simulations implement galaxy feedback (Qu et al. 2017). In IllustrisTNG, more efficient stellar winds and stronger AGN feedback suppress in-situ star formation more effectively than in Illustris, leading to higher predicted ex-situ fractions at a given stellar mass (Tacchella et al. 2019). Future explorations of this discrepancy will require a detailed comparison of different observational proxies for ex-situ fractions (e.g., Gilhuly et al. 2022; Xu et al. 2024).

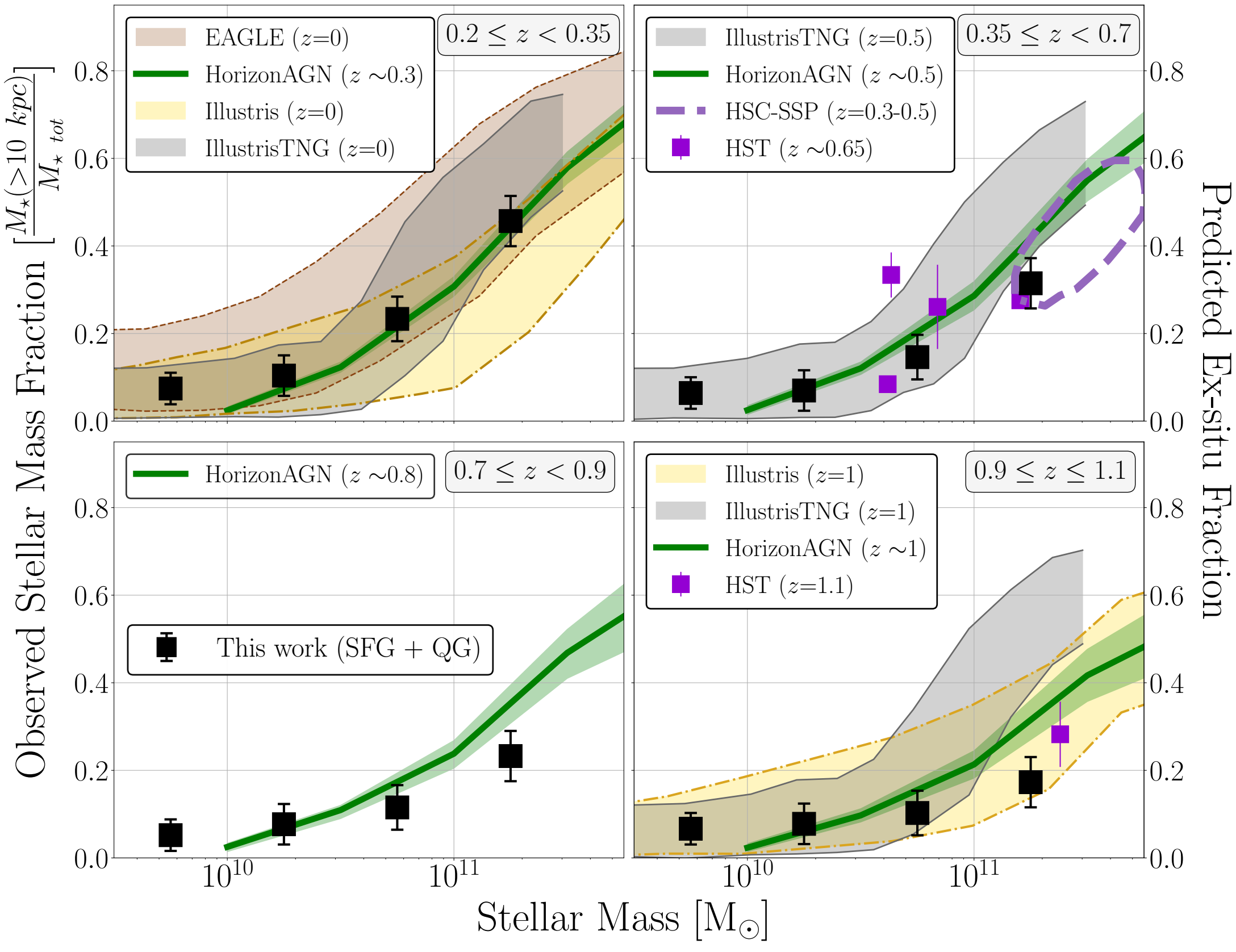

**Figure 10.** Fraction of stellar mass beyond 10 kpc as a function of galaxy stellar mass in our four redshift bins (different panels). In each panel, black squares represent all galaxies within a given $M_\star$ bin and are placed at the centers of our four $M_\star$ bins. Purple squares and the purple dashed contour region represent galaxies from the observational samples of Buitrago et al. (2017) and Huang et al. (2018), respectively. All other coloured bands represent predicted ranges of ex-situ fractions from various cosmological simulations. Green bands represent galaxies from Dubois et al. 2016 (Horizon-AGN), brown regions show predictions from Davison et al. 2020 (EAGLE), and grey and gold bands represent galaxies from Tacchella et al. 2019 (IllustrisTNG) and Rodriguez-Gomez et al. 2016 (Illustris), respectively. Error bars on mass fractions represent uncertainties on luminosity fractions computed by propagating 25/75 percentiles on median $\mu_g$ profiles via standard error propagation rules for division.

In conclusion, we find good agreement between the median ex-situ fractions of our total galaxy sample and the predicted fractions from several cosmological simulations. This agreement demonstrates that the stellar mass fraction beyond 10 kpc is a good observational proxy for the ex-situ fraction in galaxies with $M_\star \geq 10^{9.5} M_\odot$ over $0.2 \leq z \leq 1.1$ (i.e. larger ranges than previously probed in the literature).

#### 6.2.2. *Observed Trends in Ex-situ Growth*

Next, we divide our galaxy sample once again into SFGs and QGs (Sec. 2.2) to study how both populations contribute to the increase in median galaxy ex-situ fractions (black squares in Fig. 10) over $0.2 \leq z \leq 1.1$. Fig. 11 shows the fractional increase in median ex-situ fractions over the full redshift range (i.e., from highest to lowest redshift bins, or $M_{ex,\ z\sim0.3}/M_{ex,\ z\sim1}$) divided into SFG (blue stars) and QG (red circles) subpopulations.

In the low-mass sample (two lower $M_\star$ bins, Fig. 11), the median ex-situ fractions of SFGs increase by a factor of $\sim$1.1-1.3 over $0.2 \leq z \leq 1.1$, and the median ex-situ fractions of low-mass QGs increase by a factor of $\sim$1.4-1.8. In the high-mass sample (two higher $M_\star$ bins, Fig. 11), median ex-situ fractions of SFGs and QGs increase by a factor of $\sim$2.0-2.3 and $\sim$2.6-2.8, respectively. These results indicate that QGs of $M_\star \geq 10^{9.5} M_\odot$ grow more through accretion than SFGs at fixed stellar mass over $0.2 \leq z \leq 1.1$. This aligns with our results from the previous subsection (Sec. 6.1), where in-situ contributions to stellar halo growth are found to be lower in QGs than SFGs at fixed $M_\star$ (Fig. 9).

Our results support a minor merger-driven evolution for galaxies of $M_\star \geq 10^{9.5} M_\odot$ over $0.2 \leq z \leq 1.1$. Based on simulations, major mergers are predicted to produce much larger ex-situ fractions than we estimate in Fig. 10 (e.g., Rodriguez-Gomez et al. 2016; Tacchella et al. 2019; Huško et al. 2022). Major mergers are also predicted to add stellar mass to both the inner regions and outskirts of galaxies, while minor mergers predominantly deposit material throughout galaxy outskirts (e.g., Lambas et al. 2012; Hilz et al. 2013; Montenegro-Taborda et al. 2023). If major mergers were the dominant merger channel for galaxies over our redshift range, we would expect larger fractions of $\mu_g$ profile growth to occur in inner galaxy regions than we measure (first panel in Fig. 7).

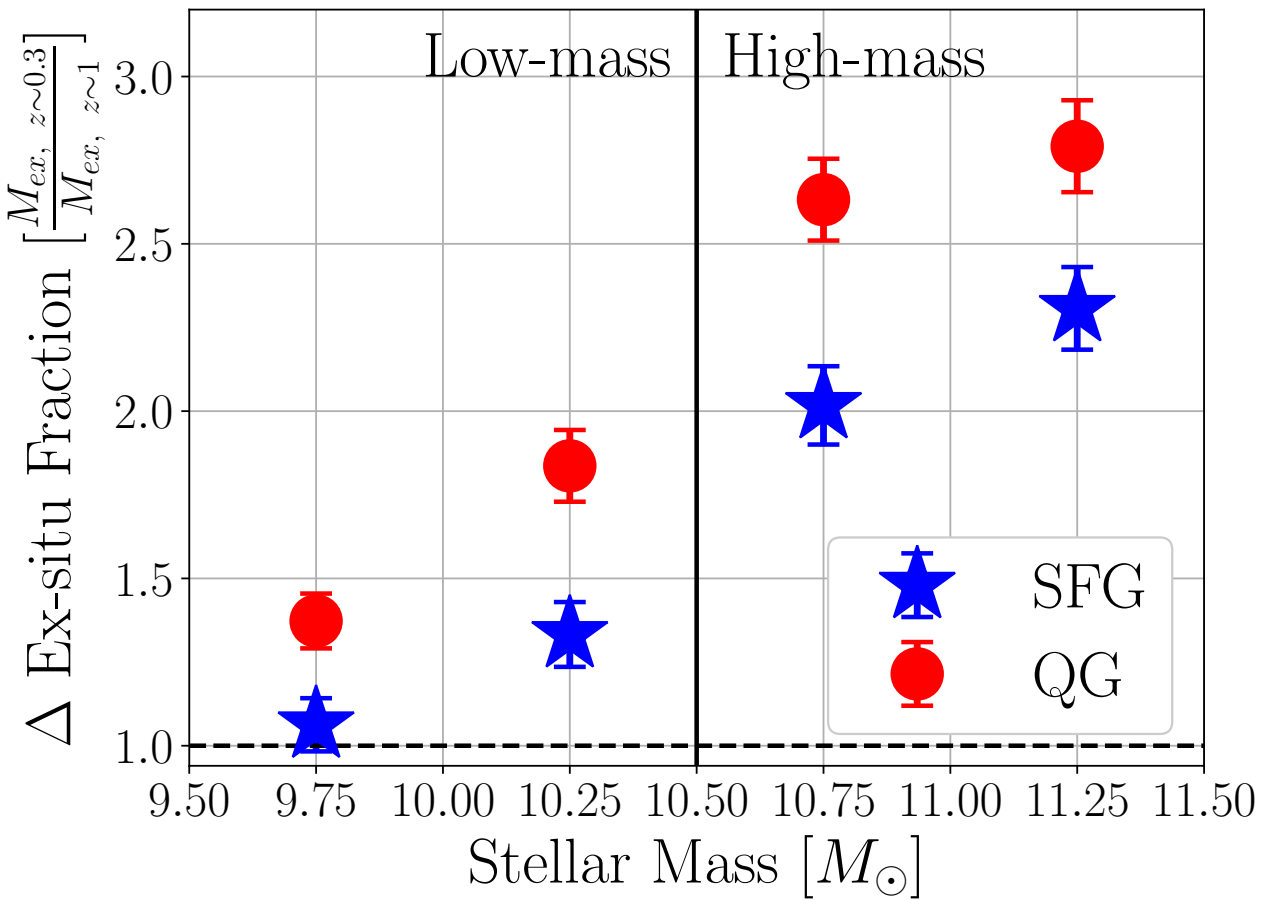

**Figure 11.** Fractional change in median ex-situ fractions ($M_{ex}$) over 0.2 $\leq z \leq$ 1.1 as a function of galaxy stellar mass. Stars and circles represent the SFG and QG samples, respectively. Data points are placed at the centers of the four $M_\star$ bins, and the vertical black line separates the low-mass and high-mass galaxy samples. Error bars on mass fractions represent uncertainties on luminosity fractions computed by propagating 25/75 percentiles on median $\mu_g$ profiles via standard error propagation rules for division.

In summary, our results confirm predictions that stellar mass growth through merger-driven accretion is increased in more massive galaxies over $0.2 \leq z \leq 1.1$, particularly in galaxies above the pivot mass range ($M_p \sim 10^{10.5} M_\odot$, Sec. 4.1) identified in size-mass relation studies. For galaxies of $M_\star \geq 10^{9.5} M_\odot$, QGs exhibit larger fractional increases to their ex-situ fractions over $0.2 \leq z \leq 1.1$ compared to SFGs at fixed $M_\star$ (Fig. 11). Our combined results support the scenario where minor mergers drive this ex-situ-related growth in high-mass ($M_\star \geq 10^{10.5} M_\odot$) SFGs and QGs of $M_\star \geq 10^{9.5} M_\odot$ over the full redshift range we study.

### 6.3. *Influence From Additional Physical Processes*

Based on the analysis presented in Sec. 6.1-6.2, merger-driven accretion is the most likely explanation for the evolution observed in the median $\mu_g$ profiles (and the underlying buildup of stellar mass) in high-mass SFGs ($M_\star \geq 10^{10.5} M_\odot$) and QGs of $M_\star \geq 10^{9.5} M_\odot$ over $0.2 \leq z \leq 1.1$. However, previous theoretical work has demonstrated that other physical processes can also contribute to the growth of galaxy $\mu$ profiles over cosmic time (e.g., Hopkins et al. 2010; Hilz et al. 2013).

One such process is dynamical heating, where a merger or close encounter injects energy into a galaxy's existing stars pushing them onto larger, more extended orbits (e.g., Hopkins et al. 2010; Tissera et al. 2013; Zhu et al. 2022). Similarly, stellar migration — the outward movement of stars formed in inner regions due to internal dynamical processes — could also redistribute stars to larger radii (e.g., Loebman et al. 2011, 2016; Debattista et al. 2017). Both effects would lead to higher $\mu$ levels at large galactic radii, which may be misinterpreted as the accretion of additional stellar material into the stellar halo regions. However, theoretical work suggests the impact on galaxy surface mass density profiles from effects such as dynamical heating is significantly smaller than the contribution from merger-driven accretion (e.g., $\lesssim 10\%$ of the mass increase expected from minor mergers; Hopkins et al. 2010). Furthermore, Boecker et al. (2023) find that $\sim 23\%$ of the innermost stars in simulated $M_\star \sim 10^{8.5-12} M_\odot$ galaxies at $z = 0$ originate from their outer regions. Outward stellar migration is thus at least partially offset by inward migration.

If a galaxy loses stellar mass from its inner regions adiabatically (e.g., through AGN or stellar feedback), the galaxy's stellar mass profile can expand due to the weaker gravitational potential in its core. This "puffing up" scenario due to adiabatic expansion can account for some of the growth in galaxy sizes ($R_e$) over time (e.g., Damjanov et al. 2009; Trujillo et al. 2011). However, Hopkins et al. (2010) demonstrates that this mechanism mostly impacts the inner regions of galaxy mass profiles, reducing the surface mass density in central regions of galaxies and causing an increase in measured galaxy sizes. The effect of adiabatic expansion on the outer portions of galaxy profiles is negligible (see their Fig. 2), and thus we can rule it out as a driver of the stellar halo growth in our sample (e.g., Fig. 8).

Some of the increase in surface brightness measured in the median $\mu_g$ profiles of our sample over $0.2 \leq z \leq 1.1$ may be due to the evolution of stellar populations in the form of changing stellar mass-to-light ratios in the rest-frame $g$-band ($M_\star/L_g$). As the stellar population of a galaxy ages its $M_\star/L$ will increase as more of its total light becomes dominated by the fainter, long-lived, lower-mass stars which emit more light at longer wavelengths (e.g., $g$-band emission; Pagel & Edmunds 1981; Bruzual & Charlot 2003; Renzini 2006). However, in Hopkins et al. (2010, their Fig. 2 and Fig. 3) the authors demonstrate that differences in $M_\star/L$ gradients caused by stellar population age gradients result in negligible evolution in the outskirts of galaxy surface mass density profiles (i.e., $\lesssim$1% of the profile change expected from minor merger-driven accretion).

In conclusion, while other physical processes have been shown to affect the evolution of galaxy $\mu$ profiles, none of them are expected to dominate over accretion when it comes to driving the growth of $\mu$ profiles at large radii.

### 6.4. *Limitations and Caveats*

Tracing the emission from galaxies over the same rest-frame wavelengths across different redshifts typically requires a $K$ correction to galaxy fluxes to account for differences between observed filter bandpasses (Hogg et al. 2002; Blanton & Roweis 2007; Huertas-Company et al. 2009; Chilingarian et al. 2010). In this work, we do not apply $K$ corrections to extracted rest-frame $g$-band surface brightness values (Sec. 4.1) in galaxy $\mu_g$ profiles. To accurately apply $K$ corrections to a galaxy light profile would require knowledge of the radial colour profile of the galaxy, as the $K$ correction should change at each radius due to the galaxy's varying colour and surface brightness. The extraction of galaxy rest-frame colour profiles is beyond the scope of this project. We will extend the analysis to include galaxy rest-frame $U-g$ colour profiles in upcoming work on the environmental dependence of galaxy stellar halo evolution (Williams et al. in prep.).

For thoroughness, we investigate the impact on our results from applying singular $K$ corrections across entire median $\mu_g$ profiles (i.e., the same correction value at each radius). To calculate $K$ corrections for individual galaxies in different subpopulations ($M_\star$ and redshift bin combinations, Table 2) we use the `kcorrect` python package and six input magnitudes ($U + grizy$) to reconstruct galaxy SEDs during template fitting. After applying these global $K$ corrections our results shown throughout Sec. 5 and Sec. 6 exhibit either no change or shift by less than $1\sigma$ which represents the 25/75 interquartile range on median $\mu_g$ profiles. Hence, we conclude that neglecting $K$ corrections to galaxy $\mu_g$ profiles does not impact the observed trends or conclusions presented in our study.

The presence of dust in galaxies could be obscuring their stellar light and affecting the extraction of individual galaxy $\mu$ profiles. González Delgado et al. (2015) extracted radial dust ($A_V$) profiles from galaxies of $M_\star \sim 10^{9-11.8} M_\odot$ at $z = 0.005 - 0.03$ in the CALIFA survey. The authors show that galaxies have negative dust gradients (i.e. the level of extinction decreases with increasing radius), and that spiral galaxies exhibit higher levels of extinction than elliptical galaxies. However, the study demonstrates that the overall amount of extinction is fairly small ($A_V \lesssim 0.6$ mag) in all galaxy types and would produce changes in our median $\mu_g$ profiles that are much smaller than the 25/75 interquartile range we show throughout our results. The amount of dust extinction in galaxies will be larger at higher redshifts, but we expect this increase to be minor over our redshift range of 0.2 $\leq z \leq$ 1.1 (e.g., Riess et al. 2000; Calzetti 2001). Based on median $A_V$ profiles obtained from a small sample of galaxies ($M_\star > 10^{10} M_\odot$) at higher redshifts ($z \sim 2$; Tacchella et al. 2018), we expect the impact from dust extinction on the evolution measured in galaxy $\mu$ profile outskirts to still be negligible (or fall within the interquartile ranges on our median $\mu_g$ profiles).

We define the stellar halo as the region spanning 2-10$R_e$ (Sec. 4.3), to enable direct comparisons with simulation-based predictions. Recent observations of disk galaxies ($M_\star \geq 10^7 M_\odot$) at $z < 0.09$ and $z \sim 1$ suggest that the inner portion of this range ($\sim$ 2–3$R_e$) may include light from in-situ stellar disks, potentially contributing to some of the stellar halo growth we measure in our SFG sample (e.g., Trujillo et al. 2020; Chamba et al. 2022; Buitrago & Trujillo 2024). Quantifying this contribution from extended disk light would require multi-component photometric decompositions of all SFGs in our sample (e.g., Simard et al. 2011; Meert et al. 2015; Spavone et al. 2021), which is beyond the scope of this study. This contamination is not a concern for QGs as they lack stellar disks.

## 7. SUMMARY AND CONCLUSIONS

We study galaxy evolution and stellar halo assembly over $0.2 \leq z \leq 1.1$ in a large mass-complete sample ($M_\star \geq 10^{9.5} M_\odot$) of 242,456 star-forming (SFG) and 88,421 quiescent (QG) galaxies from the HSC-SSP and CLAUDS surveys. We use deep multi-wavelength photometric observations ($grizy$ broadband filters, rest-frame $\sim$4000-10000 Å) to extract rest-frame $g$-band surface brightness ($\mu_g$) profiles of galaxies, enabling us to trace and analyze the extended emission throughout galaxy outskirts. Under the assumption of constant $M_\star/L$ ratios throughout galaxies, we translate the buildup in galaxy light profiles to the growth in their stellar mass. We study trends in galaxy assembly and light profile evolution by computing median $\mu_g$ profiles for different SFG and QG subpopulations ($M_\star$ and redshift bin combinations, Table 2) and comparing the evolution in their gradients and integrated quantities with decreasing redshift. We investigate the contribution to galaxy stellar halo growth from various stellar mass assembly mechanisms (i.e., in-situ vs. ex-situ stellar mass growth) and use our results to test predictions of merger-driven growth from modern cosmological simulations.

Based on our analysis of galaxy stellar halo assembly over a wider range of redshift and stellar mass than previously studied in the literature, our main finding is that accretion via minor mergers is driving stellar halo growth in QGs of $M_\star \geq 10^{9.5} M_\odot$ and high-mass SFGs ($M_\star \geq 10^{10.5} M_\odot$) over 0.2 $\leq z \leq$ 1.1. The contribution of minor mergers is a function of galaxy stellar mass: more massive galaxies grow larger fractions of their stellar halo mass through these accretions than lower mass systems.

We summarize our results in more detail in the following list:

1. The majority of growth (i.e. increase in luminosity) in the median $\mu_g$ profiles of galaxies over $0.2 \leq z \leq 1.1$ occurs

throughout the extended stellar halo regions (2-10$R_e$) of galaxies (Fig. 7). On average across our full stellar mass range ($M_\star \geq 10^{9.5} M_\odot$), $\sim$64% of the growth in SFGs and $\sim$71% of the growth in QGs since $z = 1.1$ occurs in these outer regions. In SFGs, more of this extended growth is seen in the inner regions of stellar haloes (2-4$R_e$, $\sim$50%) compared to outer stellar halo regions (4-10$R_e$, $\sim$14%). In comparison, QGs exhibit similar growth throughout both subdivided stellar halo regions ($\sim$40% and $\sim$31%).

2. More massive galaxies build up their stellar halo luminosity more rapidly than less massive galaxies over $0.2 \leq z \leq 1.1$ redshift interval in both the SFG and QG populations ($L_{halo}$, Fig. 8). In galaxies of $M_\star \geq 10^{9.5} M_\odot$, QGs assemble a larger fraction of the stellar halo material than SFGs with equivalent stellar mass (factor of $\sim$ 1.2).

3. The surface brightness gradients ($\nabla\mu_g$, Fig. 6) of QGs and SFGs follow different evolutionary trends with redshift throughout their inner galaxy (1-2$R_e$) and stellar halo regions (2-4$R_e$). By $z = 0.2$ more massive QGs exhibit flatter gradients than less massive QGs, while in the SFG sample, this trend with stellar mass is reversed. The gradients of SFGs grow slightly steeper with decreasing redshift over $0.2 \leq z \leq 1.1$ in both regions, with larger changes occurring at lower stellar masses. In contrast, the gradients of high-mass ($M_\star \geq 10^{10.5} M_\odot$) QGs grow flatter with decreasing redshift in both regions, while those of low-mass ($10^{9.5} M_\odot \leq M_\star < 10^{10.5} M_\odot$) QGs grow flatter over time only in the stellar halo region. At $z \sim 0.2$ QGs (of $M_\star \geq 10^{9.5} M_\odot$) exhibit flatter stellar halo gradients than SFGs at fixed $M_\star$. In the inner galaxy regions, this trend is only present in high-mass ($M_\star \geq 10^{10.5} M_\odot$) galaxies.

4. We estimate that stellar mass assembly through in-situ star formation can potentially fully explain the stellar halo growth observed in low-mass SFGs ($10^{9.5} M_\odot \leq M_\star < 10^{10.5} M_\odot$) over $0.2 \leq z \leq 1.1$ (Fig. 9). In contrast, in high-mass SFGs, in-situ growth can account for only $\sim$52% ($10^{10.5} M_\odot \leq M_\star < 10^{11} M_\odot$ bin) and $\sim$23% ($M_\star \geq 10^{11} M_\odot$ bin) of their total stellar mass buildup over our full redshift range. In the QG sample, we estimate that the addition of recently quenched galaxies (i.e. newcomers) to the population can account for $\sim$71%, $\sim$36%, $\sim$10%, and $\sim$15% (low to high $M_\star$ bins) of their observed stellar halo mass growth over $0.2 \leq z \leq 1.1$.

5. Based on comparisons with predictions from cosmological simulations, our results suggest accretion via minor mergers is driving the stellar halo growth in high-mass SFGs ($M_\star \geq 10^{10.5} M_\odot$) and QGs of $M_\star \geq 10^{9.5} M_\odot$ over $0.2 \leq z \leq 1.1$. We find larger contributions from merger-driven accretion (i.e., ex-situ fractions) in more massive galaxies (Fig. 10), particularly in galaxies above the pivot mass ($M_p \sim 10^{10.5} M_\odot$) identified in studies of galaxy size-stellar mass relations. Furthermore, QGs exhibit larger fractional increases to their ex-situ fractions over $0.2 \leq z \leq 1.1$ than do SFGs at fixed $M_\star$ (factor of $\sim$ 1.3).

Our study highlights the benefits of using deep observations from modern large-area imaging surveys such as HSC-SSP to study the faint stellar emission in galaxy outskirts. Our results also demonstrate the effectiveness of using the evolution in median surface brightness profiles to study trends in galaxy assembly over large cosmic timescales. In future work, we will investigate how stellar halo buildup is impacted by a galaxy's cosmic environment. We will identify galaxies within our CLAUDS+HSC-SSP sample that reside in cluster and low-density field environments, and compare the evolution in their surface brightness and rest-frame colour profiles with decreasing redshift.

## ACKNOWLEDGEMENTS

The research of D.J.W., I.D., and M.S. is supported by the Natural Sciences and Engineering Council (NSERC) of Canada. We utilize computational resources from ACENET and The Digital Research Alliance of Canada. We thank members of the extragalactic research group at Saint Mary's University, Canada, for their valuable insights and suggestions.

This work is based on data obtained and processed as part of the CFHT Large Area U-band Deep Survey (CLAUDS), which is a collaboration between astronomers from Canada, France, and China described in Sawicki et al. (2019). CLAUDS data products can be accessed from https://www.clauds.net. CLAUDS is based on observations obtained with MegaPrime/ MegaCam, a joint project of CFHT and CEA/DAPNIA, at the CFHT which is operated by the National Research Council (NRC) of Canada, the Institut National des Science de l'Univers of the Centre National de la Recherche Scientifique (CNRS) of France, and the University of Hawaii. CLAUDS uses data obtained in part through the Telescope Access Program (TAP), which has been funded by the National Astronomical Observatories, the Chinese Academy of Sciences, and the Special Fund for Astronomy from the Ministry of Finance of China. CLAUDS uses data products from TERAPIX and the Canadian Astronomy Data Centre (CADC) and was carried out using resources from Compute Canada and the Canadian Advanced Network For Astrophysical Research (CANFAR).

This paper is also based on data collected at the Subaru Telescope and retrieved from the HSC data archive system, which is operated by the Subaru Telescope and Astronomy Data Center (ADC) at the National Astronomical Observa-tory of Japan. Data analysis was in part carried out with the cooperation of the Center for Computational Astrophysics (CfCA), National Astronomical Observatory of Japan. The Hyper Suprime-Cam (HSC) collaboration includes the astronomical communities of Japan and Taiwan, and Princeton University, USA. The Hyper Suprime-Cam (HSC) collaboration includes the astronomical communities of Japan and Taiwan, and Princeton University. The HSC instrumentation and software were developed by the National Astronomical Observatory of Japan (NAOJ), the Kavli Institute for the Physics and Mathematics of the Universe (Kavli IPMU), the University of Tokyo, the High Energy Accelerator Research Organization (KEK), the Academia Sinica Institute for Astronomy and Astrophysics in Taiwan (ASIAA), and Princeton University.

*Facilities:* Subaru Hyper Suprime-Cam, CFHT MegaPrime / MegaCam

*Software:* NumPy (Harris et al. 2020), SciPy (Virtanen et al. 2020), Astropy (Astropy Collaboration et al. 2013), Photutils (Bradley et al. 2022), PetroFit (Geda et al. 2022), Scikit-learn (van der Walt et al. 2014), Matplotlib (Hunter 2007), GalPRIME (Souchereau et al. in prep.)

# APPENDIX

## A. VARYING MASKING AND BACKGROUND SUBTRACTION PARAMETERS

In this work, we employ `GalPRIME`'s source masking + 2D background subtraction routines when extracting galaxy light profiles (Sec. 3.1.1). To use `GalPRIME`'s feature of performing large batches of light profile extractions in parallel, a configuration file containing a fixed set of masking parameters is required. The choice of parameters used for masking pixels that belong to galaxies will affect the estimation of surface brightness levels in galaxy outskirts. Our choice of fixed parameters (Sec. 3.1.1) is informed by simulation-based tests (Souchereau et al. in prep.). Here we explore the effects that the changes in masking parameters have on extracted light profiles.

In Fig. 12 we highlight galaxies from the extremes of our galaxy sample and demonstrate the differences in their extracted light profiles from using more- or less-severe masking. For each galaxy, we measure the integrated stellar halo luminosity ($L_{halo}$, Sec. 5.4) and compute a relative offset ($\Delta L_{halo}$) between the profile using our chosen fixed parameters (black profiles in Fig. 12) and the profiles using the more-aggressive (gold profiles) and less-aggressive masking (green profiles).

Fig. 12 demonstrates that the relative offsets in $L_{halo}$ of *individual* galaxies are relatively small ($\pm$ $0.01 - 0.1$) when using the most extreme masking parameters compared to our choice of fixed parameters (described in Sec. 3.1.1). We do not find any significant trends between $\Delta L_{halo}$ and galaxy stellar mass, redshift, or star-forming status. Thus, our choice of fixed masking parameters introduces only minor changes in the resulting *median* galaxy light profiles, and does not impact our conclusions or the global trends we report in this study.

## B. PSF CORRECTION PROCEDURE TESTS

### B.1. *Extending HSC-SSP $grizy$ PSF Models*

As discussed in Sec. 3.1.2, we fit a three-component `Astropy` model to the raw HSC-SSP $grizy$ PSFs to extend these PSFs. This extension is required as without an accurate characterization of the PSF at large radial distances, we can significantly underestimate the contribution from the PSF to the apparent $\mu$ levels in galaxy stellar halo regions (e.g., de Jong 2008; Sandin 2014; Trujillo & Fliri 2016; Gilhuly et al. 2022).

The HSC-SSP $grizy$ PSFs obtained from the PSF Picker tool (Aihara et al. 2022) initially start as $42 \times 42$ pixel ($\sim 7'' \times 7''$) images. To accurately measure the contribution from the PSF in galaxy outskirts, we require PSF models that extend to similar sizes as galaxy images (de Jong 2008; Sandin 2014; Trujillo & Fliri 2016; Gilhuly et al. 2022). We extend these initial PSF models to match a given galaxy cutout size ($42'' \times 42''$ for $0.35 < z \leq 1.1$ galaxies, and $50.4'' \times 50.4''$ for $0.2 \leq z < 0.35$ galaxies, see Sec. 3.2) by fitting a combination of a `Moffat2D` and two `Gaussian2D` models. We choose this combination of models as they produce the best fit to the original PSFs in all $grizy$ bands. Figure 13 illustrates the fit and residuals between the three-component model and a raw $i$-band PSF.

We test the fit of this three-component model to the PSFs using 3000 different PSFs from each band (1000 from each Deep field, Sec. 2.3). In each test, we compute the RMSE and relative flux offset between the raw input PSF and the generated model. In Fig. 14 we summarize the results of all tests performed and show the mean flux offset (left panel, blue points) and mean RMSE (right panel, red points) obtained for each $grizy$ filter (separate points in each panel). Our PSF modelling and extension procedure performs equally well in all $grizy$ filters with an average RMSE of $\sim 10^{-4}$ and flux offset of $\sim 1\%$.

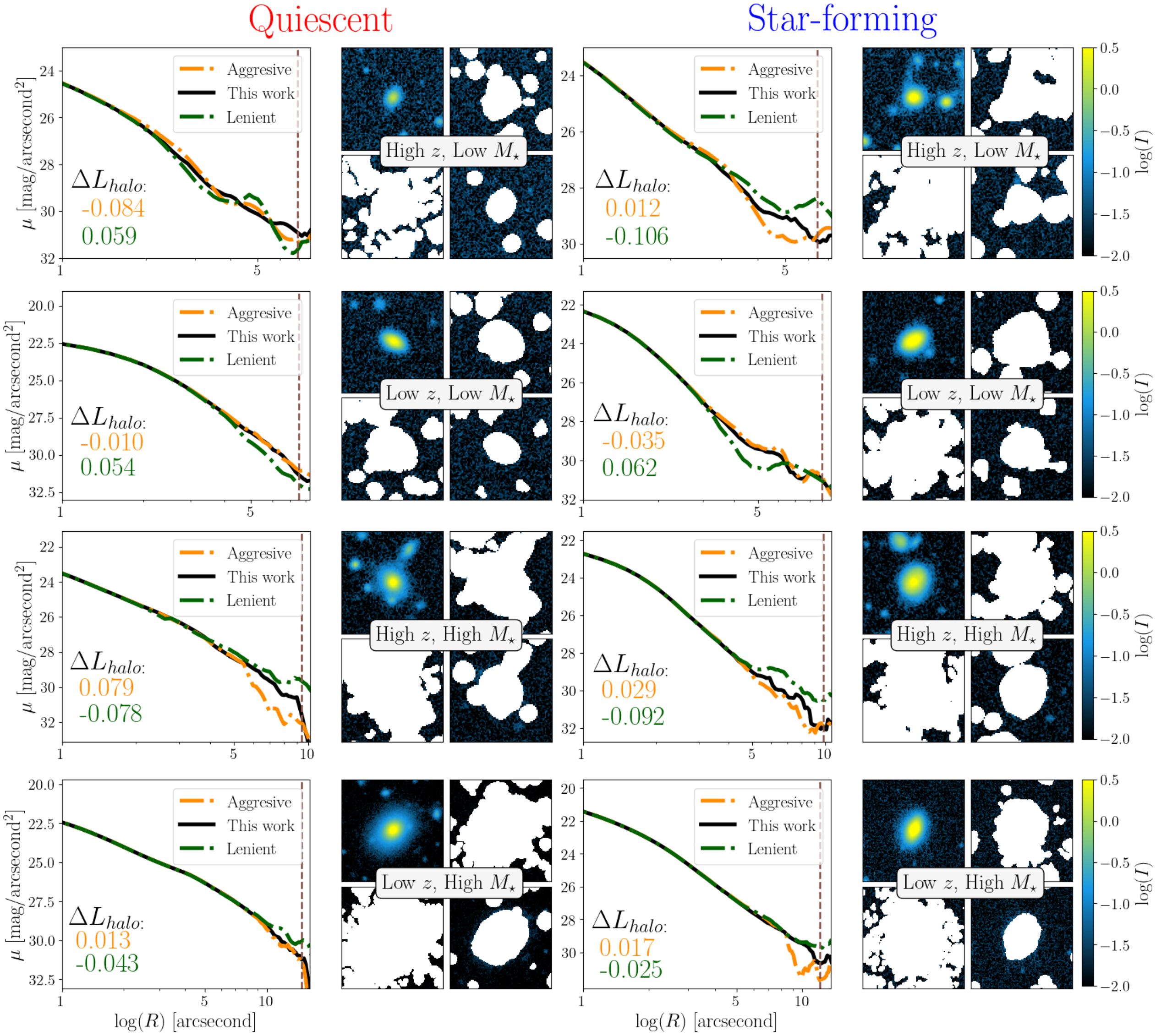

**Figure 12.** Effect on extracted galaxy light profiles from using more- or less-aggressive masking parameters in `GalPRIME`. Left and right columns show QGs and SFGs from the extremes of our galaxy sample, respectively. The top two rows show low-mass galaxy examples from our highest and lowest redshift bins (Table 2). The bottom two rows show high-mass galaxies from the same redshift bins. For each galaxy, we include a $2 \times 2$ matrix of thumbnails that include: the galaxy's image (top left), the image using our chosen masking parameters (top right), and the images using more- and less-aggressive masking (bottom left and right, respectively). To the left of each image matrix, we show the light profiles extracted from the different masked images. Gold and green text in the profile subplots show $\Delta L_{halo}$ (Sec. A) between the profile using our chosen parameters (black) and the profiles using the more-aggressive (gold) and less-aggressive (green) masking. Brown dashed vertical lines represent $10R_e$ for the galaxies.

### B.2. *Testing the PSF Correction Using Simulated Galaxies*

To ensure the robustness of our PSF correction procedure (Sec. 3.1.2), we perform tests using simulated galaxy images instead of real HSC-SSP observations so that we have knowledge of the underlying galaxy $\mu$ profile before the convolution with the PSF. This initial "ground truth" profile can be compared to the final PSF-corrected profile to quantify the performance of the procedure.

We create simulated galaxy models using the `Sersic2D` package from `Astropy`. The input parameters used for these models are based on structural parameters from the 2D bulge+disk models for SDSS galaxies by Simard et al. 2011 (their free $n_b$ bulge + disk catalogue). We first filter the catalogue parameters by apparent magnitude to match that of our CLAUDS+HSC-SSP galaxy sample (i.e. $18 \leq m_i \leq 26$). Next, we perform multidimensional KDE sampling using `KernelDensity` from `sklearn` to create distributions from which we draw input parameters. We convolve each simulated galaxy model with an extended *grizy* PSF (Appendix B.1) and add it to an intrinsic HSC-SSP background cutout in the same photometric band. In Fig. 15 we show examples of one disk- and one bulge-dominated galaxy $\mu$ profile (upper and lower rows, respectively) convolved with a PSF representing each $grizy$ filter (separate columns).

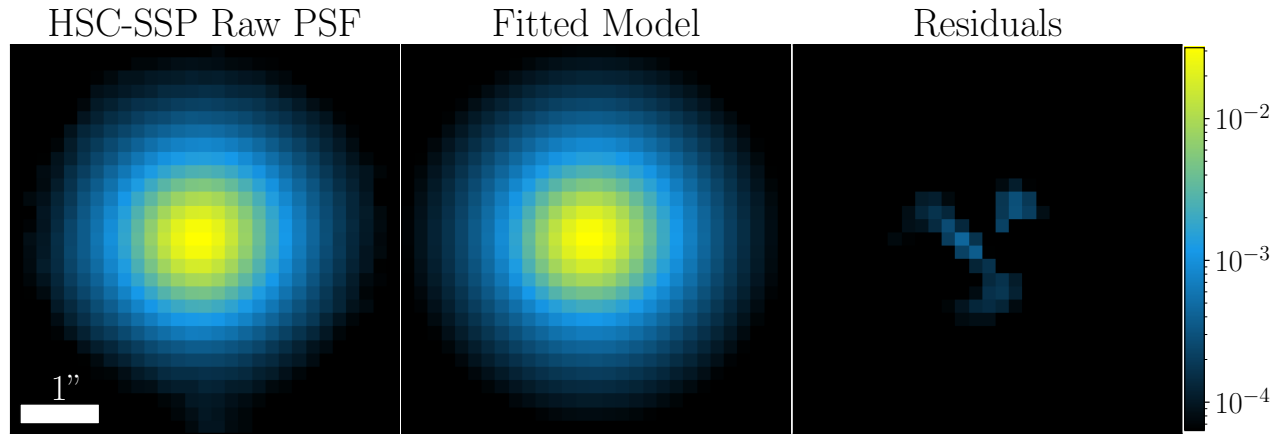

**Figure 13.** Left: Raw $i$-band PSF image obtained from the HSC-SSP PDR3 PSF picker tool (Aihara et al. 2022). Middle: Three-component `Astropy` model which is fit to the input PSF to create an extended PSF model for use in our PSF correction procedure (Sec. 3.1.2). Right: Residuals of the fit between the three-component model and raw $i$-band PSF. . The panel sizes and image scaling have been adjusted to more easily visualize the core of the PSF distribution and the small residuals that remain.

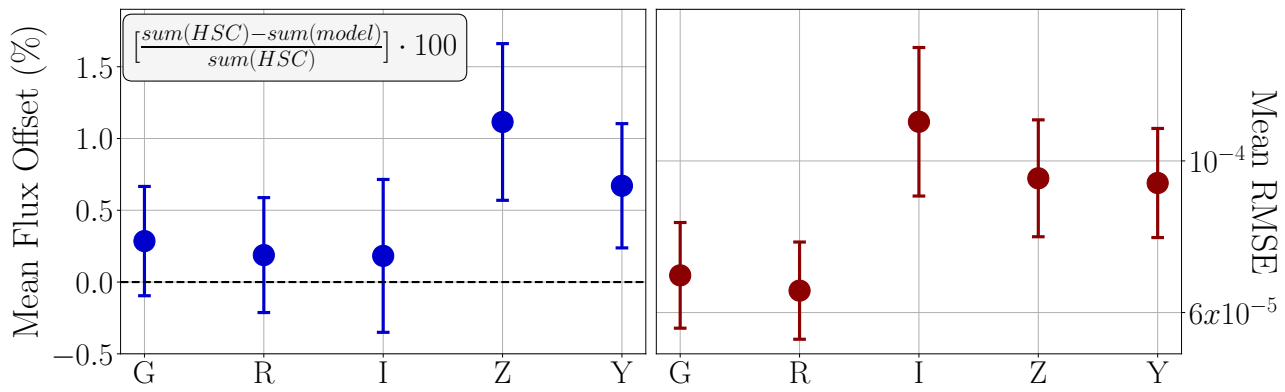

**Figure 14.** Results of testing our PSF modelling and extension procedure using 3000 different PSFs from each $grizy$ band. Left: Mean flux offset between the input and generated PSF models. The grey label shows the formula for calculating y-axis values. Right: Mean RMSE between the input and generated PSF models. Error bars represent 25/75 quartiles.

Based on RMSE values between the true (red) and PSF convolved (black dashed) galaxy $\mu$ profiles in Fig. 15, the offset in $\mu$ due to PSF convolution is comparable between different $grizy$ filters (i.e. when comparing columns along a row). However, there are clear differences in how this light is redistributed across the disk- and bulge-dominated $\mu$ profiles (i.e. when comparing top and bottom rows).

With disk-dominated $\mu$ profiles (top row), there is a smaller amount of light suppressed in inner regions ($R < 1''$) than in the bulge-dominated case (i.e. a smaller change in the central surface brightness). This redistributed light creates substantial outer wings in disk-dominated $\mu$ profiles that are significantly offset from the true light distribution (see differences between profiles in the top row at $R \gtrsim 7''$). With bulge-dominated $\mu$ profiles, the enhancement of outer wings caused by the PSF convolution is less pronounced (i.e. smaller increase in surface brightness), but the offset between the true and PSF convolved profiles is noticeable at $R \gtrsim 2''$.

Using the PSF convolved simulated galaxies as inputs, we then perform the steps outlined in Sec. 3.1.2, and calculate the RMSE between the ground truth and final PSF-corrected $\mu$ profiles obtained. We repeat this test 3000 times for each $grizy$ band (i.e., with 15000 different simulated galaxies), using different galaxy model input parameters in each test. Figure 16 shows the results of the entire set of tests for all simulated galaxies, with 2D histograms showing how RMSE varies with different input galaxy parameters.

We find that our PSF correction procedure performs very well at recovering the true $\mu$ distributions of simulated galaxies across a large sample, resulting in a median RMSE of $\sim 0.10^{+0.06}_{-0.04}$ mag/arcsec$^2$. The procedure achieves similar success in all $grizy$ filters, showing only a minor increase in median RMSE in redder filters (e.g., $\sim$0.086 vs. $\sim$0.121 mag/arcsec$^2$ in HSC-$g$ and HSC-$y$). Without the PSF correction, the extracted $\mu$ profiles of galaxies would be fairly offset from the true light distributions ($\gtrsim$1 mag/arcsec$^2$, see RMSE values in Fig. 15).

To identify any trends between RMSE and particular input galaxy parameters, we compute Pearson correlation coefficients using `pearsonr` from `scipy` and report the resulting statistics and p-values in grey text boxes in each panel of Fig. 16. From these statistical tests, we find very weak correlations between RMSE and $R_e$ ($r \sim$ -0.046), bulge component Sérsic index ($n$; $r \sim$ -0.045), and ellipticity ($r \sim$ +0.049). Based on 90% confidence intervals, errors on these correlation coefficients are $\sim$0.015. We find no correlations between RMSE and position angle ($\phi$), galaxy apparent $g$-band magnitude ($m_g$), or galaxy bulge-to-total ratios ($B/T$).

## C. MEDIAN SIZE COMPARISONS

Here we report the median rest-frame $g$-band effective radii ($R_e$) of our SFG and QG samples. Individual sizes are measured through a curve of growth procedure (Eq. 1) using the $\mu_g$ profiles of individual galaxies. As discussed in Sec. 3.2, we use geometric mean light profiles and thus our galaxy sizes are circularized effective radii.

In Fig. 17 we show our median sizes as a function of redshift with each panel showing a different $M_\star$ bin (Table 2) from our SFG (first column, bluer colours) and QG samples (second column, redder colours). Shown for comparison in each panel as different grey symbols are median circularized galaxy $R_e$ reported by previous studies in the literature.

Our median sizes are in good agreement ($\lesssim$ 1-2$\sigma$) with those from previous studies (e.g., van der Wel et al. 2014; Roy et al. 2018; George et al. 2024) and follow similar trends with decreasing redshift. Small differences in sizes between studies are likely due to varying methodologies used to fit the surface brightness distributions of galaxies and calculate effective radii (e.g., a 1D curve of growth vs. 2D image fitting approach).

We note our sizes are slightly larger (factor of $\sim$1.1-1.4 depending on the galaxy subpopulation) than those in George et al. (2024), who also uses the CLAUDS+HSC-SSP datasets. Some of this size offset may be because the au-thors only study galaxies found in the central 1.6 deg$^2$ of the COSMOS field, where UltraDeep data is available. However, some degree of offset is expected given our inclusion of extended emission from galaxy outskirts in our size measurements. The offset between our sizes and those in George et al. (2024) is larger in more massive galaxies and in QGs over SFGs at fixed $M_\star$. Based on results in our study (e.g., Fig. 10), we interpret these trends as an increasing influence of accreted stellar material in the outskirts of more massive (and quiescent) galaxies.

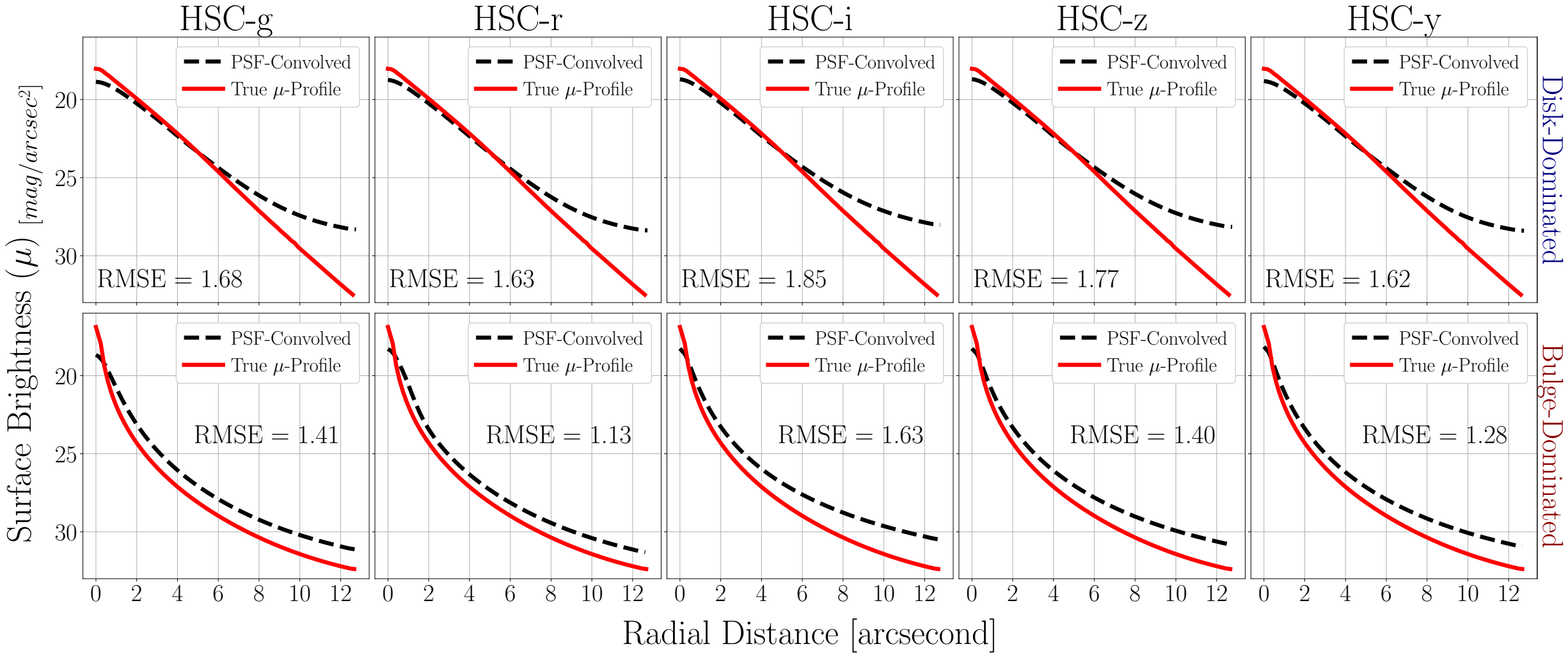

**Figure 15.** Effect of PSF convolution on two different simulated galaxy surface brightness profiles. The top row shows a disk-dominated profile, and the bottom row shows a bulge-dominated profile. Different columns show convolutions with PSFs from different HSC-SSP $grizy$ filters. The intrinsic galaxy profile is shown in red in each panel, and the PSF-convolved profile is in black (dashed).

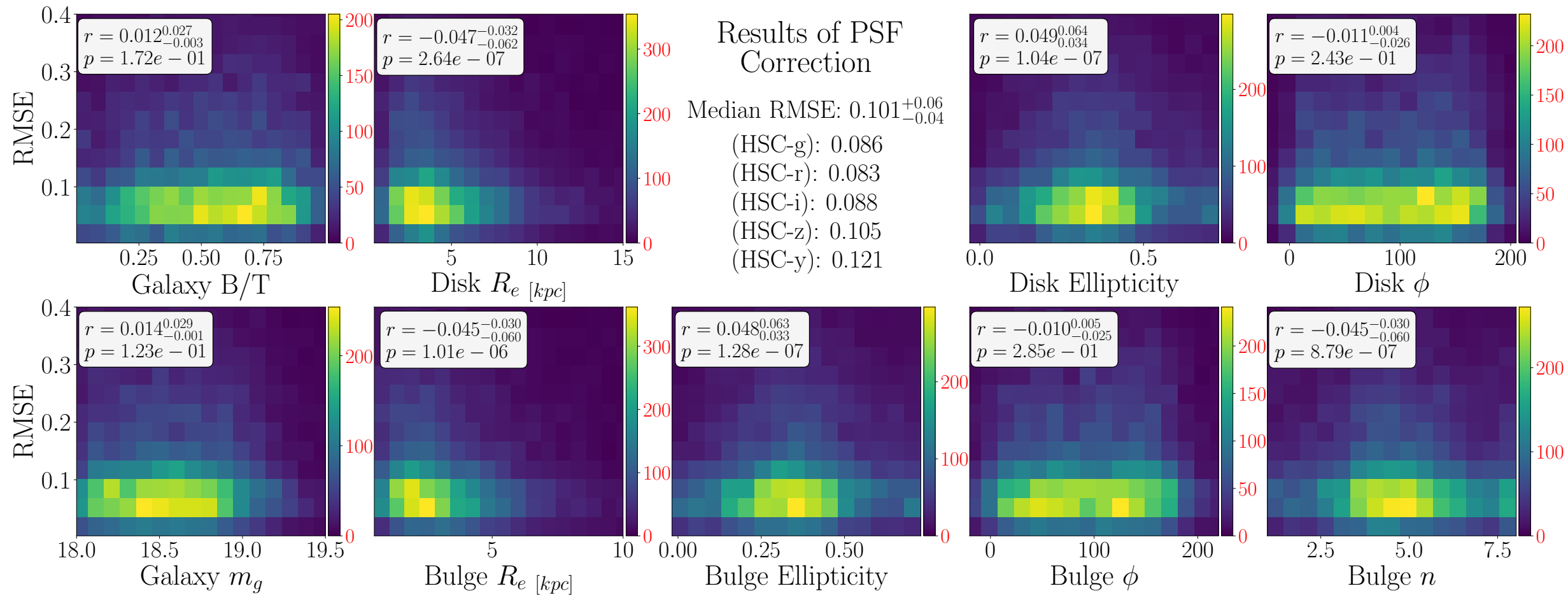

**Figure 16.** Results of the simulation-based tests of our PSF correction procedure. 2D histograms show RMSE between ground truth and corrected surface brightness profiles as a function of different input galaxy parameters (from the catalogue of Simard et al. 2011). Pearson correlation coefficients and p-values are displayed in grey text boxes in each panel. The middle panel in the top row displays the median RMSE obtained across all 15000 simulated galaxy tests, with the median RMSE in different $grizy$ filters listed underneath.

## REFERENCES

Aihara, H., Arimoto, N., Armstrong, R., et al. 2018, PASJ, 70, S4, doi: 10.1093/pasj/psx066

Aihara, H., AlSayyad, Y., Ando, M., et al. 2019, PASJ, 71, 114, doi: 10.1093/pasj/psz103

—. 2022, PASJ, 74, 247, doi: 10.1093/pasj/psab122

Ann, H. B., & Park, H. W. 2018, Journal of Korean Astronomical Society, 51, 73, doi: 10.5303/JKAS.2018.51.4.73

Annunziatella, M., Sajina, A., Stefanon, M., et al. 2023, AJ, 166, 25, doi: 10.3847/1538-3881/acd773

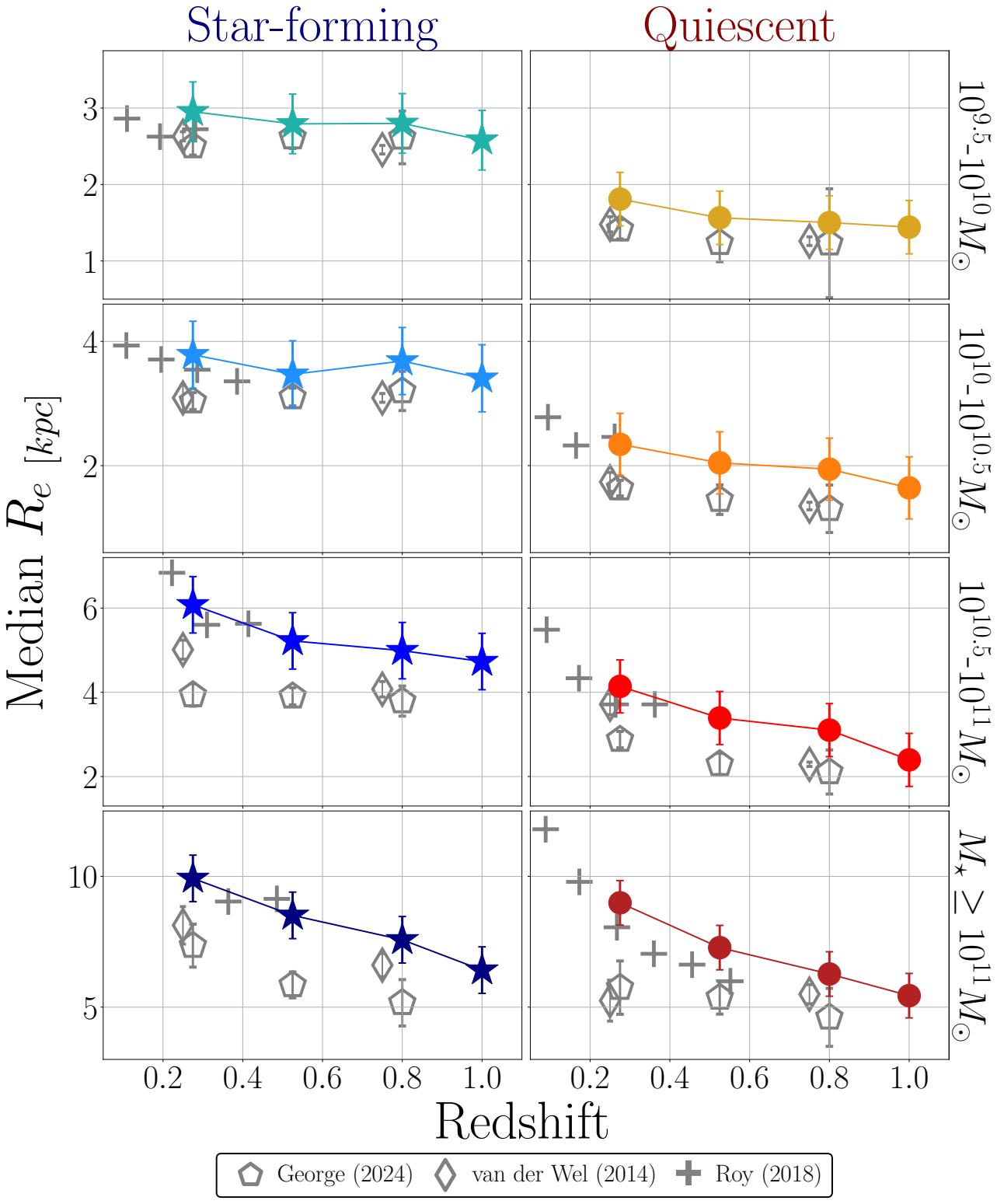

**Figure 17.** Median $R_e$ (rest-frame $g$-band) as a function of redshift for our SFG (left-hand column) and QG (right-hand column) samples. Data points are placed in the center of our four redshift ranges (Table 2). Different rows show different stellar mass ranges (right-hand labels). Error bars represent 25/75 percentiles on median sizes, while bootstrapped errors on median values are smaller than the size of symbols shown. Grey symbols in each panel represent sizes from previous studies shown for comparison - George et al. 2024 (grey pentagons), van der Wel et al. 2014 (grey diamonds), and Roy et al. 2018 (grey plus symbols).

Arnouts, S., Le Floc'h, E., Chevallard, J., et al. 2013, A&A, 558, A67, doi: 10.1051/0004-6361/201321768

Astropy Collaboration, Robitaille, T. P., Tollerud, E. J., et al. 2013, A&A, 558, A33, doi: 10.1051/0004-6361/201322068

Athanassoula, E. 2005, MNRAS, 358, 1477, doi: 10.1111/j.1365-2966.2005.08872.x

Bell, E. F., Naab, T., McIntosh, D. H., et al. 2006, ApJ, 640, 241, doi: 10.1086/499931

Bergin, E. A., & Tafalla, M. 2007, ARA&A, 45, 339, doi: 10.1146/annurev.astro.45.071206.100404

Bernardi, M., Roche, N., Shankar, F., & Sheth, R. K. 2011, MNRAS, 412, L6, doi: 10.1111/j.1745-3933.2010.00982.x

Bezanson, R., van Dokkum, P. G., Tal, T., et al. 2009, ApJ, 697, 1290, doi: 10.1088/0004-637X/697/2/1290

Blanton, M. R., & Roweis, S. 2007, AJ, 133, 734, doi: 10.1086/510127

Boecker, A., Neumayer, N., Pillepich, A., et al. 2023, MNRAS, 519, 5202, doi: 10.1093/mnras/stac3759

Borlaff, A., Eliche-Moral, M. C., Beckman, J. E., et al. 2017, A&A, 604, A119, doi: 10.1051/0004-6361/201630282

Bosch, J., Armstrong, R., Bickerton, S., et al. 2018, PASJ, 70, S5, doi: 10.1093/pasj/psx080

Boselli, A., Fossati, M., & Sun, M. 2022, A&A Rv, 30, 3, doi: 10.1007/s00159-022-00140-3

Bottrell, C., Yesuf, H. M., Popping, G., et al. 2024, MNRAS, 527, 6506, doi: 10.1093/mnras/stad2971

Bouwens, R. J., & Silk, J. 1996, ApJL, 471, L19, doi: 10.1086/310329

Boylan-Kolchin, M., Ma, C.-P., & Quataert, E. 2008, MNRAS, 383, 93, doi: 10.1111/j.1365-2966.2007.12530.x

Bradley, L., Sipőcz, B., Robitaille, T., et al. 2022, astropy/photutils: 1.5.0, 1.5.0, Zenodo, Zenodo, doi: 10.5281/zenodo.6825092

Bruzual, G., & Charlot, S. 2003, MNRAS, 344, 1000, doi: 10.1046/j.1365-8711.2003.06897.x

Buitrago, F., & Trujillo, I. 2024, A&A, 682, A110, doi: 10.1051/0004-6361/202346133

Buitrago, F., Trujillo, I., Curtis-Lake, E., et al. 2017, MNRAS, 466, 4888, doi: 10.1093/mnras/stw3382

Bundy, K., Fukugita, M., Ellis, R. S., et al. 2009, ApJ, 697, 1369, doi: 10.1088/0004-637X/697/2/1369

Calzetti, D. 2001, PASP, 113, 1449, doi: 10.1086/324269

Cannarozzo, C., Leauthaud, A., Oyarzún, G. A., et al. 2022, arXiv e-prints, arXiv:2210.08109. https://arxiv.org/abs/2210.08109

Carollo, C. M., Bschorr, T. J., Renzini, A., et al. 2013, ApJ, 773, 112, doi: 10.1088/0004-637X/773/2/112

Chabrier, G. 2003, PASP, 115, 763, doi: 10.1086/376392

Chamba, N., Trujillo, I., & Knapen, J. H. 2022, A&A, 667, A87, doi: 10.1051/0004-6361/202243612

Chen. 2019, Master's Thesis, Saint Mary's University. https://library2.smu.ca/handle/01/29025

Chen et al. in prep.

Chilingarian, I. V., Melchior, A.-L., & Zolotukhin, I. Y. 2010, MNRAS, 405, 1409, doi: 10.1111/j.1365-2966.2010.16506.x

Conselice, C. J., Mundy, C. J., Ferreira, L., & Duncan, K. 2022, ApJ, 940, 168, doi: 10.3847/1538-4357/ac9b1a

Cook, B. A., Conroy, C., Pillepich, A., Rodriguez-Gomez, V., & Hernquist, L. 2016, ApJ, 833, 158, doi: 10.3847/1538-4357/833/2/158

Cook, M., Barausse, E., Evoli, C., Lapi, A., & Granato, G. L. 2010, MNRAS, 402, 2113, doi: 10.1111/j.1365-2966.2009.15875.x

Cooper, A. P., Cole, S., Frenk, C. S., et al. 2010, MNRAS, 406, 744, doi: 10.1111/j.1365-2966.2010.16740.x

Daddi, E., Renzini, A., Pirzkal, N., et al. 2005, ApJ, 626, 680, doi: 10.1086/430104

Damjanov, I., Sohn, J., Geller, M. J., Utsumi, Y., & Dell'Antonio, I. 2022, arXiv e-prints, arXiv:2210.01129. https://arxiv.org/abs/2210.01129

Damjanov, I., Zahid, H. J., Geller, M. J., et al. 2019, ApJ, 872, 91, doi: 10.3847/1538-4357/aaf97d

Damjanov, I., McCarthy, P. J., Abraham, R. G., et al. 2009, ApJ, 695, 101, doi: 10.1088/0004-637X/695/1/101

Davison, T. A., Norris, M. A., Pfeffer, J. L., Davies, J. J., & Crain, R. A. 2020, MNRAS, 497, 81, doi: 10.1093/mnras/staa1816

de Jong, R. S. 2008, MNRAS, 388, 1521, doi: 10.1111/j.1365-2966.2008.13505.x

De Lucia, G., Hirschmann, M., & Fontanot, F. 2019, MNRAS, 482, 5041, doi: 10.1093/mnras/sty3059

Debattista, V. P., Ness, M., Gonzalez, O. A., et al. 2017, MNRAS, 469, 1587, doi: 10.1093/mnras/stx947

Desprez, G., Picouet, V., Moutard, T., et al. 2023, A&A, 670, A82, doi: 10.1051/0004-6361/202243363

Dey, A., Najita, J. R., Koposov, S. E., et al. 2023, ApJ, 944, 1, doi: 10.3847/1538-4357/aca5f8

Driver, S. P., & Robotham, A. S. G. 2010, MNRAS, 407, 2131, doi: 10.1111/j.1365-2966.2010.17028.x

D'Souza, R., Kauffman, G., Wang, J., & Vegetti, S. 2014, MNRAS, 443, 1433, doi: 10.1093/mnras/stu1194

Dubois, Y., Peirani, S., Pichon, C., et al. 2016, MNRAS, 463, 3948, doi: 10.1093/mnras/stw2265

Elias, L. M., Sales, L. V., Creasey, P., et al. 2018, MNRAS, 479, 4004, doi: 10.1093/mnras/sty1718

Ellison, S. L., Catinella, B., & Cortese, L. 2018, MNRAS, 478, 3447, doi: 10.1093/mnras/sty1247

Ellison, S. L., Mendel, J. T., Patton, D. R., & Scudder, J. M. 2013, MNRAS, 435, 3627, doi: 10.1093/mnras/stt1562

Ellison, S. L., Thorp, M. D., Pan, H.-A., et al. 2020, MNRAS, 492, 6027, doi: 10.1093/mnras/staa001

Euclid Collaboration, Desprez, G., Paltani, S., et al. 2020, A&A, 644, A31, doi: 10.1051/0004-6361/202039403

Franx, M., van Dokkum, P. G., Kelson, D., & Illingworth, G. D. 2000, in ASP Conference Series, Vol. 197, Dynamics of Galaxies, 231

Gan, J.-L., Kang, X., Hou, J.-L., & Chang, R.-X. 2010, Research in Astronomy and Astrophysics, 10, 1242, doi: 10.1088/1674-4527/10/12/005

Geda, R., Crawford, S. M., Hunt, L., et al. 2022, AJ, 163, 202, doi: 10.3847/1538-3881/ac5908

Genina, A., Deason, A. J., & Frenk, C. S. 2023, MNRAS, 520, 3767, doi: 10.1093/mnras/stad397

George, A., Damjanov, I., Sawicki, M., et al. 2024, MNRAS, 528, 4797, doi: 10.1093/mnras/stae154

Géron, T., Smethurst, R. J., Lintott, C., et al. 2024, ApJ, 973, 129, doi: 10.3847/1538-4357/ad66b7

Gilhuly, C., Merritt, A., Abraham, R., et al. 2022, ApJ, 932, 44, doi: 10.3847/1538-4357/ac6750

Giri, G., Barway, S., & Raychaudhury, S. 2023, MNRAS, 520, 5870, doi: 10.1093/mnras/stad474

González Delgado, R. M., García-Benito, R., Pérez, E., et al. 2015, A&A, 581, A103, doi: 10.1051/0004-6361/201525938

Graham, A. W. 2023, MNRAS, 518, 6293, doi: 10.1093/mnras/stac3173

Graham, A. W., & Driver, S. P. 2005, PASA, 22, 118, doi: 10.1071/AS05001

Guedes, J., Callegari, S., Madau, P., & Mayer, L. 2011, ApJ, 742, 76, doi: 10.1088/0004-637X/742/2/76

Gwyn, S. D. J. 2008, PASP, 120, 212, doi: 10.1086/526794

Harris, C. R., Millman, K. J., van der Walt, S. J., et al. 2020, Nature, 585, 357, doi: 10.1038/s41586-020-2649-2

Hickson, P. 1997, ARA&A, 35, 357, doi: 10.1146/annurev.astro.35.1.357

Hilz, M., Naab, T., & Ostriker, J. P. 2013, MNRAS, 429, 2924, doi: 10.1093/mnras/sts501

Hirschmann, M., Naab, T., Ostriker, J. P., et al. 2015, MNRAS, 449, 528, doi: 10.1093/mnras/stv274

Hogg, D. W., Baldry, I. K., Blanton, M. R., & Eisenstein, D. J. 2002, arXiv e-prints, astro. https://arxiv.org/abs/astro-ph/0210394

Hopkins, P. F., Bundy, K., Hernquist, L., Wuyts, S., & Cox, T. J. 2010, MNRAS, 401, 1099, doi: 10.1111/j.1365-2966.2009.15699.x

Huang, S., Ho, L. C., Peng, C. Y., Li, Z.-Y., & Barth, A. J. 2013, ApJL, 768, L28, doi: 10.1088/2041-8205/768/2/L28

Huang, S., Leauthaud, A., Greene, J. E., et al. 2018, MNRAS, 475, 3348, doi: 10.1093/mnras/stx3200

Huertas-Company, M., Tasca, L., Rouan, D., et al. 2009, A&A, 497, 743, doi: 10.1051/0004-6361/200811255

Hunter, J. D. 2007, Computing in Science and Engineering, 9, 90, doi: 10.1109/MCSE.2007.55

Huško, F., Lacey, C. G., & Baugh, C. M. 2022, arXiv e-prints, arXiv:2207.07139. https://arxiv.org/abs/2207.07139

Ilbert, O., McCracken, H. J., Le Fèvre, O., et al. 2013, A&A, 556, A55, doi: 10.1051/0004-6361/201321100

Iodice, E., Capaccioli, M., Grado, A., et al. 2016, ApJ, 820, 42, doi: 10.3847/0004-637X/820/1/42

Iodice, E., Spavone, M., Capaccioli, M., et al. 2017, ApJ, 839, 21, doi: 10.3847/1538-4357/aa6846

Jackson, R. A., Kaviraj, S., Martin, G., et al. 2022, MNRAS, 511, 607, doi: 10.1093/mnras/stac058

Jackson, R. A., Martin, G., Kaviraj, S., et al. 2019, MNRAS, 489, 4679, doi: 10.1093/mnras/stz2440

—. 2020, MNRAS, 494, 5568, doi: 10.1093/mnras/staa970

Jarvis, M. J., Bonfield, D. G., Bruce, V. A., et al. 2013, MNRAS, 428, 1281, doi: 10.1093/mnras/sts118

Jedrzejewski, R. I. 1987, MNRAS, 226, 747, doi: 10.1093/mnras/226.4.747

Jensen, J., Hayes, C. R., Sestito, F., et al. 2024, MNRAS, 527, 4209, doi: 10.1093/mnras/stad3322

Kaviraj, S. 2014, MNRAS, 440, 2944, doi: 10.1093/mnras/stu338
Kaviraj, S., Huertas-Company, M., Cohen, S., et al. 2014, MNRAS, 443, 1861, doi: 10.1093/mnras/stu1220
Kawinwanichakij, L., Silverman, J. D., Ding, X., et al. 2021, ApJ, 921, 38, doi: 10.3847/1538-4357/ac1f21
Khoperskov, S., & Bertin, G. 2017, A&A, 597, A103, doi: 10.1051/0004-6361/201629032
Kormendy, J. 1993, in IAU Symposium, Vol. 153, Galactic Bulges, ed. H. Dejonghe & H. J. Habing, 209
Kormendy, J., & Kennicutt, Robert C., J. 2004, ARA&A, 42, 603, doi: 10.1146/annurev.astro.42.053102.134024
Lambas, D. G., Alonso, S., Mesa, V., & O'Mill, A. L. 2012, A&A, 539, A45, doi: 10.1051/0004-6361/201117900
Lange, R., Driver, S. P., Robotham, A. S. G., et al. 2015, MNRAS, 447, 2603, doi: 10.1093/mnras/stu2467
Li, J., Huang, S., Leauthaud, A., et al. 2021, arXiv e-prints, arXiv:2111.03557. https://arxiv.org/abs/2111.03557
Li, W., Nair, P., Rowlands, K., et al. 2023, MNRAS, 523, 720, doi: 10.1093/mnras/stad1473
Loebman, S. R., Debattista, V. P., Nidever, D. L., et al. 2016, ApJL, 818, L6, doi: 10.3847/2041-8205/818/1/L6
Loebman, S. R., Roškar, R., Debattista, V. P., et al. 2011, ApJ, 737, 8, doi: 10.1088/0004-637X/737/1/8
Man, A., & Belli, S. 2018, Nature Astronomy, 2, 695, doi: 10.1038/s41550-018-0558-1
Martig, M., Bournaud, F., Teyssier, R., & Dekel, A. 2009, ApJ, 707, 250, doi: 10.1088/0004-637X/707/1/250
Martínez-Delgado, D., Peñarrubia, J., Gabany, R. J., et al. 2008, ApJ, 689, 184, doi: 10.1086/592555
Matharu, J. 2019, doi: https://doi.org/10.17863/CAM.44899
McCracken, H. J., Milvang-Jensen, B., Dunlop, J., et al. 2012, A&A, 544, A156, doi: 10.1051/0004-6361/201219507
McKee, C. F., & Ostriker, E. C. 2007, ARA&A, 45, 565, doi: 10.1146/annurev.astro.45.051806.110602
Meert, A., Vikram, V., & Bernardi, M. 2015, MNRAS, 446, 3943, doi: 10.1093/mnras/stu2333
Merrifield, M. R., Gerssen, J., & Kuijken, K. 2001, in ASP Conference Series, Vol. 230, Disk Galaxies, 221–224, doi: 10.48550/arXiv.astro-ph/0008290
Merritt, A., Pillepich, A., van Dokkum, P., et al. 2020, MNRAS, 495, 4570, doi: 10.1093/mnras/staa1164
Merritt, A., van Dokkum, P., Abraham, R., & Zhang, J. 2016a, ApJ, 830, 62, doi: 10.3847/0004-637X/830/2/62
—. 2016b, ApJ, 830, 62, doi: 10.3847/0004-637X/830/2/62
Montenegro-Taborda, D., Rodriguez-Gomez, V., Pillepich, A., et al. 2023, MNRAS, 521, 800, doi: 10.1093/mnras/stad586
Moore, B., Katz, N., Lake, G., Dressler, A., & Oemler, A. 1996, Nature, 379, 613, doi: 10.1038/379613a0
Moutard, T., Malavasi, N., Sawicki, M., Arnouts, S., & Tripathi, S. 2020, MNRAS, 495, 4237, doi: 10.1093/mnras/staa1434
Moutard, T., Sawicki, M., Arnouts, S., et al. 2018, MNRAS, 479, 2147, doi: 10.1093/mnras/sty1543
Moutard, T., Arnouts, S., Ilbert, O., et al. 2016, A&A, 590, A103, doi: 10.1051/0004-6361/201527294
Mowla, L., van der Wel, A., van Dokkum, P., & Miller, T. B. 2019a, ApJL, 872, L13, doi: 10.3847/2041-8213/ab0379
Mowla, L. A., van Dokkum, P., Brammer, G. B., et al. 2019b, ApJ, 880, 57, doi: 10.3847/1538-4357/ab290a
Naab, T., Johansson, P. H., & Ostriker, J. P. 2009, ApJL, 699, L178, doi: 10.1088/0004-637X/699/2/L178
Oser, L., Ostriker, J. P., Naab, T., Johansson, P. H., & Burkert, A. 2010, ApJ, 725, 2312, doi: 10.1088/0004-637X/725/2/2312
Ownsworth, J. R., Conselice, C. J., Mortlock, A., et al. 2014, MNRAS, 445, 2198, doi: 10.1093/mnras/stu1802
Pagel, B. E. J., & Edmunds, M. G. 1981, ARA&A, 19, 77, doi: 10.1146/annurev.aa.19.090181.000453
Patel, S. G., van Dokkum, P. G., Franx, M., et al. 2013, ApJ, 766, 15, doi: 10.1088/0004-637X/766/1/15
Perrotta, S., Coil, A. L., Rupke, D. S. N., et al. 2024, ApJ, 975, 263, doi: 10.3847/1538-4357/ad7b0c
Peschken, N., Łokas, E. L., & Athanassoula, E. 2020, MNRAS, 493, 1375, doi: 10.1093/mnras/staa299
Pillepich, A., Madau, P., & Mayer, L. 2015, ApJ, 799, 184, doi: 10.1088/0004-637X/799/2/184
Pillepich, A., Vogelsberger, M., Deason, A., et al. 2014, MNRAS, 444, 237, doi: 10.1093/mnras/stu1408
Qu, Y., Helly, J. C., Bower, R. G., et al. 2017, MNRAS, 464, 1659, doi: 10.1093/mnras/stw2437
Renzini, A. 2006, ARA&A, 44, 141, doi: 10.1146/annurev.astro.44.051905.092450
Riess, A. G., Filippenko, A. V., Liu, M. C., et al. 2000, ApJ, 536, 62, doi: 10.1086/308939
Rodriguez-Gomez, V., Pillepich, A., Sales, L. V., et al. 2016, MNRAS, 458, 2371, doi: 10.1093/mnras/stw456
Roy, N., Napolitano, N. R., La Barbera, F., et al. 2018, MNRAS, 480, 1057, doi: 10.1093/mnras/sty1917
Saglia, R. P., Sánchez-Blázquez, P., Bender, R., et al. 2016, A&A, Volume 596, id.C1, 3 pp., doi: 10.1051/0004-6361/201014703e
Sandin, C. 2014, A&A, 567, A97, doi: 10.1051/0004-6361/201423429
—. 2015, A&A, 577, A106, doi: 10.1051/0004-6361/201425168
Sawicki, M., Arnouts, S., Huang, J., et al. 2019, MNRAS, 489, 5202, doi: 10.1093/mnras/stz2522
Sérsic, J. L. 1963, Boletin de la Asociacion Argentina de Astronomia La Plata Argentina, 6, 41
Sestito, F., Zaremba, D., Venn, K. A., et al. 2023, MNRAS, 525, 2875, doi: 10.1093/mnras/stad2427
Shi, K., Malavasi, N., Toshikawa, J., & Zheng, X. 2024, ApJ, 961, 39, doi: 10.3847/1538-4357/ad11d7

Simard, L., Mendel, J. T., Patton, D. R., Ellison, S. L., & McConnachie, A. W. 2011, ApJS, 196, 11, doi: 10.1088/0067-0049/196/1/11

Souchereau et al. in prep., GalPRIME. https://pypi.org/project/galprime/

Spavone, M., Krajnović, D., Emsellem, E., Iodice, E., & den Brok, M. 2021, A&A, 649, A161, doi: 10.1051/0004-6361/202040186

Spavone, M., Capaccioli, M., Napolitano, N., et al. 2017, Galaxies, 5, 31, doi: 10.3390/galaxies5030031

Spavone, M., Iodice, E., van de Ven, G., et al. 2020, A&A, 639, A14, doi: 10.1051/0004-6361/202038015

Strateva, I., Ivezić, Ž., Knapp, G. R., et al. 2001, AJ, 122, 1861, doi: 10.1086/323301

Szomoru, D., Franx, M., & van Dokkum, P. G. 2012, ApJ, 749, 121, doi: 10.1088/0004-637X/749/2/121

Szomoru, D., Franx, M., van Dokkum, P. G., et al. 2013, ApJ, 763, 73, doi: 10.1088/0004-637X/763/2/73

—. 2010, ApJL, 714, L244, doi: 10.1088/2041-8205/714/2/L244

Tacchella, S., Carollo, C. M., Förster Schreiber, N. M., et al. 2018, ApJ, 859, 56, doi: 10.3847/1538-4357/aabf8b

Tacchella, S., Diemer, B., Hernquist, L., et al. 2019, MNRAS, 487, 5416, doi: 10.1093/mnras/stz1657

Tal, T., & van Dokkum, P. G. 2011, ApJ, 731, 89, doi: 10.1088/0004-637X/731/2/89

Tissera, P. B., Scannapieco, C., Beers, T. C., & Carollo, D. 2013, MNRAS, 432, 3391, doi: 10.1093/mnras/stt691

Trujillo, I., Aguerri, J. A. L., Cepa, J., & Gutiérrez, C. M. 2001, MNRAS, 321, 269, doi: 10.1046/j.1365-8711.2001.03987.x

Trujillo, I., & Bakos, J. 2013a, MNRAS, 431, 1121, doi: 10.1093/mnras/stt232

—. 2013b, MNRAS, 431, 1121, doi: 10.1093/mnras/stt232

Trujillo, I., Chamba, N., & Knapen, J. H. 2020, MNRAS, 493, 87, doi: 10.1093/mnras/staa236

Trujillo, I., Conselice, C. J., Bundy, K., et al. 2007, MNRAS, 382, 109, doi: 10.1111/j.1365-2966.2007.12388.x

Trujillo, I., Ferreras, I., & de La Rosa, I. G. 2011, MNRAS, 415, 3903, doi: 10.1111/j.1365-2966.2011.19017.x

Trujillo, I., & Fliri, J. 2016, ApJ, 823, 123, doi: 10.3847/0004-637X/823/2/123

van der Walt, S., Schönberger, J. L., Nunez-Iglesias, J., et al. 2014, arXiv e-prints. https://arxiv.org/abs/1407.6245

van der Wel, A., Franx, M., van Dokkum, P. G., et al. 2014, ApJ, 788, 28, doi: 10.1088/0004-637X/788/1/28

van Dokkum, P. G., & Franx, M. 1996, MNRAS, 281, 985, doi: 10.1093/mnras/281.3.985

—. 2001, arXiv, astro, doi: 10.48550/arXiv.astro-ph/0101468

van Dokkum, P. G., Whitaker, K. E., Brammer, G., et al. 2010, ApJ, 709, 1018, doi: 10.1088/0004-637X/709/2/1018

Virtanen, P., Gommers, R., Oliphant, T. E., et al. 2020, Nature Methods, 17, 261, doi: 10.1038/s41592-019-0686-2

Wang, W., Han, J., Sonnenfeld, A., et al. 2019, MNRAS, 487, 1580, doi: 10.1093/mnras/stz1339

Weaver, J. R., Davidzon, I., Toft, S., et al. 2023, A&A, 677, A184, doi: 10.1051/0004-6361/202245581

Whitney, A., Conselice, C. J., Duncan, K., & Spitler, L. R. 2020, ApJ, 903, 14, doi: 10.3847/1538-4357/abb824

Wilkinson, S., Ellison, S. L., Bottrell, C., et al. 2022, MNRAS, 516, 4354, doi: 10.1093/mnras/stac1962

Williams, R. J., Quadri, R. F., Franx, M., van Dokkum, P., & Labbé, I. 2009, ApJ, 691, 1879, doi: 10.1088/0004-637X/691/2/1879

Willmer, C. N. A. 2018, ApJS, 236, 47, doi: 10.3847/1538-4365/aabfdf

Xu, S., Huang, S., Leauthaud, A., et al. 2024, arXiv e-prints, arXiv:2412.03406, doi: 10.48550/arXiv.2412.03406

Zhu, L., Pillepich, A., van de Ven, G., et al. 2022, A&A, 660, A20, doi: 10.1051/0004-6361/202142496